\documentclass{bmvc2k}

\title{Check Your Other Door! Creating Backdoor Attacks in the Frequency Domain}

\addauthor{Hasan Abed Al Kader Hammoud}{hasanabedalkader.hammoud@kaust.edu.sa}{1}
\addauthor{Bernard Ghanem}{bernard.ghanem@kaust.edu.sa}{1}

\addinstitution{
 King Abdullah University of Science and Technology (KAUST)\\
 Thuwal, Saudi Arabia
}

\runninghead{Hammoud, Ghanem}{Check Your Other Door!}

\begin{document}

\maketitle

\begin{abstract}
Deep Neural Networks (DNNs) are ubiquitous and span a variety of applications ranging from image classification to real-time object detection. As DNN models become more sophisticated, the computational cost of training these models becomes a burden. For this reason, outsourcing the training process has been the go-to option for many DNN users. Unfortunately, this comes at the cost of vulnerability to backdoor attacks. These attacks aim to establish hidden backdoors in the DNN so that it performs well on clean samples, but outputs a particular target label when a trigger is applied to the input. Existing backdoor attacks either generate triggers in the spatial domain or naively poison frequencies in the Fourier domain. In this work, we propose a pipeline based on \emph{Fourier heatmaps} to generate a spatially dynamic 
and invisible backdoor attack in the frequency domain. The proposed attack is extensively evaluated on various datasets and network architectures. Unlike most existing backdoor attacks, the proposed attack can achieve high attack success rates with low poisoning rates and little to no drop in performance while remaining imperceptible to the human eye.
Moreover, we show that the models poisoned by our attack are resistant to various state-of-the-art (SOTA) defenses, so we contribute two possible defenses that can evade the attack.
\end{abstract}

\section{Introduction}
\label{sec:intro}

Deep neural networks (DNNs) play a crucial role in various applications such as facial recognition systems \cite{Parkhi2015DeepFR}, medical image analysis \cite{Litjens2017ASO}, autonomous driving \cite{Sallab2017DeepRL}, among others \cite{Graves2013SpeechRW,Jumper2021HighlyAP}. As the tasks become more difficult, the need for more sophisticated and complex models arises. Such models are generally harder to train and might require extensive hyperparameter tuning to achieve the required performance. Recently, and due to the limited access to computational power for most individuals and small companies, \textit{outsourced training} and the use of out-of-the-box pre-trained models became popular \cite{Liu2017NeuralT}.

\begin{figure}[t]
\centering
  \begin{subfigure}[t]{0.18\linewidth}
    \includegraphics[width=\textwidth]{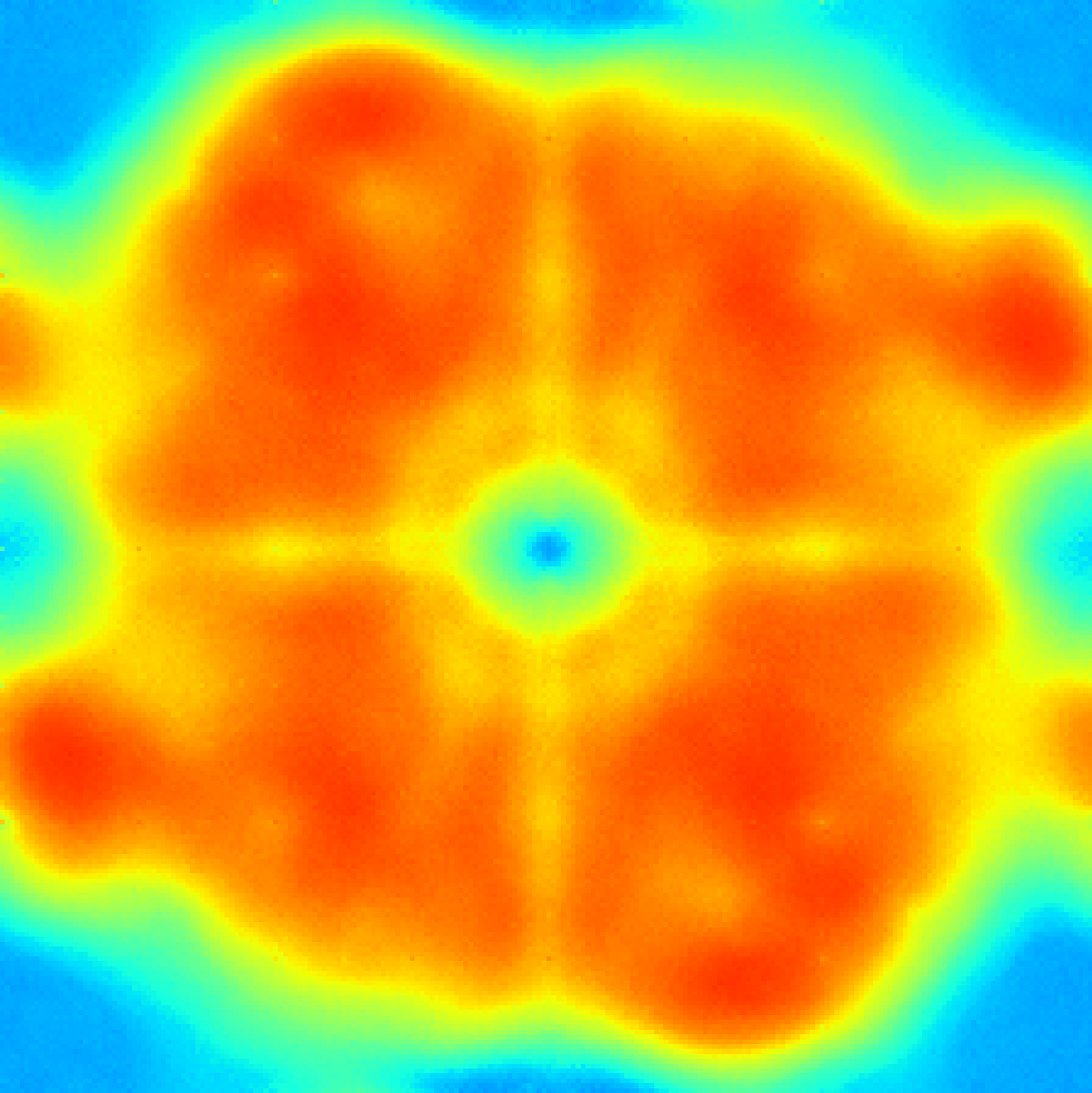}
    \caption{}
    \label{fig-a-pull}
  \end{subfigure}
  \begin{subfigure}[t]{0.18\linewidth}
    \includegraphics[width=\textwidth]{"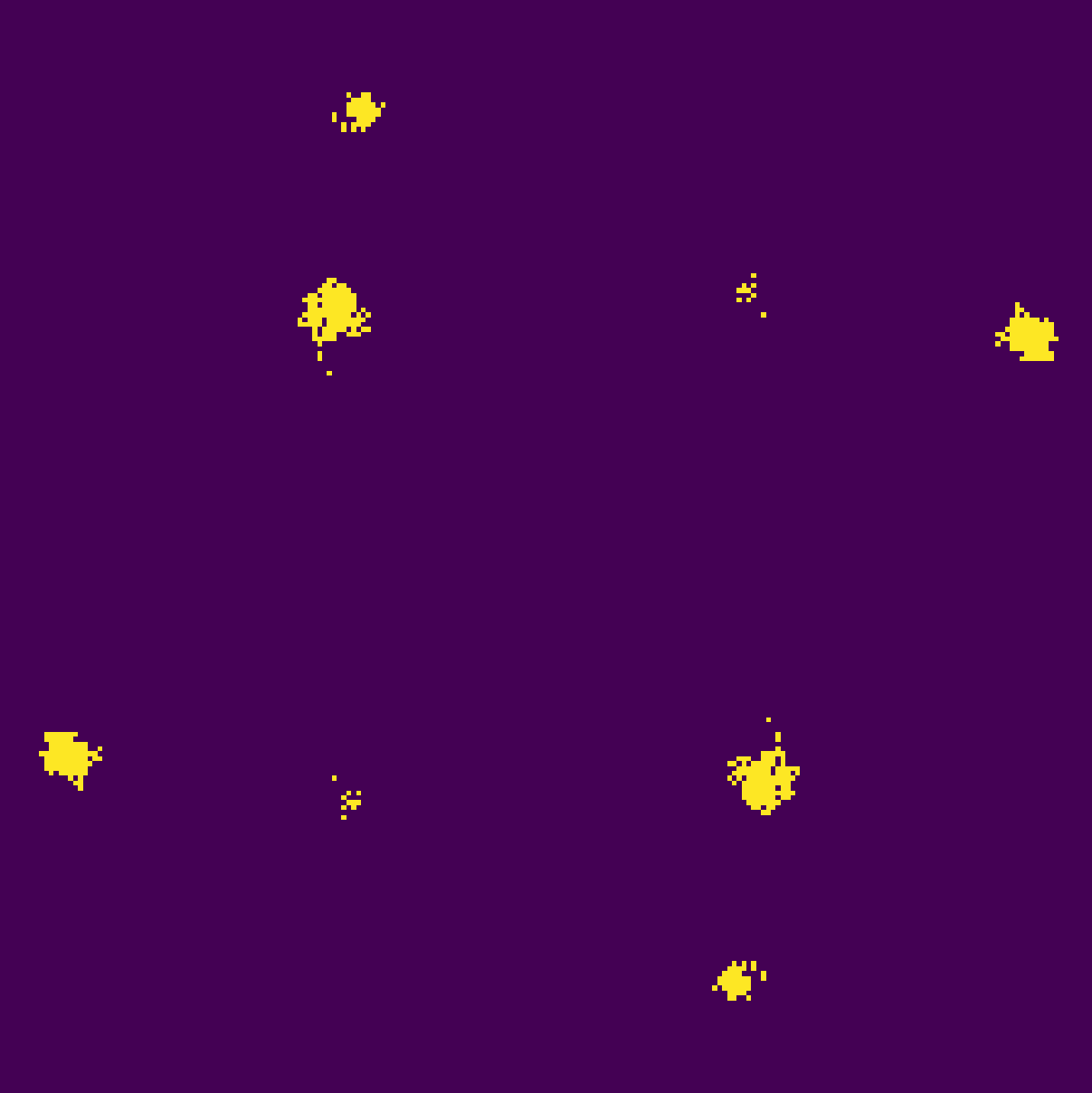"}
    \caption{}
    \label{fig-b-pull}
  \end{subfigure}
  \begin{subfigure}[t]{0.18\linewidth}
    \includegraphics[width=\textwidth]{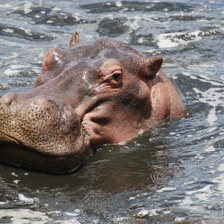}
    \caption{}
    \label{fig-c-pull}
  \end{subfigure}
    \begin{subfigure}[t]{0.18\linewidth}
    \includegraphics[width=\textwidth]{"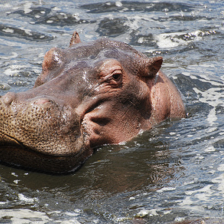"}
    \caption{}
    \label{fig-d-pull}
   \end{subfigure}
      \begin{subfigure}[t]{0.18\linewidth}
    \includegraphics[width=\textwidth]{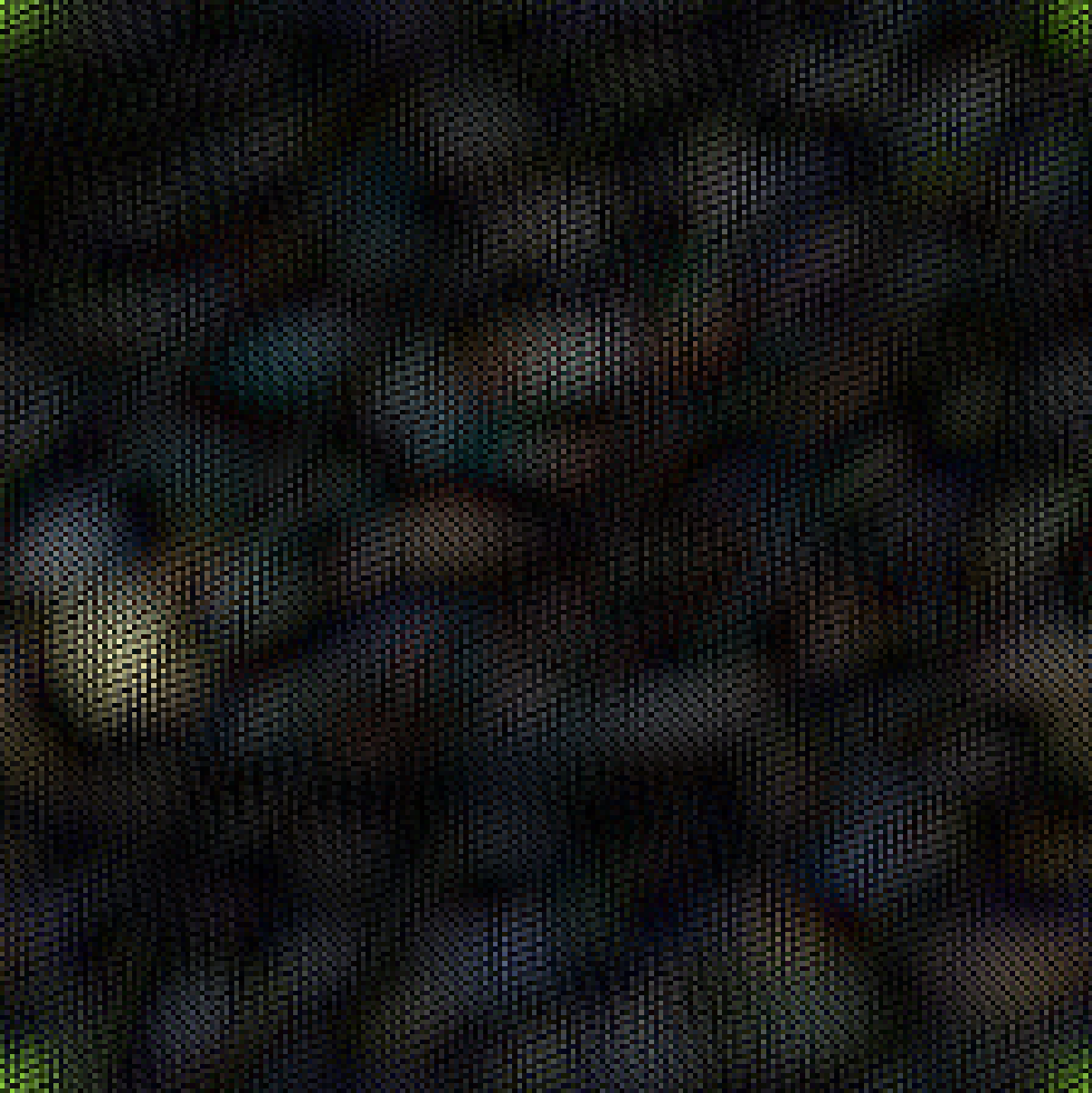}
    \caption{}
    \label{fig-e-pull}
  \end{subfigure}
\vspace{3pt}
\caption{\textbf{Backdoor Attacks in the Frequency Domain.} Frequency-based backdoor attacks exploit the frequency sensitivity of a network, \textit{i.e}~the sensitivity of its performance to variations in individual frequency components in the Fourier domain. Our proposed attacks focus on poisoning the most sensitive frequencies. (a) ResNet50's sensitivity Fourier heatmap (red regions are highly sensitive, while blue regions are less sensitive); (b) Top-$k$ selected frequencies, into which backdoor attacks are embedded; (c) Clean image; (d) Poisoned image;
(e) Scaled absolute difference ($\times 20$) between the poisoned and clean images.
}
\vspace{-0.2cm}
\end{figure}

Outsourced training creates a set of serious vulnerabilities, as it involves several stages that the outsourcer could exploit, including data collection, data pre-processing, and model deployment \cite{Li2020BackdoorLA,Gu2019BadNetsEB,Liu2018TrojaningAO,Gao2020BackdoorAA}. An important threat that could be exploited during training is called \textit{a backdoor attack}. Backdoor attacks create an association between an attacker-defined pattern, called the trigger, and a chosen target label in such a way that the malicious actor can instigate the trigger at will without degrading the model's performance on clean samples. 
This association is usually created through \textit{training data poisoning} \cite{Li2020BackdoorLA,Gu2019BadNetsEB,Liu2018TrojaningAO,Biggio2012PoisoningAA}, where the adversary applies a trigger to a set of images from the training set and then switches their ground truth label to a chosen target class before model training begins. \looseness=-1000

Most existing backdoor attacks \cite{Gu2019BadNetsEB,zhang2021PoisonIR,Chen2017TargetedBA,Barni2019ANB,Liu2020ReflectionBA,Liu2018TrojaningAO,Wenger2021BackdoorAA,Kwon2022BlindNetBA,Zhang2021AdvDoorAB,Yan2021DeHiBDH} rely on the spatial domain to generate and embed the trigger. For example, \cite{Gu2019BadNetsEB} applies a white square stamp on the corner of some training images 
to poison the data. Other methods such as \cite{Liu2018TrojaningAO} rely on an optimization-based approach to generate optimal trigger values. These attacks experience a sharp trade-off between the amount of poisoned data, the invisibility and success of the attack, and the performance of the model on the original task. On the other hand, most backdoor defenses rely on the spatial domain or properties of this domain to detect and mitigate attacks \cite{Guo2019TABORAH,Wang2019NeuralCI,Qiao2019DefendingNB,Gao2019STRIPAD}. Since most backdoor attack techniques tend to be visible and static, \emph{i.e}~the same spatial trigger is applied to all poisoned images, defense techniques in the spatial domain, such as reversed trigger construction \cite{Wang2019NeuralCI,Guo2019TABORAH} and fine-pruning \cite{Liu2018FinePruningDA}, easily succeed in detecting, reverse engineering, and mitigating the embedded backdoor trigger. 

Recently, \cite{Feng2021FIBAFB,Wang2021BackdoorAT} have proposed creating backdoor attacks in the frequency domain. However, both proposed attacks naively select the frequency components to poison.

\vspace{3pt}\noindent
\textbf{Contributions}. Given the weaknesses associated with developing backdoor attacks in the spatial domain and the limitations of existing frequency attacks, in this work, we propose a backdoor attack that utilizes Fourier heatmaps to design a sophisticated backdoor poisoning attack in the frequency domain. Unlike previous attacks, our frequency-based attack does not face the aforementioned trade-offs. We also show two potential ways to defend against frequency-based backdoor attacks and possible ways for the attacker to bypass these defenses. The proposed method is extensively evaluated on multiple models and datasets.
 

\vspace{-10pt}
\section{Related Work}
\label{sec:related}

\noindent \textbf{Backdoor Attacks.} Backdoor attacks were first introduced in \cite{Gu2019BadNetsEB} as a possible security breach that could be exploited in DNNs. They showed that adding a simple patch to the corner of a subset of the training images creates a backdoor that could be maliciously triggered to output a predefined target label. Later, several works were introduced, such as \cite{Liu2018TrojaningAO}, where the values of a predefined mask were optimized to obtain an optimal trigger. On another track, \cite{Chen2017TargetedBA} realized the importance of having invisible or imperceptible triggers to evade possible human inspection. The authors proposed blending the backdoor trigger and the clean images together, replacing the previously used stamping technique. 
Along these lines, other invisible attacks were proposed, such as 
\cite{Li2021InvisibleBA} which used least-significant bit (LSB) algorithm from the steganography literature to generate an invisible attack, \cite{Nguyen2021WaNetI} which utilized image warping to poison data samples, and \cite{zhang2021PoisonIR} which proposed having input-aware trigger patterns that poison the edges of the image. \cite{Doan2021LIRALI} highlighted the importance of learning the trigger-generating transformation to achieve a high attack success rate. \cite{Doan2021BackdoorAW} proposed utilizing the latent space representation to generate imperceptible backdoor triggers by minimizing the Wasserstein distance between the representations of clean and poisoned samples. 
\cite{Li2021InvisibleBA2}, also inspired by steganography, generated sample-specific triggers by encoding an attacker-specified ``string" into clean samples using an autoencoder network. 
\cite{Zeng2021RethinkingTB} analyzed the characteristics of spatial backdoor attacks in the frequency domain and proposed a technique to create smooth but visible spatial backdoor triggers. Recently, \cite{Feng2021FIBAFB,Wang2021BackdoorAT} introduced a simple way to apply trojan attacks in the frequency domain. Specifically, \cite{Feng2021FIBAFB} blends the low-frequency content of a trigger image with that of the target images, and \cite{Wang2021BackdoorAT} arbitrarily poisons a high-frequency component and a mid-frequency component.

\begin{tcolorbox}[colback = BoxBlue, colframe=gray,top=3pt,left=0pt,right=0pt,bottom=3pt]
\textbf{Contribution. }Our work adds to the literature an invisible frequency backdoor attack. Unlike existing frequency backdoor attacks \cite{Feng2021FIBAFB, Wang2021BackdoorAT}, our attack poisons the data by altering \underline{\emph{well-chosen}} frequency components based on the model's frequency sensitivity.  \looseness=-10000
\end{tcolorbox}

\vspace{3pt}\noindent \textbf{Backdoor defenses. } Early defense mechanisms such as fine-pruning \cite{Liu2018FinePruningDA} relied on neuron activations to mitigate backdoors embedded in a DNN. In particular, pruning the least active neurons on clean images and then fine-tuning the model on clean samples can reverse the backdoor attack. 
\cite{Tran2018SpectralSI} and \cite{Chen2019DetectingBA} used robust statistics and analysis of neural network activations, respectively, to thwart and detect backdoor attacks. Later, more sophisticated optimization-based methods, such as Neural Cleanse (NC) \cite{Wang2019NeuralCI}, TABOR \cite{Guo2019TABORAH} and ABS \cite{Liu2019ABSSN}, were developed to mitigate backdoor attacks. NC computes an anomaly index, which indicates whether an abnormally short distance exists between a particular class and all other classes. If the anomaly index exceeds a threshold, NC finds a reverse engineered trigger that is used to fine-tune the model on poisoned but correctly labeled samples. \cite{Doan2020FebruusIP} relied on computing class activation maps using Grad-CAM \cite{Selvaraju2019GradCAMVE} to find the regions the network is attending to in hopes of detecting the attacker-triggered region, which is then replaced through image restoration. \cite{Zheng2021TopologicalDO} adopted persistent homology from topological data analysis to discover structural abnormalities in poisoned models. TOP \cite{Huster2021TOPBD} showed that adversarial perturbations transfer better from image to image in poisoned models compared to clean ones, which can be used to detect poisoned models. STRIP \cite{Gao2019STRIPAD} observes that when a poisoned image is blended with a clean one, the backdoor is still activated, which allows for detecting backdoor attacks by analyzing the entropy of the prediction vectors. SPECTRE \cite{Hayase2021SPECTREDA} explores robust covariance estimation to amplify the spectral signal \emph{i.e} the signature of poisoned data. \looseness=-1

\vspace{-0.1cm}
\begin{tcolorbox}[colback = BoxBlue, colframe=gray,top=3pt,left=0pt,right=0pt,bottom=3pt]
\textbf{Contribution. } Our work proposes two defenses that alter the frequency spectrum of the input, to mitigate the adverse effects of frequency-based backdoor attacks.   \looseness=-10000
\end{tcolorbox}


\vspace{-15pt}
\section{Preliminaries}
\label{sec:preliminaries}

\vspace{-5pt}
To clearly detail our proposed frequency-based approach, we briefly review the concept of \textit{Fourier heatmaps} that was first introduced in \cite{Yin2019AFP}. Fourier heatmaps provide a tool for analyzing the sensitivity of a DNN to a specific Fourier frequency basis by analyzing how this DNN performs when subjected to input perturbations in this basis \cite{Yin2019AFP}.

\noindent \textbf{Notation.} We denote the 2D Discrete Fourier Transform of an image $X \in \mathbb{R}^{d_1 \times d_2}$ by $\mathcal{F}: \mathbb{R}^{d_1 \times d_2} \rightarrow \mathbb{C}^{d_1 \times d_2}$ and its inverse by $\mathcal{F}^{-1}$ (both operations are applied per channel). By default, we assume that the frequency components are shifted toward the center of the Fourier spectrum, \emph{i.e}~low frequencies are set about the center. 

\noindent \textbf{2D Fourier Basis.} Let $\mathcal{U}_{i,j}$ be a real valued matrix in $\mathbb{R}^{d_1
\times d_2}$ with the following properties. \textbf{(1)} It has a Frobenius norm $\norm{\mathcal{U}_{i,j}}_F = 1$; \textbf{(2)}  $\mathcal{F}(\mathcal{U}_{i,j})$ has up to two non-zero elements located at $(i,j)$ and its conjugate symmetric component (symmetric relative to the origin of the spectrum). We refer to such a matrix $\mathcal{U}_{i,j}$ as a 2D Fourier basis at $(i,j)$.

\noindent \textbf{Fourier Heatmaps.} We denote a batch of $B$ images as $\mathcal{I}$, the Fourier basis perturbation factor by $\alpha$, and a uniformly and randomly sampled matrix from $\{-\mathds{1},\mathds{1}\}$ by $\mathbf{r}$, where $\mathds{1}$ is the matrix of all ones in $\mathbb{R}^{d_1\times d_2}$. Let $\Tilde{\mathcal{I}}$ denote the perturbed batch of images, where
    $\Tilde{\mathcal{I}} = \mathcal{I} + \alpha (\mathbf{r} \odot \mathcal{U}_{i,j})$
, $\odot$ is the Hadamard product. Note that the addition is performed across all channels of images in the batch. 
To measure the sensitivity of a classification DNN to the frequency basis at $(i,j)$, we forward pass the perturbed batch $\Tilde{\mathcal{I}}$ through the DNN and compute its output error rate w.r.t.~the ground truth image labels for the specified $(i,j)$ basis. When repeated for all $(i,j)$ pairs, we can visualize the DNN's sensitivity to all 2D Fourier bases through a matrix denoted as a Fourier heatmap \cite{Yin2019AFP} (see Figure \ref{fig-a-pull} for an example).


\vspace{-15pt}
\section{Proposed Method}
\label{sec:ourmethod}


\vspace{-5pt}

Following \cite{Gu2019BadNetsEB,Xiong2020EscapingBA,Salem2020DynamicBA,Chen2020BadNLBA,Zhang2021BackdoorAT,Liu2018TrojaningAO}, we consider the threat model, in which the victim outsources the training process to a trainer that has access to: \textbf{(1)} the victim's network architecture and \textbf{(2)} their training dataset. The victim accepts the model provided by the adversary if its classification accuracy on the validation set is satisfactory.

\begin{figure*}[t]
    \centering
    \includegraphics[width=0.9\textwidth]{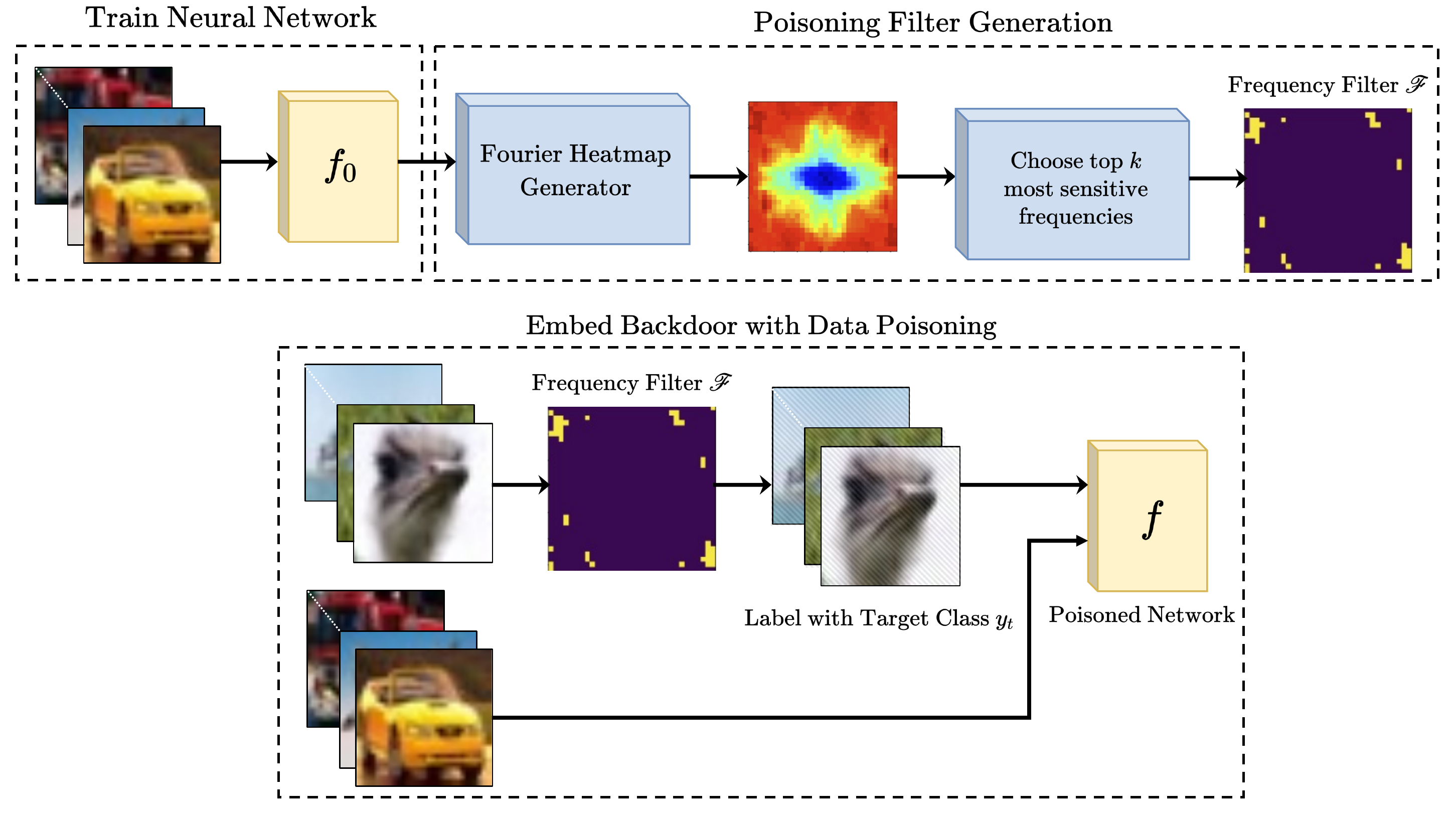}\vspace{-6pt}
    \caption{\textbf{Pipeline.} We illustrate the pipeline for our proposed frequency-based data poisoning method. After training a network naturally, the Fourier heatmap for this model is generated and the top-$k$ most sensitive frequencies are selected as a poisoning filter. This filter is then used to poison a subset of the training dataset before training the poisoned model.
    }
    \label{fig:pipeline}
\end{figure*}

Now we provide a detailed explanation of the proposed frequency-based backdoor attack pipeline. As explained in Section \ref{sec:preliminaries}, Fourier heatmaps provide a tool for analyzing the sensitivity of a DNN to input perturbations in particular 2D Fourier bases. Knowledge of the network's sensitive frequencies allows the attacker to design an attack that exploits these frequencies to embed a frequency-based backdoor that maintains a good performance on the original classification task, embeds a strong backdoor trigger that activates the target class at will, and is both invisible and achievable with small poisoning rates. 
Figure \ref{fig:pipeline} visualizes the proposed pipeline. Below we summarize the recipe for creating frequency-based backdoors. 

\vspace{3pt} \noindent \textbf{Stage 1: Poisoning Filter Generation: } \textbf{1.}  Train a neural network on the clean dataset and the architecture provided by the user. We denote this clean network by $f_0$.
\textbf{2.} Generate the Fourier heatmap for $f_0$ and store the indices of the top-$k$ most sensitive frequencies, $\mathbb{I}_k$, and then generate a binary mask $\mathcal{M}$ as shown in equation \ref{eqn2}.
    \textbf{3.} Generate three additive frequency masks one for each channel ($\mathcal{A}_R$, $\mathcal{A}_G$ and $\mathcal{A}_B$) as shown in equation \ref{eqn3}.
    The values for additive masks $\mathcal{A}_{i,j}$ for $(i,j) \in \mathbb{I}_k $ should be selected to satisfy the 
    invisibility requirement at hand (discussed later in Section \ref{Sec:OurAttack}). For a simple yet flexible design, we set the nonzero values in any individual additive mask to be the same, but different from one mask to another.

\noindent\begin{minipage}{.5\linewidth}
  \begin{align}\label{eqn2}
    \mathcal{M}_{i,j} = \begin{cases} 1 \quad (i,j) \in \mathbb{I}_k \\0 \quad \text{otherwise}
    \end{cases}
  \end{align}
\end{minipage}%
\begin{minipage}{.5\linewidth}
  \begin{align}\label{eqn3}
    \mathcal{A}_{\{R,G,B\}_{i,j}}  \begin{cases} \neq 0 \quad (i,j) \in \mathbb{I}_k \\ = 0 \quad \text{otherwise}
    \end{cases}
  \end{align}
\end{minipage}

\vspace{3pt}\noindent\textbf{Stage 2: Creating the Backdoor through Data Poisoning:} 
\textbf{1.} Specify a set of samples to poison and denote it by $\mathcal{I}_P$. The cardinality of $\mathcal{I}_P$ is denoted by $|\mathcal{I}_P|$ and refers to the number of poisoned samples. The poisoning rate is defined as the ratio of the number of poisoned samples to the total number of samples in the training set. \textbf{2.} For each sample $\mathcal{S} \in \mathcal{I}_P$, and for each channel, apply the following operations: 
    \begin{align}
    \mathcal{S}_{\{R,G,B\}}  := \mathcal{F}^{-1}( \mathcal{F} (\mathcal{S}_{\{R,G,B\}}) \odot (\mathds{1}-\mathcal{M}) + \mathcal{A}_{\{R,G,B\}})
    \end{align}   

\vspace{-0.2cm}
    
    \noindent where each channel is treated separately.
\textbf{3.} Change the label of the samples in $\mathcal{I}_P$ to the specific target label $t$. 
\textbf{4.} Proceed with training the neural network on the poisoned training dataset to obtain a backdoored or poisoned model $f$.

It should be noted that the operations carried out on the Fourier transformed channels could be thought of as simply changing the values of the components of the top-$k$ most sensitive 2D Fourier bases by different values that carry the poisoning information. This could be thought of as a frequency-based version of spatial trigger stamping. 
Section \ref{sec:ablation} discusses the importance of choosing the top-$k$ values rather than random or bottom-$k$ elements. The supplementary material contains variants of the proposed method. It includes experiments, where additive masks have \textbf{(1)} varying random values for each channel and \textbf{(2)} the same values across all channels. We also consider adopting a binary mask $(\mathcal{M})$ generated for one architecture and applying it as a poisoning mask for another. Additionally, we discuss two possible variations of the pipeline that \textbf{(1)} extend the applicability of our attack to the multi-target attack regime; \textbf{(2)} allow for an efficient end-to-end frequency backdoor attack.



\begin{figure*}[t!]
 \vspace{-3pt}
\centering
  \includegraphics[width=0.8\textwidth]{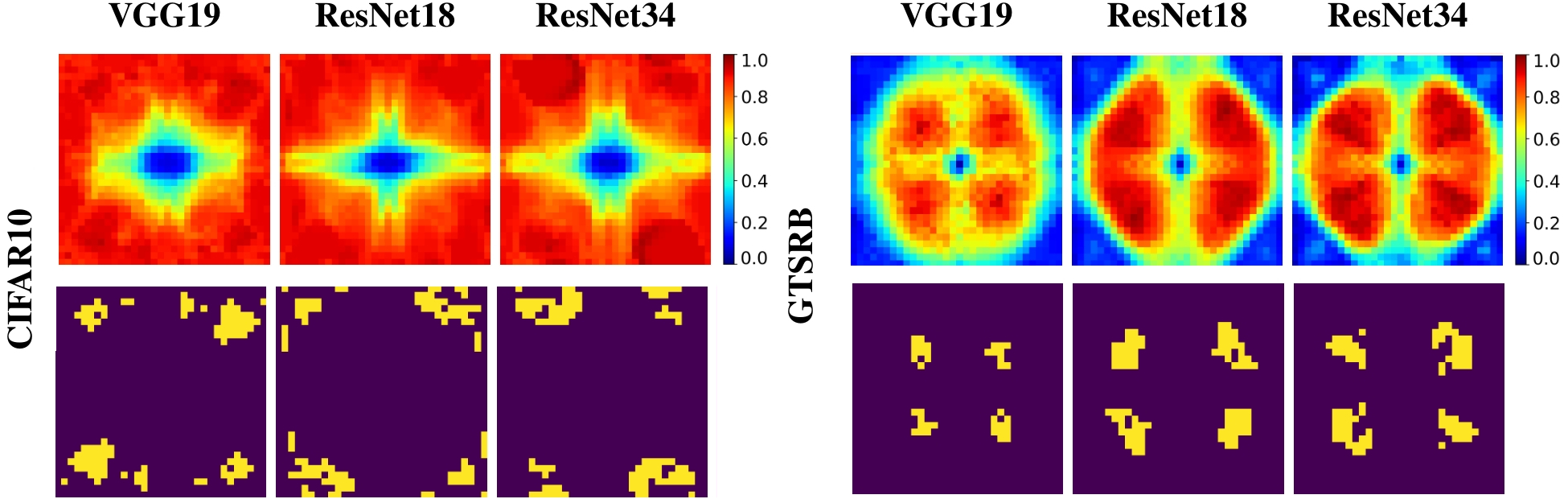}

  \caption{\textbf{Fourier Heatmaps and Top-$k$ Masks.} Rows 1 shows the heatmaps of various architectures trained on CIFAR10 and GTSRB, respectively. Rows 2 shows the respective binary mask ($\mathcal{M}$), which defines the $k$ most sensitive frequencies in the heatmap. } 
  \label{fig:Masks_GTSRB}
  \vspace{-18pt}
\end{figure*}
\section{Experiments}
\label{sec:experiments}

\vspace{-0.3cm}
In this section, we present the details of our implementation and experiments to evaluate our proposed attack mechanism on various datasets and network architectures. Afterwards, we evaluate our attacked models against three state-of-the-art defenses (three more are found in the supplementary). Finally, we show two defenses against frequency-based backdoor attacks and potential ways for the attacker to defend against them.

\subsection{Implementation Details}

\ \begin{wraptable}[24]{r}{0.4\textwidth}
\renewcommand{\arraystretch}{0.95}
\centering
\vspace{-1cm}
\scalebox{0.7}{
\begin{tabular}{@{}lccccccccc}
\specialrule{1.25pt}{0pt}{1pt}
                                    & \textbf{Poisoning Rate}  & \textbf{CDA(\%)} & \textbf{ASR(\%)}                            \\ \hline
                                    & 0.0\%                   & 93.92            & \cellcolor[HTML]{EFEFEF}-        \\ \hhline{~---------} 
                                    & 0.1\%                   & 94.00            & \cellcolor[HTML]{EFEFEF}1.54     \\ \hhline{~---------} 
                                    & 0.2\%                   & 94.14            & \cellcolor[HTML]{EFEFEF}72.31    \\\hhline{~---------}
                                    & 0.4\%                   & 94.20            & \cellcolor[HTML]{EFEFEF}85.05    \\ \hhline{~---------} 
\multirow{-5}{*}{\rotatebox{90}{\textbf{CIFAR10}}}  & 1.0\%                   & 94.38            & \cellcolor[HTML]{EFEFEF}99.44 \\ \hhline{~---------}
                                    & 3.0\%                   & 94.31            & \cellcolor[HTML]{EFEFEF}99.79   \\ \hline
                                    & 0.0\%                   & 75.95            & \cellcolor[HTML]{EFEFEF}-     \\ \hhline{~---------} 
                                    & 0.1\%                   & 75.76            & \cellcolor[HTML]{EFEFEF}60.57    \\ \hhline{~---------} 
                                    & 0.2\%                   & 75.75            & \cellcolor[HTML]{EFEFEF}92.78     \\ \hhline{~---------} 
                                    & 0.4\%                   & 75.92            & \cellcolor[HTML]{EFEFEF}96.49    \\\hhline{~---------} 
\multirow{-5}{*}{\rotatebox{90}{\textbf{CIFAR100}}} & 1.0\%                   & 76.05            & \cellcolor[HTML]{EFEFEF}98.99     \\ \hhline{~---------}
                                    & 3.0\%                   & 75.36            & \cellcolor[HTML]{EFEFEF}99.93   \\ \hline
                                    & 0.0\%                   & 97.11            & \cellcolor[HTML]{EFEFEF}-       \\ \hhline{~---------} 
                                    & 0.1\%                   & 97.09            & \cellcolor[HTML]{EFEFEF}71.12    \\ \hhline{~---------} 
                                    & 0.2\%                   & 97.19            & \cellcolor[HTML]{EFEFEF}89.59  \\  \hhline{~---------}
                                    & 0.4\%                   & 97.33            & \cellcolor[HTML]{EFEFEF}98.04    \\ \hhline{~---------} 
\multirow{-5}{*}{\rotatebox{90}{\textbf{GTSRB}}}    & 1.0\%                   & 97.25            & \cellcolor[HTML]{EFEFEF}98.62   \\ \hhline{~---------} 
                                    & 3.0\%                   & 97.47            & \cellcolor[HTML]{EFEFEF}99.80    \\ \hline
                                    & 0.0\%                   & 67.51            & \cellcolor[HTML]{EFEFEF}-      \\ \hhline{~---------} 
                                    & 0.5\%                   & 67.38            & \cellcolor[HTML]{EFEFEF}0.17    \\ \hhline{~---------} 
                                    & 1.0\%                   & 67.13            & \cellcolor[HTML]{EFEFEF}87.74   \\ \hhline{~---------} 
\multirow{-3}{*}{\rotatebox{90}{\textbf{\ ImageNet}}} & 2.0\%                   & 67.26            & \cellcolor[HTML]{EFEFEF}98.01   \\ \hhline{~---------} 
                                    & 3.0\%                   & 67.26            & \cellcolor[HTML]{EFEFEF}98.32    \\ \bottomrule
\end{tabular}}
\vspace{6pt}
 \caption{\textbf{Evaluation of the proposed backdoor attack.} We benchmark our proposed attack for ResNet18 trained on various datasets and poisoning rates. Our attack can maintain CDA, while registering high ASR even with small poisoning rates \textit{( full table in suppl.). }
 }
 \label{Table:Main}
\end{wraptable}
\noindent Following \cite{zhang2021PoisonIR,Nguyen2021WaNetI,Doan2021LIRALI,Schwarzschild2021JustHT} we evaluate our attack on various datasets, network architectures, and poisoning rates.


\noindent \underline{\textcolor{Klein_Blue}{\emph{\textbf{Datasets.}}}} We evaluate our proposed pipeline on commonly used datasets: CIFAR10 \cite{Krizhevsky2009LearningML}, CIFAR100 \cite{Krizhevsky2009LearningML}, GTSRB \cite{Houben-IJCNN-2013}, and ImageNet \cite{ILSVRC15}. \underline{\textcolor{Klein_Blue}{\emph{\textbf{Network Architectures.}}}} We study six network architectures of different complexity: ResNet18, ResNet34, ResNet50 \cite{He2016DeepRL}, DenseNet121 \cite{Huang2017DenselyCC}, VGG19 \cite{Simonyan2015VeryDC}, and WideResNet34 \cite{Zagoruyko2016WideRN}. \underline{\textcolor{Klein_Blue}{\emph{\textbf{Network}}}} \underline{\textcolor{Klein_Blue}{\emph{\textbf{Performance Metrics.}}}} To evaluate the performance of backdoored models, we use two common metrics: Clean Data Accuracy (CDA), which measures the performance of the network on clean samples, and Attack Success Rate (ASR), which measures the effectiveness of the backdoor attack in triggering the target label.  \underline{\textcolor{Klein_Blue}{\emph{\textbf{Invisibility Metrics.}}}} Following other papers \cite{zhang2021PoisonIR,Li2021InvisibleBA,Nguyen2021WaNetI,Wang2021BackdoorAT,Liu2020ReflectionBA,Hu2021ArtificialIS}, we evaluate the invisibility of the proposed attack using three metrics: Peak Signal-to-Noise-Ratio (PSNR), Structural SIMilarity (SSIM), and Learned Perceptual Image Patch Similarity (LPIPS). Invisibility is a crucial metric for backdoor attacks, as it is required to fool any possible human inspection that may detect the applied trigger.



\subsection{Frequency-Based Backdoor Attacks}
\label{Sec:OurAttack}
\ 

\noindent \textbf{Backdoored Network Performance.}
As discussed in Section
\ref{sec:ourmethod}, we first train baseline networks on each dataset and compute the corresponding Fourier heatmaps and binary masks. The accuracies of the baseline models ($f_0$) are shown in Table \ref{Table:Main} (0\% Poisoning Rate). The heatmaps and masks for various architectures trained on CIFAR10 and GTSRB are shown in Figure \ref{fig:Masks_GTSRB}, respectively. The remaining filters and heatmaps are provided in the supplementary material. In our experiments, the choice of $k$, which defines the number of nonzero indices of $\mathcal{M}$ and the corresponding values for the additive masks $\mathcal{A}_{\{R,G,B\}}$, is made such that: \textbf{(1)} the $\ell_2$ norm of the attack (\emph{i.e}~the $\ell_2$ norm of the absolute difference of the image before and after poisoning) does not exceed, on average, a threshold $\delta_P$ ($\delta_P = 2.0$ for ImageNet and $\delta_P = 1.0$ for all other datasets), and \textbf{(2)} the invisibility metrics (PSNR, SSIM, LPIPS) reach satisfactory values. Table \ref{Table:Main} shows the CDA of the backdoored model ($f$) and the ASR of frequency-based triggers for CIFAR10, CIFAR100, GTSRB, and ImageNet and for ResNet18 with different poisoning rates. Similar to \cite{zhang2021PoisonIR}, we also highlight the effect of changing the poisoning rate on the CDA and ASR metrics. As observed, even with a low poisoning rate, we can embed a backdoor attack with a high ASR with little or no drop in CDA. The target label was arbitrarily chosen as the first class of each dataset. Since the datasets are class-balanced, any target label will lead to a similar performance. \begin{wraptable}[16]{r}{0.4\linewidth}
\centering
\vspace{0.25cm}
\renewcommand{\arraystretch}{0.95}
\scalebox{0.65}{
\begin{tabular}{@{}llll@{}}
\toprule
\textbf{Method} & \textbf{PSNR$\uparrow$} & \textbf{SSIM$\uparrow$} & \textbf{LPIPS$\downarrow$} \\ \hline
BadNets \cite{Gu2019BadNetsEB}         & 27.03         & 0.9921        & 0.0149        \\
Blend \cite{Chen2017TargetedBA}        & 19.18         & 0.7291        & 0.2097        \\
SIG \cite{Barni2019ANB}      & 25.12         & 0.8988        & 0.0532        \\
Refool \cite{Liu2020ReflectionBA}            & 16.59         & 0.7701        & 0.2461        \\ 
SPM     \cite{Liao2020BackdoorEI}       & 38.65        & 0.9665        & 0.0022        \\ 
Poison Ink \cite{zhang2021PoisonIR}            & 41.62         & 0.9915        & 0.0020        \\ 
FTrojan \cite{Wang2021BackdoorAT} & 44.87 & 0.9942 & 0.0005\\
FIBA \cite{Feng2021FIBAFB} & 18.05 & 0.8077 & 0.1113\\
\hline
Ours (ResNet18) & 47.26         & \textbf{0.9998}        & 0.0006        \\ 
Ours (ResNet34) & 47.55         & \textbf{0.9998}        & 0.0004        \\ 
Ours (ResNet50) & 46.90         & \textbf{0.9998}        & 0.0009       \\ 
Ours (DenseNet121) & 47.21        & \textbf{0.9998}        & \textbf{0.0001}\\
Ours (VGG19) & 46.19        & \textbf{0.9998}        & 0.0008\\ 

\hline
\end{tabular}}
\vspace{4pt}
\caption{\textbf{Comparing Invisibility Metrics of Backdoor Attacks on ImageNet.} Our attack achieves the best invisibility scores compared to other existing methods. }
\label{tab:invsmetrics}

\end{wraptable}

Table \ref{table:compare} compares our method with existing spatial and frequency backdoor attacks. The results for SIG, Refool, SPM, and Poison Ink  are taken from \cite{zhang2021PoisonIR}. Our frequency-based backdoor attack achieves SOTA results in almost all scenarios considered. Note that the training setup adopted to generate our results is the same for all other methods. A further comparison with other backdoor attacks is provided in the supplementary.




\noindent \textbf{Invisibility of the Proposed Attack.} Table \ref{tab:invsmetrics} compares our proposed frequency-based backdoor attack with other attacks based on their invisibility metrics (PSNR,SSIM,LPIPS). The results of the other methods are taken from \cite{zhang2021PoisonIR} (except for \cite{Wang2021BackdoorAT, Feng2021FIBAFB}). Our proposed attack achieves the highest PSNR and SSIM, and the lowest LPIPS compared to other backdoor attacks. The PSNR of our method could be further improved at the cost of ASR by selecting fewer frequencies to poison; however, the invisibility metrics (PSNR, SSIM, LPIPS) ``saturate'' beyond a certain point where further improvements become insignificant and unneeded.



\begin{table*}[b!]
\caption{\textbf{Comparison between the Proposed Attack and Backdoor Attacks in the Literature.} Our frequency-based attack achieves SOTA ASR, CDA, PSNR, and LPIPS metrics. The results shown are for VGG19 trained on CIFAR10. Legend: \textbf{First Best}, \underline{Second Best}}
\vspace{6pt}
\scalebox{0.56}{
\begin{tabular}{cc|c|c|c|c|c|c|c|c}
\hline
\multicolumn{1}{l|}{\textbf{Metric}}                   & \textbf{Ratio}        & \textbf{SIG}    & \textbf{Refool}   & \textbf{SPM}     & \textbf{WaNet}   & \textbf{FIBA}& \textbf{FTrojan}  & \textbf{Poison Ink}   &  \textbf{Ours} \\ \hline
\multicolumn{1}{l|}{\multirow{3}{*}{\textbf{CDA/ASR}}} & 3\%                   & 89.74 / 99.23   &  89.20 / 87.16    & 88.89 / 58.53    & \underline{91.86} / 32.86    & 90.92 / 90.10 & 91.31 / \textbf{99.99}   & 89.65 / 94.22         &  \textbf{92.31} / \underline{99.43}          \\ \cline{2-10} 
\multicolumn{1}{l|}{}                                  & 5\%                   & 89.64 / 99.47   & 89.16 / 89.79     & 88.90 / 57.69    & 91.47 / 88.15    & 90.69 / 95.06 & \underline{91.64} / \underline{99.10}   & 89.69 / 93.58                  &  \textbf{ 91.88 / 99.88}        \\ \cline{2-10} 
\multicolumn{1}{l|}{}                                  & 10\%                  & 89.45 / 99.40   & 88.80 / 92.80     & 89.07 / 57.33    & \underline{91.22} / 96.96    & 90.41 / 95.86 & 90.93 / \textbf{100.00}   & 89.47 / 93.67                   &  \textbf{92.10} / \underline{99.97}        \\ \hline
\multicolumn{2}{l|}{\textbf{PSNR$\uparrow$/LPIPS$\downarrow$}}                 & 25.12 / 0.0400  & 19.38 / 0.0397    & 38.94 / \underline{0.0001}   & 31.53 / 0.0047    & 19.40 / 0.0180 & 41.01 / \underline{0.0001}   & \underline{42.95} / \underline{0.0001}                  &  \textbf{43.15} / \textbf{0.00001}          \\ \hline
\end{tabular}}  
\label{table:compare}
\vspace{-0.25cm}
\end{table*}

\begin{figure}[t!]
\centering
  \caption{\textbf{Evaluation of defenses:} Evaluation of various SOTA defenses against the proposed frequency-based attack shows the power of the proposed method in evading the defenses. (a) Grad-CAM shows high similarity in the attention regions for poisoned and non-poisoned models; (b) Pruning the poisoned model maintains high ASR even after significant drop in CDA. (c) Neural Cleanse anomaly indices fall below the anomaly threshold (2.0). 
  } 
  \vspace{0.15cm}
  \begin{subfigure}[t]{0.3\linewidth}
    \centering
    \includegraphics[height=2.2cm]{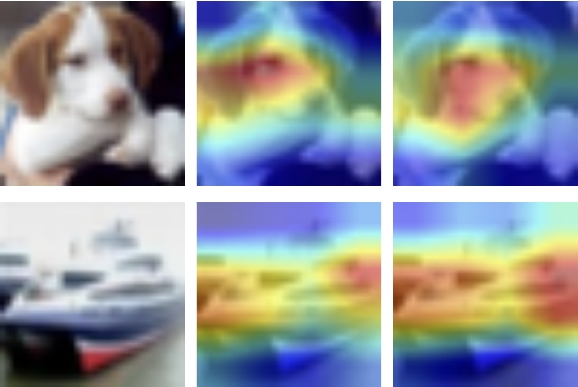}
    \caption{Grad-CAM \cite{Selvaraju2019GradCAMVE}}
    \label{defa2}
  \end{subfigure}
  \begin{subfigure}[t]{0.3\linewidth}
      \centering
    \includegraphics[height=2.5cm]{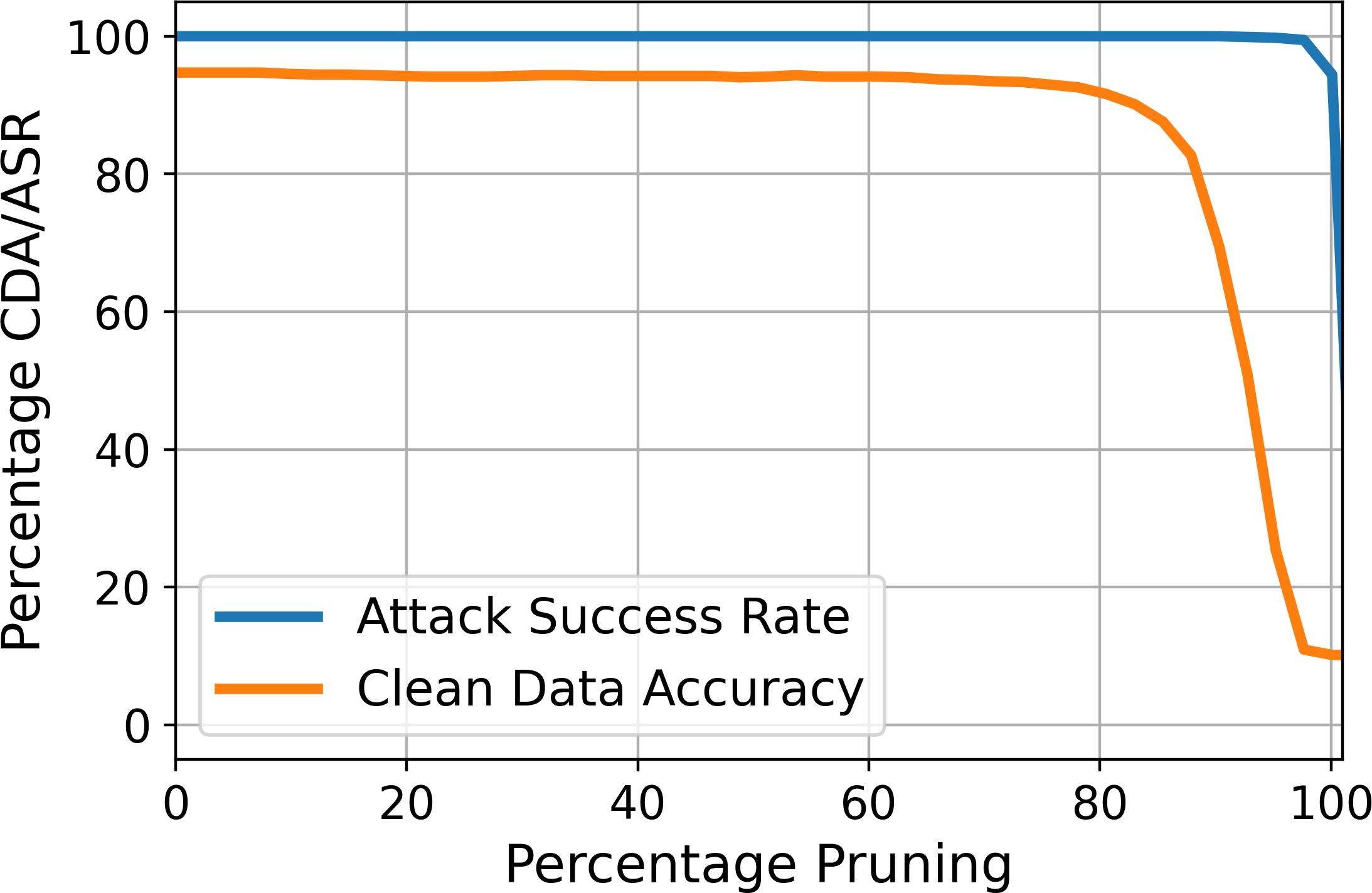}
    \caption{Pruning \cite{Liu2018FinePruningDA}}
    \label{defb2}
  \end{subfigure}
    \begin{subfigure}[t]{0.3\linewidth}
    \centering
    \includegraphics[height=2.5cm]{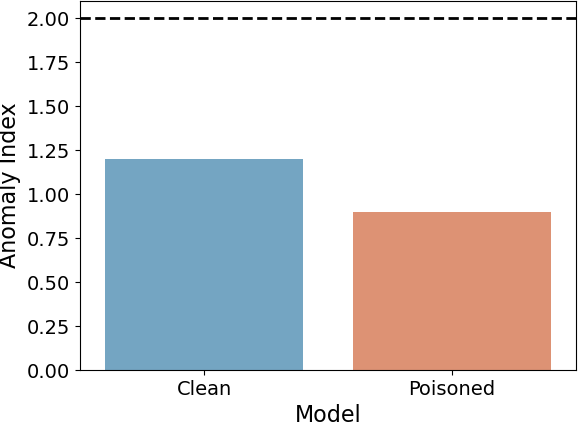}
    \caption{Neural Cleanse \cite{Wang2019NeuralCI}}
    \label{defc2}
  \end{subfigure}
  \label{fig:evaldef}
  \vspace{-0.4cm}
\end{figure}

\vspace{-0.4cm}
\subsection{Evaluation Against Backdoor Defenses}

We evaluate our attacked models against three SOTA backdoor defenses, namely, Neural Cleanse \cite{Wang2019NeuralCI}, Grad-CAM \cite{Selvaraju2019GradCAMVE}, and Pruning \cite{Liu2018FinePruningDA}.  Being invisible and dynamic in the spatial domain, frequency-based backdoor attacks can easily evade SOTA defenses. The results of the three defenses against our attacked ResNet18 model trained on CIFAR10 with $1\%$ poisoning rate are shown in Figure \ref{fig:evaldef}.
Figure \ref{defa2} shows the Grad-CAM \cite{Selvaraju2019GradCAMVE} results for two images and their backdoor attacked versions using our frequency-based approach. 
Grad-CAM uses gradients of a particular class to visualize where the network is looking/focusing at to make its prediction. As shown in Figure \ref{defa2}, our frequency-based backdoor attacks do not introduce an observable change in the ``attention" of the network. For each of the two samples presented (first column), we compute the Grad-CAM by passing the clean samples into the clean network ($f_0$) (middle column), and then show the Grad-CAM for passing the poisoned samples into the backdoored model ($f$) (third column). 

Since the network still focuses on the same parts of the input image, methods like Februus \cite{Doan2020FebruusIP}  fail to remove the embedded backdoor, as observed by \cite{zhang2021PoisonIR}. Figure \ref{defb2} shows the performance of our attack against the pruning defense in \cite{Liu2018FinePruningDA}, which prunes the least active neurons (on clean samples) and then fine-tunes the network on clean samples. 

\begin{table}[b!]
\centering

\renewcommand{\arraystretch}{0.95}
\scalebox{0.8}{
\begin{tabular}{cccccccc}
\specialrule{1.25pt}{0pt}{1pt}
                                   &                         & \multicolumn{2}{c}{\textbf{JPEG}}        & \multicolumn{2}{c}{\textbf{Autoencoder}}               & \multicolumn{2}{c}{\textbf{JPEG+Autoencoder}} \\ \cline{2-8}  
                                   & \textbf{Poisoning Rate} & \textbf{CDA}              & \textbf{ASR} & \textbf{CDA}              & \textbf{ASR}               & \textbf{CDA}          & \textbf{ASR}          \\ \hline
\multirow{5}{*}{\rotatebox{90}{\textbf{CIFAR10}}}  & 0.1\%                   & 94.19                     & \cellcolor[HTML]{EFEFEF}1.76         & 93.73                     & \cellcolor[HTML]{EFEFEF}0.22                       & 94.65                 & \cellcolor[HTML]{EFEFEF} 0.66                  \\ \hhline{~-------}
                                   & 0.2\%                   & 94.37                     & \cellcolor[HTML]{EFEFEF}18.02        & 94.38                     & \cellcolor[HTML]{EFEFEF}22.86                      & 94.22                 & \cellcolor[HTML]{EFEFEF}3.08                  \\ \hhline{~-------}  
                                   & 0.5\%                   & 93.94                     & \cellcolor[HTML]{EFEFEF}83.52        & 94.17                     & \cellcolor[HTML]{EFEFEF}73.85                      & 94.49                 & \cellcolor[HTML]{EFEFEF}36.48                 \\ \hhline{~-------} 
                                   & 1.0\%                   & 94.28                     & \cellcolor[HTML]{EFEFEF}96.48        & 94.61                     & \cellcolor[HTML]{EFEFEF}93.63                      & 94.24                 & \cellcolor[HTML]{EFEFEF}90.11                 \\ \hhline{~-------}  
                                   & 3.0\%                   & 94.26                     & \cellcolor[HTML]{EFEFEF}99.34        & 94.13                     & \cellcolor[HTML]{EFEFEF}98.90                      & 94.32                 & \cellcolor[HTML]{EFEFEF}98.46                 \\ \hline
\multirow{5}{*}{\rotatebox{90}{\textbf{CIFAR100}}}& 0.1\%                   & 76.57                     & \cellcolor[HTML]{EFEFEF}14.26        & \multicolumn{1}{r}{76.19} & \cellcolor[HTML]{EFEFEF}14.06                      & 76.05                 & \cellcolor[HTML]{EFEFEF}2.57                  \\ \hhline{~-------}  
                                   & 0.2\%                   & 77.14                     & \cellcolor[HTML]{EFEFEF}75.25        & \multicolumn{1}{r}{75.96} & \cellcolor[HTML]{EFEFEF}83.76                      & 75.40                 & \cellcolor[HTML]{EFEFEF}32.08                 \\ \hhline{~-------}  
                                   & 0.5\%                   & 75.86                     &\cellcolor[HTML]{EFEFEF} 95.25        & \multicolumn{1}{r}{76.07} & \cellcolor[HTML]{EFEFEF}94.06                      & 76.35                 & \cellcolor[HTML]{EFEFEF}95.05                 \\ \hhline{~-------}  
                                   & 1.0\%                   & 75.43                     &\cellcolor[HTML]{EFEFEF} 99.21        & \multicolumn{1}{r}{75.57} & \cellcolor[HTML]{EFEFEF}97.82                      & 76.16                 & \cellcolor[HTML]{EFEFEF}96.83                 \\ \hhline{~-------}  
                                   & 3.0\%                   & 75.07                     &\cellcolor[HTML]{EFEFEF} 99.80        & \multicolumn{1}{r}{76.26} & \cellcolor[HTML]{EFEFEF}99.54                     & 75.51                 & \cellcolor[HTML]{EFEFEF}98.81                 \\ \hline
\multirow{5}{*}{\rotatebox{90}{\textbf{GTSRB}}}     & 0.1\%                   & 97.27 & \cellcolor[HTML]{EFEFEF}52.46        & 97.45                     & \cellcolor[HTML]{EFEFEF} 69.55  & 96.97                 & \cellcolor[HTML]{EFEFEF}48.13                 \\ \hhline{~-------} 
                                   & 0.2\%                   & 96.79 &\cellcolor[HTML]{EFEFEF} 74.07        & 97.39                     & \cellcolor[HTML]{EFEFEF}81.14  & 97.09                 & \cellcolor[HTML]{EFEFEF}73.87                 \\ \hhline{~-------}  
                                   & 0.5\%                   & 97.25 & \cellcolor[HTML]{EFEFEF}90.18        & 97.14                     & \cellcolor[HTML]{EFEFEF}94.50  & 96.84                 & \cellcolor[HTML]{EFEFEF}95.09                 \\ \hhline{~-------}
                                   & 1.0\%                   & 94.34 & \cellcolor[HTML]{EFEFEF}86.44        & 97.00                     & \cellcolor[HTML]{EFEFEF}99.02  & 95.56                 & \cellcolor[HTML]{EFEFEF}94.89                 \\ \hhline{~-------}  
                                   & 3.0\%                   & 93.72 & \cellcolor[HTML]{EFEFEF}98.43        & 97.25                     & \cellcolor[HTML]{EFEFEF}99.78 & 92.99                 & \cellcolor[HTML]{EFEFEF}97.64                 \\ \hline
\end{tabular}}
\vspace{7pt}
\caption{\textbf{Augmentation Defense:} CDA and ASR of backdoored ResNet18 trained on various datasets with JPEG compression and Autoencoder augmentation. The ASR and CDA are maintained even when no preprocessing technique is used.}
\label{Augment1}
\end{table}
We see that pruning our backdoored model does not eliminate the backdoor. This is mainly attributed to the fact that frequency-based poisoning is of low norm and therefore gets embedded into most weights of the network rather than hidden into particular neurons. Figure \ref{defc2} shows the anomaly index computed by Neural Cleanse \cite{Wang2019NeuralCI} for both the baseline and our backdoored/poisoned model. Since the anomaly index of the poisoned model is less than the anomaly index threshold defined by Neural Cleanse (2.0), 
Neural Cleanse fails to detect that the frequency-based backdoored model is actually poisoned. Further evaluation of these defenses and evaluation of additional defenses, namely, STRIP \cite{Gao2019STRIPAD}, Spectral Signatures \cite{Tran2018SpectralSI}, and Activation Clustering \cite{Chen2019DetectingBA} on different models and datasets is provided in the supplementary.

\vspace{-0.3cm}
\subsection{Defenses Against Frequency-based Backdoors}

Since the additive mask values could be 
\textit{arbitrarily} chosen, a simple inspection of the Fourier transforms of the input may not be successful in detecting the poisoned samples. Therefore, we discuss two possible ways to defend against 
frequency backdoor attacks. 

\begin{wrapfigure}[18]{r}{6cm}
    \centering
        \includegraphics[width=\linewidth]{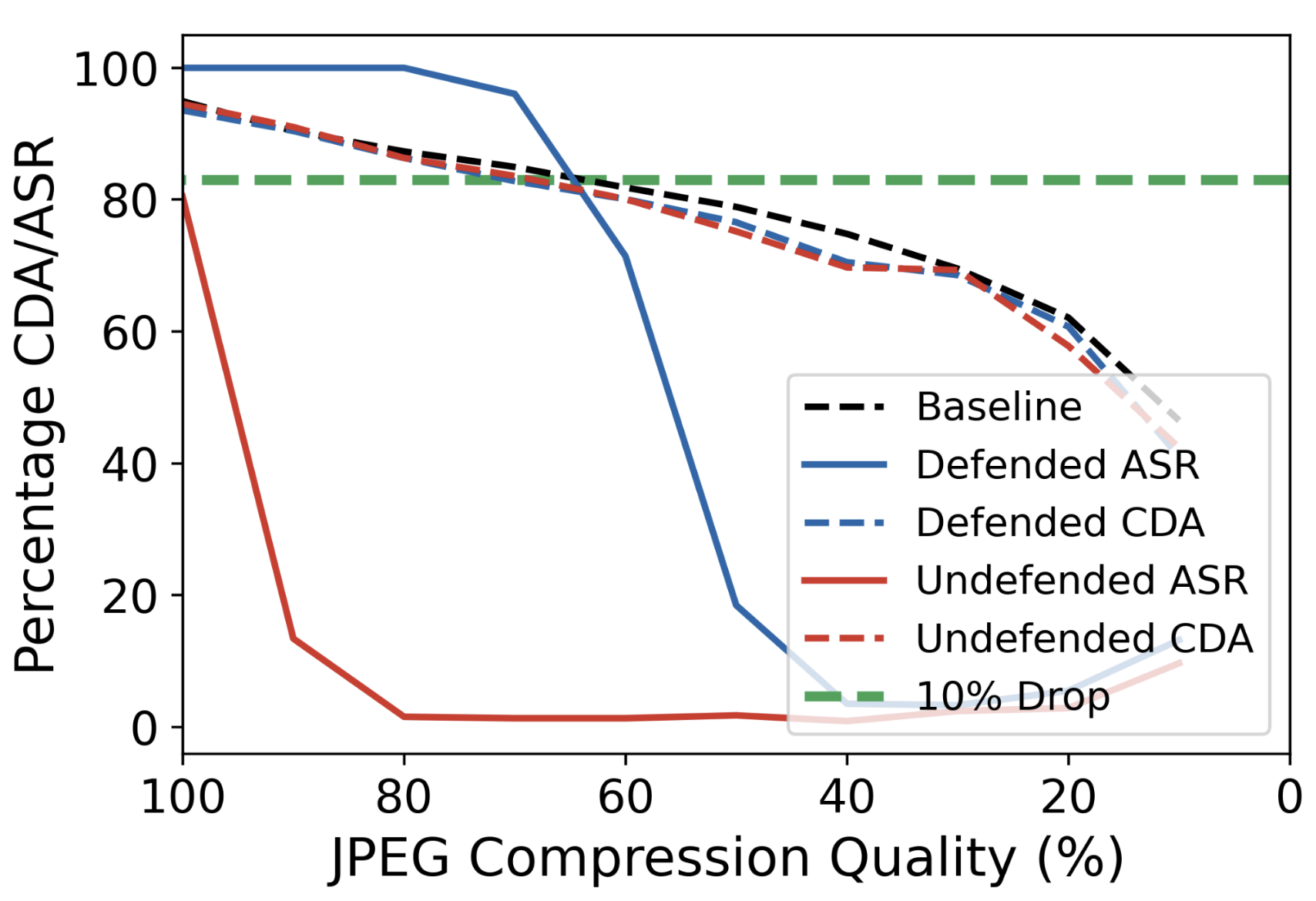}
\caption{\textbf{Defending with JPEG Augmentation.} Training on JPEG compressed images maintains a high ASR even after a drop of 10\% in CDA. The baseline denotes the CDA of the baseline model evaluated on compressed images.}
    \label{augmentationresults}
\end{wrapfigure}

For a successful defense, the defender should manipulate the frequency spectrum of the input images to break the backdoor trigger while maintaining a satisfactory CDA. We show that this is possible using two techniques: (1) passing the image through an autoencoder and (2) compressing the image. These two methods are used in the robustness literature and have proven to be useful in protecting DNNs from adversarial attacks \cite{Das2017KeepingTB,Das2018SHIELDFP}. Autoencoders have also been used as a preprocessing mechanism to disable backdoor triggers \cite{Liu2017NeuralT}. Applying an autoencoder trained on CIFAR10 can almost completely deactivate the embedded frequency backdoor. A similar effect is observed for compression, where the ASR of the backdoored model drops to almost 0\% after 20\% of JPEG compression.

A possible solution to bypass
both of these defenses is to apply a technique similar to adversarial training \cite{Goodfellow2015ExplainingAH,Zhang2019TheoreticallyPT}. The attacker can train on compressed and/or auto-encoded versions of the poisoned images. This augmentation translates to embedding multiple versions of the backdoor into the model. Figure \ref{augmentationresults} shows the ASR and CDA for both an undefended poisoned model and a defended one. For the undefended model, \emph{i.e}~no augmentation, the backdoor immediately breaks down as compression is applied. On the other hand, the defended model can maintain an $ASR>80\%$ even beyond 25\% compression, where the CDA drops by 10\%.  Finally, we note that the above augmentations still allow us to reach a high ASR with a minimal drop in CDA for our backdoored models. Therefore, if the defender does not set a defense  mechanism, the backdoor still functions properly. The results for ResNet18 trained on CIFAR10, GTSRB, and CIFAR100 with different augmentations are shown in Table \ref{Augment1}. The results for other models and datasets are presented in the supplementary material.

\vspace{-0.45cm}
\subsection{Ablation Study}
\label{sec:ablation}




\begin{wraptable}[8]{r}{7.5cm}
\centering
\vspace{-1cm}
\scalebox{0.7}{
\begin{tabular}{c|c|c|c|}
\cline{2-4}
\textbf{}                                   & \textbf{Poisoning Rate}                   & \textbf{1\%} & \textbf{2\%} \\ \hline
\multicolumn{1}{|c|}{\textbf{Random (1)}}   & \multirow{5}{*}{\textbf{CDA(\%)/ASR(\%)}} & 67.24/53.91  & 66.83/60.49  \\ \cline{1-1} \cline{3-4} 
\multicolumn{1}{|c|}{\textbf{Random (2)}}   &                                           & 67.23/56.88  & 66.80/66.11  \\ \cline{1-1} \cline{3-4} 
\multicolumn{1}{|c|}{\textbf{Bottom-$k$ (1)}} &                                           & 67.03/22.58  & 66.80/55.96  \\ \cline{1-1} \cline{3-4} 
\multicolumn{1}{|c|}{\textbf{Bottom-$k$ (2)}} &                                           & 67.04/0.31   & 67.02/92.81  \\ \cline{1-1} \cline{3-4} 
\multicolumn{1}{|c|}{\textbf{Top-$k$ (1)}}    &                                           & 67.13/87.74  & 67.26/98.01  \\ \hline
\end{tabular}}
\vspace{5pt}
\caption{\textbf{Effect of Different Frequency Selection Schemes:} Results for frequency filters generated using least sensitive, most sensitive and random frequencies. Choosing the top-$k$ most sensitive frequencies provides the highest ASR among those options.}
\label{Table:Ablation}
\end{wraptable}

We study the effect of choosing \textbf{(i)} random frequencies and \textbf{(ii)} bottom-$k$, \emph{i.e}~least sensitive frequencies, as compared to choosing the top-$k$ frequencies from the Fourier heatmap. Table \ref{Table:Ablation} shows the results of poisoning a ResNet18 trained on ImageNet using two different random filters and two different bottom-$k$ filters (two different values were chosen for $k$ to control the PSNR), where the runs for a particular scheme are numbered in brackets. The random filters were generated using Bernoulli trials with $p=0.005$ at each Fourier basis (Random (1): PSNR = 47.62/ Random (2): PSNR = 46.62). Bottom-$k$ filters were generated by selecting the $k$-least sensitive frequencies (Bottom-$k$ (1): PSNR = 51.23 /Bottom-$k$ (2): PSNR = 31.23). In general, bottom-$k$ and random frequencies contain low frequency components, which greatly affect the invisibility of the attack.

\begin{figure}
\centering
  \begin{subfigure}[t]{0.17\linewidth}
    \includegraphics[width=\textwidth]{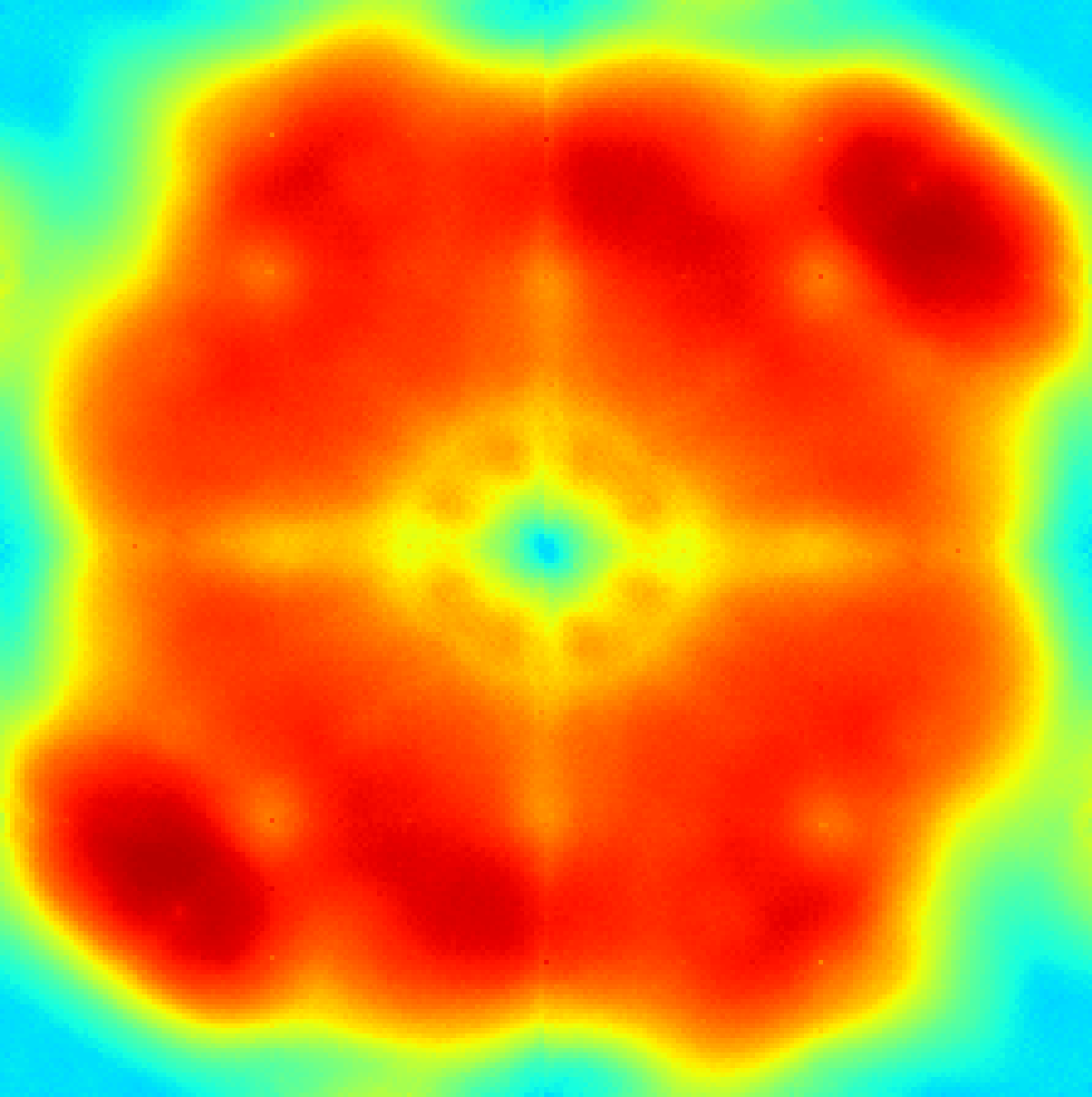}
    \caption{}
    \label{fig-a-ablate}
  \end{subfigure}\hspace{0.5cm}
  \begin{subfigure}[t]{0.17\linewidth}
    \includegraphics[width=\textwidth]{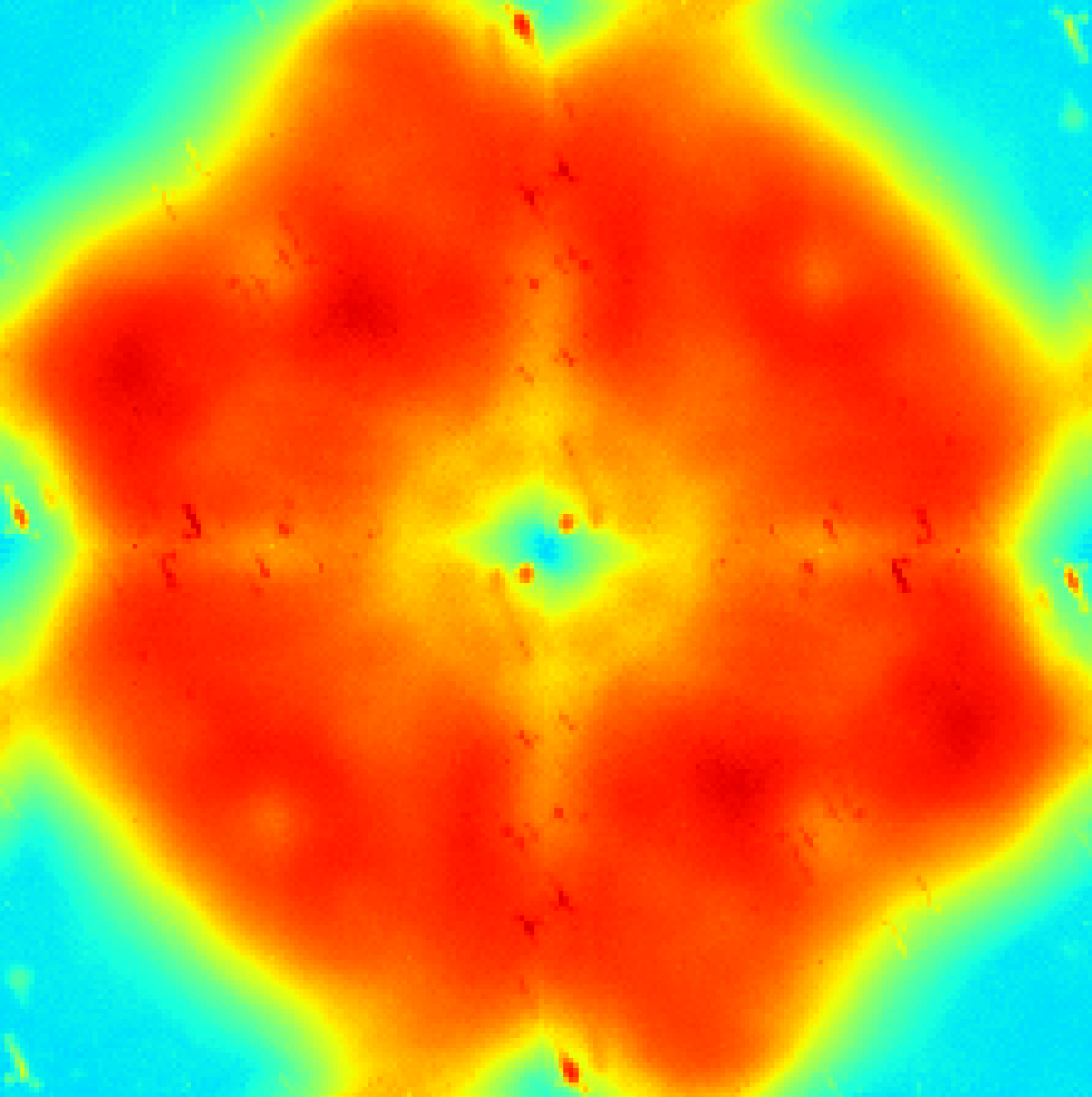}
    \caption{}
    \label{fig-b-ablate}
  \end{subfigure}\hspace{0.5cm}
   \begin{subfigure}[t]{0.17\linewidth}
    \includegraphics[width=\textwidth]{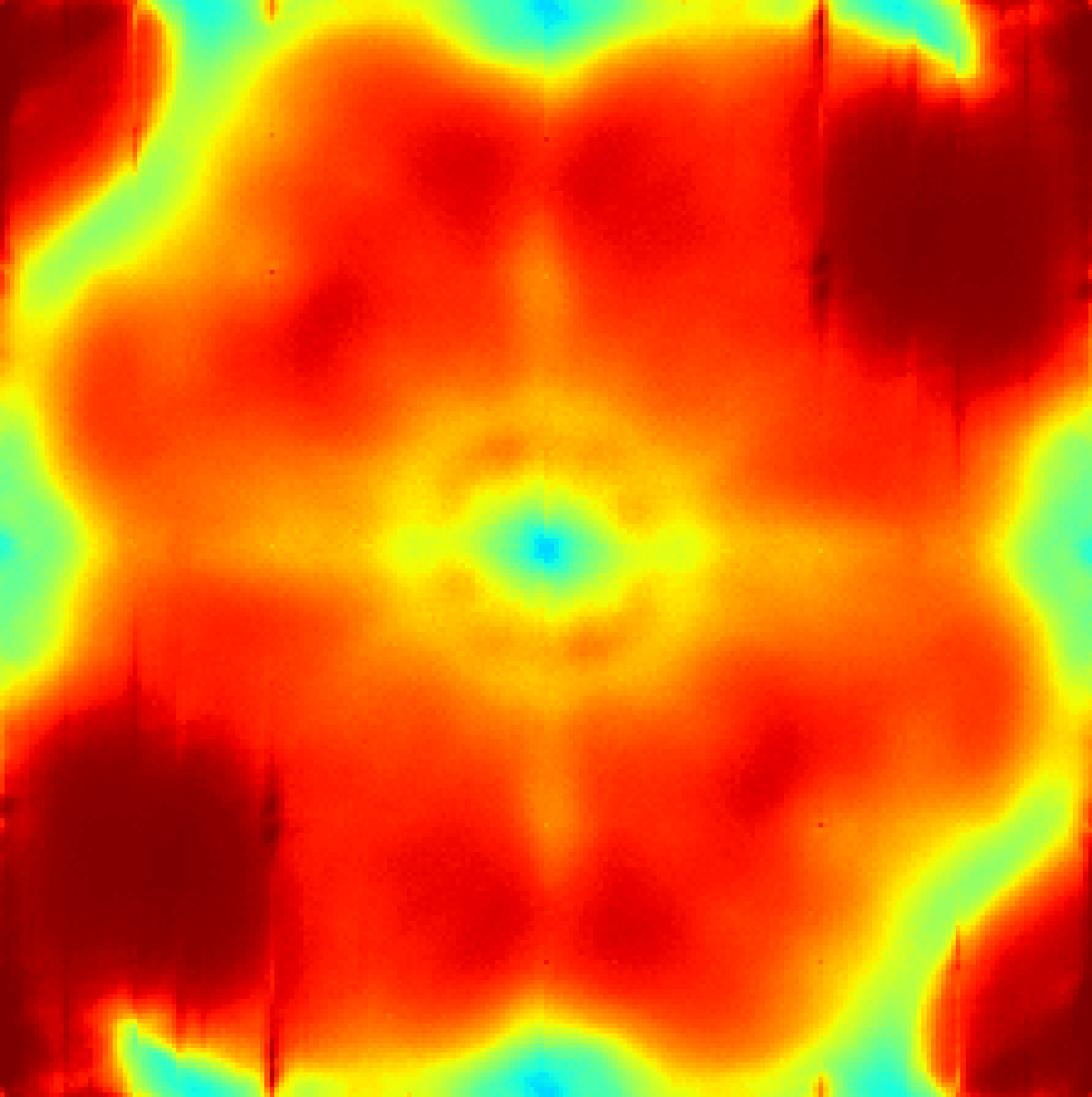}
    \caption{}
    \label{fig-c-ablate}
  \end{subfigure}\hspace{0.5cm}
    \begin{subfigure}[t]{0.17\linewidth}
    \includegraphics[width=\textwidth]{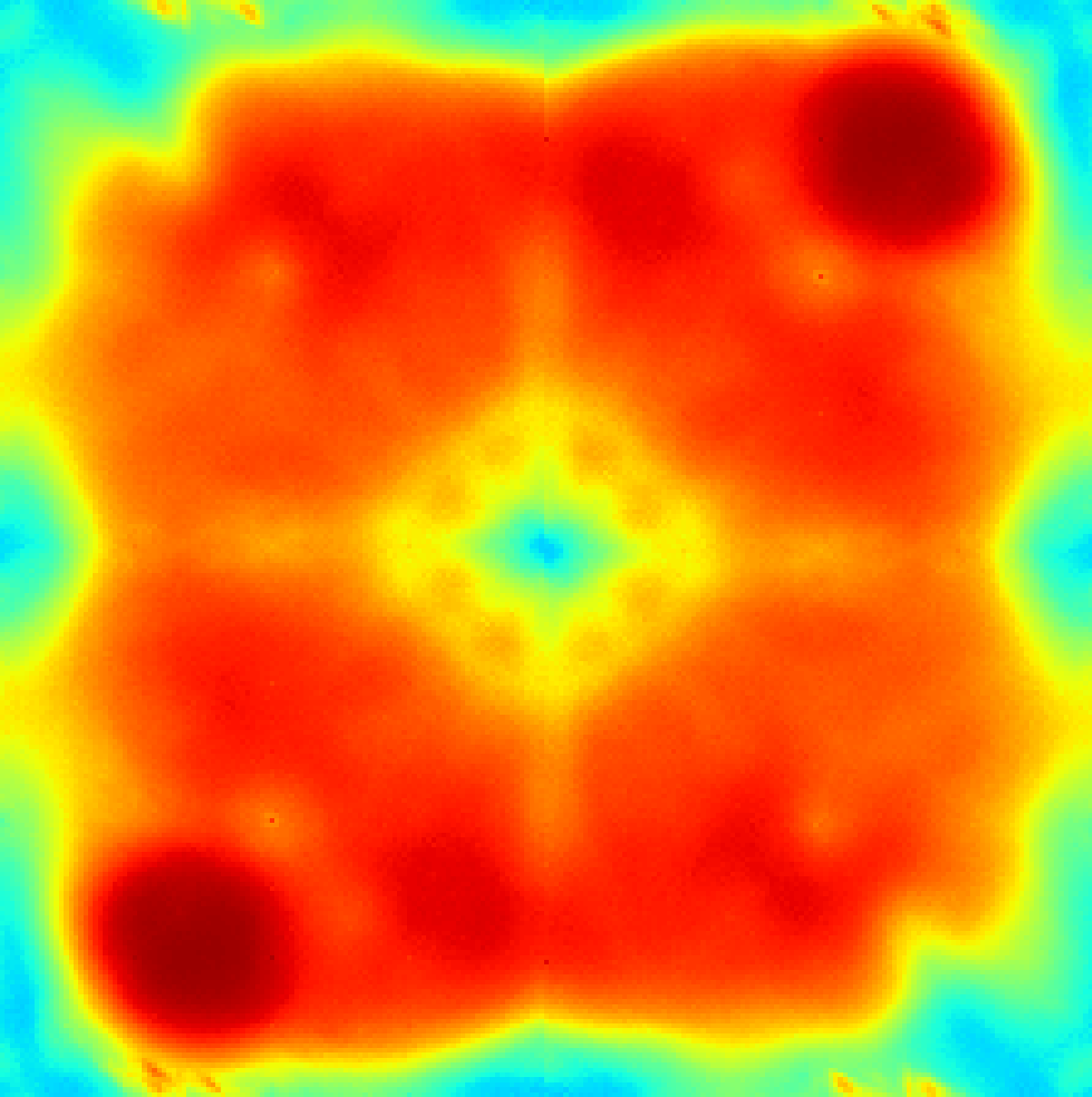}
    \caption{}
    \label{fig-d-ablate}
  \end{subfigure}
   \vspace{6pt}
  \caption{\textbf{Heatmaps of Ablated Frequency Selection:} Fourier heatmaps of frequency-based backdoor attacks with different frequency selection schemes: (a) Clean Model; (b) Random Frequency Selection; (c) Bottom-$k$ Frequency Selection; (d) Proposed Top-$k$ Frequency Selection. } 
  \label{fig:AblateFig}
  \vspace{-0.5cm}
\end{figure}

One can see the importance of choosing top-$k$ frequencies over the other two options, as it leads to a high ASR at a small poisoning rate while maintaining a high PSNR. This is attributed to the fact that the network relies on the most sensitive frequencies to perform the classification task at hand. Therefore, embedding a backdoor attack into the most sensitive frequencies allows the network to learn the backdoor trigger with little effort, compared to other frequency selection schemes.

Finally, an interesting observation can be made by looking at the Fourier heatmaps of these models. Figure \ref{fig:AblateFig} visualizes the Fourier heatmaps for Random (2), Bottom-$k$ (2), and Top-$k$ models. We can see a significant explosion in frequency sensitivity in the case of selecting the bottom-$k$ components and ``chicken-pox" like sensitivity for the random frequency selection (dotted in the positions of randomly sampled frequency bases). Our method of using the top-$k$ most sensitive frequencies is more conservative in introducing modifications to the network's 
clean heatmap; however, it also experiences mild ``sensitivity leakage" at certain frequencies. The supplementary shows the Fourier heatmaps for 
other backdoor attacks and provides a discussion about detecting backdoor attacked models using Fourier heatmaps.


\vspace{-0.4cm}
\section{Conclusion}
\label{sec:conclusion}

\vspace{-0.2cm} In this work, we proposed a new \textit{frequency} backdoor attack that takes advantage of the natural frequency sensitivity of the DNN. Through extensive experiments, we showed the effectiveness of the proposed attack in embedding imperceptible backdoors that can evade existing defenses while achieving both a high ASR and a CDA. We also laid the foundations for future defenses against frequency-based backdoor attacks through (1) data preprocessing using autoencoders and compression; and (2) Fourier heatmap visualization. 

\section{Acknowledgement}

This work was supported by the King Abdullah University of Science and Technology (KAUST) Office of Sponsored Research (OSR) under Award No. OSR-CRG2019-4033, as well as, the SDAIA-KAUST Center of Excellence in Data Science and Artificial Intelligence (SDAIA-KAUST AI).

\bibliography{egbib.bib}
\clearpage
\appendix

{ \ \vspace{0.2cm}}

\begin{center}
    \Large \textbf{Supplementary Material}
\end{center}
\section{Introduction to Supplementary Material}

In this supplementary material, we present the extended results and variants of the proposed frequency-based backdoor attack. 
Section \ref{sec1} shows the full version of Table 1 from the main paper, this includes evaluation of the proposed pipeline on additional network architectures. 
Section \ref{sec2} presents an extended evaluation of the proposed augmentation in Section 5.4 of the main paper (similar to Table 4).
Section \ref{sec3} discusses different design choices for the additive filters $\mathcal{A}_{R,G,B}$.
In section \ref{sec4} we extend the applicability of the proposed attack to the multitarget attack regime. Section \ref{Sec:Efficient} presents a more efficient variant of the proposed method.
Section \ref{sec5} shows the result of applying a binary and an additive filter generated from one model to poison another.
Section \ref{sec6} visualizes the proposed backdoor attack in the spatial domain, showing that the attack is highly dynamic.
Section \ref{sec7} displays the Fourier heatmaps and top-$k$ selected frequencies (binary masks) for various datasets and architectures.
Section \ref{sec8} shows the Fourier heatmaps for different spatial backdoor attacks, highlighting a new possible defense against backdoor attacks. Section \ref{sec9} presents a further evaluation of the spatial defenses discussed in the manuscript and presents three additional defenses, namely, STRIP \cite{Gao2019STRIPAD}, Spectral Signatures \cite{Tran2018SpectralSI}, and Activation Clustering \cite{Chen2019DetectingBA}. Section \ref{sec92}, shows an evaluation of the proposed defense against other frequency backdoor attacks. Section \ref{sec11} presents a further comparison of our proposed attack against existing spatial backdoor attacks. Finally, section \ref{sec12} provides insights about the relationship between the model's learning capacity and the capability of embedding a backdoor attack into the model.


\clearpage
\section{Evaluation of the Proposed Backdoor Attack }
\label{sec1}
\let\thefootnote\relax\footnotetext{WideResNet34 was not included for ImageNet experiments as there is no official implementation of this model in \textit{torchvision.models} .}
\begin{table*}[hbpt!]
\centering
\renewcommand{\arraystretch}{1.4}
\scalebox{0.57}{
\begin{tabular}{cccccccccccccc}
\specialrule{1.2pt}{0pt}{1pt}
                                  &                         & \multicolumn{2}{c}{\textbf{ResNet18}} & \multicolumn{2}{c}{\textbf{ResNet34}} & \multicolumn{2}{c}{\textbf{ResNet50}} & \multicolumn{2}{c}{\textbf{DenseNet121}} & \multicolumn{2}{c}{\textbf{VGG19}} & \multicolumn{2}{c}{\textbf{WideResNet34}} \\ \cmidrule(l){2-14} 
                                  & \textbf{Poisoning Rate} & \textbf{CDA}      & \textbf{ASR}      & \textbf{CDA}      & \textbf{ASR}      & \textbf{CDA}      & \textbf{ASR}      & \textbf{CDA}        & \textbf{ASR}        & \textbf{CDA}     & \textbf{ASR}    & \textbf{CDA}        & \textbf{ASR}       \\ \specialrule{0.2pt}{0pt}{1pt}
\multirow{5}{*}{\textbf{CIFAR10}} 
& 0.0\%                                    & 93.92                         &\cellcolor[HTML]{EFEFEF} -                             & 94.59                         & \cellcolor[HTML]{EFEFEF}-                             & 94.10                         & \cellcolor[HTML]{EFEFEF}-                             & 94.70                         & \cellcolor[HTML]{EFEFEF}-                             & 92.47                         & \cellcolor[HTML]{EFEFEF}-                             & 95.33                         & \cellcolor[HTML]{EFEFEF}-                             \\ \hhline{~-------------} 
                          & 0.1\%                                    & 94.00                         &\cellcolor[HTML]{EFEFEF} 1.54                          & 94.49                         & \cellcolor[HTML]{EFEFEF}0.83                          & 94.48                         & \cellcolor[HTML]{EFEFEF}53.63                         & 94.94                         & \cellcolor[HTML]{EFEFEF}86.98                          & 92.63                         & \cellcolor[HTML]{EFEFEF}0.44                          & 95.73                         & \cellcolor[HTML]{EFEFEF}84.91                         \\ \hhline{~-------------} 
                          & 0.2\%                                    & 94.14                         & \cellcolor[HTML]{EFEFEF}72.31                         & 94.26                         & \cellcolor[HTML]{EFEFEF}66.46                         & 94.45                         & \cellcolor[HTML]{EFEFEF}87.91                         & 94.54                         & \cellcolor[HTML]{EFEFEF}95.77                         & 92.39                         & \cellcolor[HTML]{EFEFEF}0.44                          & 95.42                         & \cellcolor[HTML]{EFEFEF}96.89                         \\ \hhline{~-------------} 
                          & 0.4\%                                    & 94.20                         & \cellcolor[HTML]{EFEFEF}85.05                         & 94.33                         & \cellcolor[HTML]{EFEFEF}90.97                         & 94.37                         & \cellcolor[HTML]{EFEFEF}95.38                         & 94.89                         & \cellcolor[HTML]{EFEFEF}96.48                         & 92.17                         & \cellcolor[HTML]{EFEFEF}1.62                          & 95.48                         & \cellcolor[HTML]{EFEFEF}99.34                         \\ \hhline{~-------------} 
                          & 1.0\%                                    & 94.38                         & \cellcolor[HTML]{EFEFEF}99.44                         & 94.44                         & \cellcolor[HTML]{EFEFEF}91.75                        & 94.32                         & \cellcolor[HTML]{EFEFEF}99.34                         & 94.83                         & \cellcolor[HTML]{EFEFEF}98.70                         & 91.95                         & \cellcolor[HTML]{EFEFEF}99.39                         & 95.70                         & \cellcolor[HTML]{EFEFEF}99.80                         \\ \hhline{~-------------} 
                          & 3.0\%                                    & 94.31                         &  \cellcolor[HTML]{EFEFEF}99.79                        & 94.41                         & \cellcolor[HTML]{EFEFEF}99.64                         & 94.31                         & \cellcolor[HTML]{EFEFEF}99.36                         & 94.94                         & \cellcolor[HTML]{EFEFEF}99.89                        & 91.89                         & \cellcolor[HTML]{EFEFEF}99.81                       & 95.44                         & \cellcolor[HTML]{EFEFEF}99.99                        \\ \specialrule{0.2pt}{0pt}{0pt}
\multirow{5}{*}{\textbf{CIFAR100}} 
 & 0.0\%                                    & 75.95                         & \cellcolor[HTML]{EFEFEF}-                             & 75.66                         & \cellcolor[HTML]{EFEFEF}-                             & 77.36                         & \cellcolor[HTML]{EFEFEF}-                             & 78.98                         & \cellcolor[HTML]{EFEFEF}-                             & 67.45                         & \cellcolor[HTML]{EFEFEF}-                             & 79.55                         & \cellcolor[HTML]{EFEFEF}-                             \\ \hhline{~-------------} 
                          & 0.1\%                                    & 75.76                         &\cellcolor[HTML]{EFEFEF} 60.57                         & 76.76                         & \cellcolor[HTML]{EFEFEF}65.18                         & 76.73                         & \cellcolor[HTML]{EFEFEF}42.18                         & 78.34                         & \cellcolor[HTML]{EFEFEF}73.47                         & 67.78                         & \cellcolor[HTML]{EFEFEF}0.40                          & 79.84                         & \cellcolor[HTML]{EFEFEF}43.96                         \\ \hhline{~-------------} 
                          & 0.2\%                                    & 75.75                         & \cellcolor[HTML]{EFEFEF}92.78                         & 74.79                         & \cellcolor[HTML]{EFEFEF}84.09                         & 77.87                         & \cellcolor[HTML]{EFEFEF}78.21                         & 79.1                          & \cellcolor[HTML]{EFEFEF}89.31                         & 67.72                         & \cellcolor[HTML]{EFEFEF}0.59                          & 79.24                         & \cellcolor[HTML]{EFEFEF}78.42                         \\ \hhline{~-------------} 
                          & 0.4\%                                    & 75.92                         & \cellcolor[HTML]{EFEFEF}96.49                         & 76.25                         & \cellcolor[HTML]{EFEFEF}99.29                         & 77.69                         & \cellcolor[HTML]{EFEFEF}83.96                         & 79.1                          & \cellcolor[HTML]{EFEFEF}92.67                         & 67.61                         & \cellcolor[HTML]{EFEFEF}0.20                          & 79.14                         & \cellcolor[HTML]{EFEFEF}87.33                         \\ \hhline{~-------------} 
                          & 1.0\%                                    & 76.05                         &\cellcolor[HTML]{EFEFEF} 98.99                         & 74.95                         & \cellcolor[HTML]{EFEFEF}99.44                         & 77.12                         & \cellcolor[HTML]{EFEFEF}90.49                         & 78.6                          & \cellcolor[HTML]{EFEFEF}96.44                         & 65.84                         & \cellcolor[HTML]{EFEFEF}0.40                          & 79.14                         & \cellcolor[HTML]{EFEFEF}98.02                         \\ \hhline{~-------------} 
                          & 3.0\%                                    & 75.36                         & \cellcolor[HTML]{EFEFEF}99.93                        & 76.51                         & \cellcolor[HTML]{EFEFEF}99.84                        & 76.58                         & \cellcolor[HTML]{EFEFEF}98.61                         & 78.31                         & \cellcolor[HTML]{EFEFEF}99.60                         & 67.14                         & \cellcolor[HTML]{EFEFEF}99.00                         & 78.74                         & \cellcolor[HTML]{EFEFEF}99.41                         \\ \specialrule{0.2pt}{0pt}{0pt}
\multirow{5}{*}{\textbf{GTSRB}}   
& 0.0\%                                    & 97.11                         & \cellcolor[HTML]{EFEFEF}-                             & 97.00                         & \cellcolor[HTML]{EFEFEF}-                             & 97.23                         &\cellcolor[HTML]{EFEFEF} -                             & 97.22                         &\cellcolor[HTML]{EFEFEF} -                             & 96.23                         & \cellcolor[HTML]{EFEFEF}-                             & 97.76                         & \cellcolor[HTML]{EFEFEF}-                             \\ \hhline{~-------------}  
                          & 0.1\%                                    & 97.09                         & \cellcolor[HTML]{EFEFEF}71.12                         & 96.90                         & \cellcolor[HTML]{EFEFEF}74.52                        & 97.41                         & \cellcolor[HTML]{EFEFEF}82.32                         & 97.16                         & \cellcolor[HTML]{EFEFEF}76.82                         & 96.48                         & \cellcolor[HTML]{EFEFEF}0.00                          & 97.29                         & \cellcolor[HTML]{EFEFEF}71.38                         \\ \hhline{~-------------}  
                          & 0.2\%                                    & 97.19                         & \cellcolor[HTML]{EFEFEF}89.59                         & 97.06                         & \cellcolor[HTML]{EFEFEF}83.69                         & 97.14                         & \cellcolor[HTML]{EFEFEF}86.25                         & 97.11                         & \cellcolor[HTML]{EFEFEF}99.61                         & 96.74                         & \cellcolor[HTML]{EFEFEF}0.20                          & 97.64                         & \cellcolor[HTML]{EFEFEF}88.74                         \\ \hhline{~-------------}  
                          & 0.4\%                                    & 97.33                         & \cellcolor[HTML]{EFEFEF}98.04                         & 96.73                         & \cellcolor[HTML]{EFEFEF}97.25                         & 96.95                         & \cellcolor[HTML]{EFEFEF}97.25                         & 97.43                         & \cellcolor[HTML]{EFEFEF}99.61                         & 96.01                         & \cellcolor[HTML]{EFEFEF}2.95                          & 97.43                         & \cellcolor[HTML]{EFEFEF}98.61                       \\ \hhline{~-------------}  
                          & 1.0\%                                    & 97.25                         & \cellcolor[HTML]{EFEFEF}98.62                         & 97.03                         & \cellcolor[HTML]{EFEFEF}99.61                         & 97.22                         & \cellcolor[HTML]{EFEFEF}98.04                         & 97.17                         & \cellcolor[HTML]{EFEFEF}99.61                         & 96.27                         & \cellcolor[HTML]{EFEFEF}88.41                         & 96.87                         & \cellcolor[HTML]{EFEFEF}99.76                        \\ \hhline{~-------------}  
                          & 3.0\%                                    & 97.47                         & \cellcolor[HTML]{EFEFEF}99.80                         & 96.76                         & \cellcolor[HTML]{EFEFEF}99.98                       & 96.98                         & \cellcolor[HTML]{EFEFEF}99.97                       & 97.49                         & \cellcolor[HTML]{EFEFEF}100.00                       & 96.29                         & \cellcolor[HTML]{EFEFEF}99.61                         & 97.26                         & \cellcolor[HTML]{EFEFEF}99.97                       \\ \specialrule{0.2pt}{0pt}{0pt}
\multirow{4}{*}{\textbf{ImageNet}}& 0.0\%                                    & 67.51                         & \cellcolor[HTML]{EFEFEF}-                             & 70.86                         & \cellcolor[HTML]{EFEFEF}-                             & 73.35                         & \cellcolor[HTML]{EFEFEF}-                             & 74.10                         & \cellcolor[HTML]{EFEFEF}-                             & 72.11                             & \cellcolor[HTML]{EFEFEF}-                             & -                             & \cellcolor[HTML]{EFEFEF}-                             \\ \hhline{~-------------}  
                          & 0.5\%                                    & 67.38                         & \cellcolor[HTML]{EFEFEF}0.17                          & 71.20                         & \cellcolor[HTML]{EFEFEF}84.70                         & 73.27                         & \cellcolor[HTML]{EFEFEF}96.00                         & 73.91                         & \cellcolor[HTML]{EFEFEF}95.32                         & 71.49                             & \cellcolor[HTML]{EFEFEF}91.96                             & -                             & \cellcolor[HTML]{EFEFEF}-                             \\ \hhline{~-------------}  
                          & 1.0\%                                    & 67.13                         & \cellcolor[HTML]{EFEFEF}87.74                         & 70.74                         & \cellcolor[HTML]{EFEFEF}95.96                         & 73.38                         & \cellcolor[HTML]{EFEFEF}98.03                         & 74.21                         & \cellcolor[HTML]{EFEFEF}98.05                         & 72.33                             & \cellcolor[HTML]{EFEFEF}96.64                             & -                             & \cellcolor[HTML]{EFEFEF}-                             \\ \hhline{~-------------}  
                          & 2.0\%                                    & 67.26                         & \cellcolor[HTML]{EFEFEF}98.01                         & 70.57                         & \cellcolor[HTML]{EFEFEF}98.87                         & 72.78                         & \cellcolor[HTML]{EFEFEF}98.85                         & 73.75                         & \cellcolor[HTML]{EFEFEF}99.34                         & 71.62                             & \cellcolor[HTML]{EFEFEF}95.379                            & -                             & \cellcolor[HTML]{EFEFEF}-                             \\ \hhline{~-------------}  
                          & 3.0\%                                    & 67.26                         & \cellcolor[HTML]{EFEFEF}98.32                         & 70.67                         & \cellcolor[HTML]{EFEFEF}98.95                         & 72.30                         & \cellcolor[HTML]{EFEFEF}99.25                         & 73.39                         & \cellcolor[HTML]{EFEFEF}99.85                         & 72.05                             & \cellcolor[HTML]{EFEFEF}97.51                             & -                             & \cellcolor[HTML]{EFEFEF}-                             \\ \bottomrule
\end{tabular}}
\vspace{10pt}
 \caption{\textbf{Evaluation of the proposed backdoor attack.} We benchmark our proposed frequency-based backdoor attack on different
network architectures, datasets, and poisoning rates. These results show that our attack can maintain clean data accuracy, while registering
high attack success rates even with small poisoning rates. 
 }
 \label{Table:Main}
\end{table*}

\clearpage
\section{Evaluation of Augmented Models}
\label{sec2}

The manuscript discusses two defense techniques against frequency-based backdoor attacks and a simple technique to bypass them through training data augmentation.
The results presented in the paper correspond to ResNet18 trained on CIFAR10, CIFAR100, and GTSRB.
Tables \ref{table2_a}, \ref{table2_b} and \ref{table2_c} present the results for ResNet34, WideResNet34, and VGG19 trained on the aforementioned datasets with training data augmentation.
Based on these results, training data augmentation was shown to be a viable counter-attack to the proposed backdoor defenses.

\subsection{ResNet34}
\begin{table}[htbp!]
\centering
\scalebox{0.75}{
\begin{tabular}{cccccccc}
\cline{2-8}
                                    &                         & \multicolumn{2}{c}{\textbf{Autoencoder}}                         & \multicolumn{2}{c}{\textbf{JPEG}}            & \multicolumn{2}{c}{\textbf{JPEG+Autoencoder}} \\ \cline{2-8} 
                                    & \textbf{Poisoning Rate} & \textbf{CDA} & \textbf{ASR}                                      & \textbf{CDA} & \textbf{ASR}                  & \textbf{CDA}  & \textbf{ASR}                  \\ \hline
                                    & 0.1\%                   & 94.75        & \cellcolor[HTML]{EFEFEF}0.12                      & 94.17        & \cellcolor[HTML]{EFEFEF}0.22  & 94.75         & \cellcolor[HTML]{EFEFEF}0.44  \\ \cline{2-8} 
                                    & 0.2\%                   & 94.49        & \cellcolor[HTML]{EFEFEF}0.88                      & 94.75        & \cellcolor[HTML]{EFEFEF}2.86  & 94.26         & \cellcolor[HTML]{EFEFEF}1.32  \\ \cline{2-8} 
                                    & 0.4\%                   & 94.71        & \cellcolor[HTML]{EFEFEF}80.66                     & 94.68        & \cellcolor[HTML]{EFEFEF}78.90 & 93.95         & \cellcolor[HTML]{EFEFEF}17.14 \\ \cline{2-8} 
                                    & 1.0\%                   & 94.30        & \cellcolor[HTML]{EFEFEF}95.82                     & 94.49        & \cellcolor[HTML]{EFEFEF}87.91 & 94.11         & \cellcolor[HTML]{EFEFEF}92.53 \\ \cline{2-8} 
\multirow{-5}{*}{\textbf{CIFAR10}}  & 3.0\%                   & 94.51        & \cellcolor[HTML]{EFEFEF}98.46                     & 94.47        & \cellcolor[HTML]{EFEFEF}97.58 & 94.50         & \cellcolor[HTML]{EFEFEF}94.73 \\ \hline
                                    & 0.1\%                   & 76.84        & \cellcolor[HTML]{EFEFEF}10.70                     & 77.53        & \cellcolor[HTML]{EFEFEF}61.39 & 77.52         & \cellcolor[HTML]{EFEFEF}3.56  \\ \cline{2-8} 
                                    & 0.2\%                   & 76.08        & \cellcolor[HTML]{EFEFEF}19.41                     & 76.55        & \cellcolor[HTML]{EFEFEF}84.55 & 76.39         & \cellcolor[HTML]{EFEFEF}38.81 \\ \cline{2-8} 
                                    & 0.4\%                   & 77.56        & \cellcolor[HTML]{EFEFEF}95.44                     & 76.49        & \cellcolor[HTML]{EFEFEF}95.84 & 77.12         & \cellcolor[HTML]{EFEFEF}93.27 \\ \cline{2-8} 
                                    & 1.0\%                   & 76.98        & \cellcolor[HTML]{EFEFEF}99.60                     & 77.20        & \cellcolor[HTML]{EFEFEF}96.63 & 77.31         & \cellcolor[HTML]{EFEFEF}97.22 \\ \cline{2-8} 
\multirow{-5}{*}{\textbf{CIFAR100}} & 3.0\%                   & 76.53        & \cellcolor[HTML]{EFEFEF}99.61                     & 76.26        & \cellcolor[HTML]{EFEFEF}99.60 & 76.43         & \cellcolor[HTML]{EFEFEF}99.00 \\ \hline
                                    & 0.1\%                   & 97.25        & \multicolumn{1}{c}{\cellcolor[HTML]{EFEFEF}43.81} & 97.19        & \cellcolor[HTML]{EFEFEF}59.33 & 96.96         & \cellcolor[HTML]{EFEFEF}48.33 \\ \cline{2-8} 
                                    & 0.2\%                   & 96.96        & \multicolumn{1}{c}{\cellcolor[HTML]{EFEFEF}88.02} & 97.18        & \cellcolor[HTML]{EFEFEF}86.64 & 97.05         & \cellcolor[HTML]{EFEFEF}78.78 \\ \cline{2-8} 
                                    & 0.4\%                   & 97.11        & \multicolumn{1}{c}{\cellcolor[HTML]{EFEFEF}93.91} & 97.09        & \cellcolor[HTML]{EFEFEF}88.21 & 96.78         & \cellcolor[HTML]{EFEFEF}94.50 \\ \cline{2-8} 
                                    & 1.0\%                   & 95.25        & \multicolumn{1}{c}{\cellcolor[HTML]{EFEFEF}91.16} & 96.84        & \cellcolor[HTML]{EFEFEF}97.05 & 93.99         & \cellcolor[HTML]{EFEFEF}80.35 \\ \cline{2-8} 
\multirow{-5}{*}{\textbf{GTSRB}}    & 3.0\%                   & 94.61        & \multicolumn{1}{c}{\cellcolor[HTML]{EFEFEF}98.82} & 96.92        & \cellcolor[HTML]{EFEFEF}99.02 & 94.79         & \cellcolor[HTML]{EFEFEF}88.41 \\ \hline
\end{tabular}}
\vspace{10pt}
\caption{\textbf{Augmentation Maintains Performance (CIFAR10, CIFAR100 and GTSRB):} CDA and ASR of backdoored ResNet34 trained on CIFAR10, CIFAR100 and GTSRB with JPEG compression and Autoencoder augmentation. Both ASR and CDA are maintained even when no preprocessing technique is used.}
\label{table2_a}
\end{table}

\clearpage

\subsection{WideResNet34}

\begin{table}[htbp!]
\centering
\scalebox{0.75}{
\begin{tabular}{cccccccc}
\cline{2-8}
                                    &                         & \multicolumn{2}{c}{\textbf{Autoencoder}}                         & \multicolumn{2}{c}{\textbf{JPEG}}            & \multicolumn{2}{c}{\textbf{JPEG+Autoencoder}} \\ \cline{2-8} 
                                    & \textbf{Poisoning Rate} & \textbf{CDA} & \textbf{ASR}                                      & \textbf{CDA} & \textbf{ASR}                  & \textbf{CDA}  & \textbf{ASR}                  \\ \hline
                                    & 0.1\%                   & 95.70        & \cellcolor[HTML]{EFEFEF}63.96                     & 95.40        & \cellcolor[HTML]{EFEFEF}69.01 & 95.63         & \cellcolor[HTML]{EFEFEF}1.98  \\ \cline{2-8} 
                                    & 0.2\%                   & 95.44        & \cellcolor[HTML]{EFEFEF}89.01                     & 95.30        & \cellcolor[HTML]{EFEFEF}86.59 & 95.69         & \cellcolor[HTML]{EFEFEF}80.88 \\ \cline{2-8} 
                                    & 0.4\%                   & 95.26        & \cellcolor[HTML]{EFEFEF}97.14                     & 95.57        & \cellcolor[HTML]{EFEFEF}88.57 & 95.32         & \cellcolor[HTML]{EFEFEF}85.49 \\ \cline{2-8} 
                                    & 1.0\%                   & 95.13        & \cellcolor[HTML]{EFEFEF}98.46                     & 95.56        & \cellcolor[HTML]{EFEFEF}97.36 & 95.46         & \cellcolor[HTML]{EFEFEF}95.82 \\ \cline{2-8} 
\multirow{-5}{*}{\textbf{CIFAR10}}  & 3.0\%                   & 95.48        & \cellcolor[HTML]{EFEFEF}99.12                     & 95.37        & \cellcolor[HTML]{EFEFEF}98.02 & 95.53         & \cellcolor[HTML]{EFEFEF}98.24 \\ \hline
                                    & 0.1\%                   & 79.15        & \cellcolor[HTML]{EFEFEF}25.94                     & 79.27        & \cellcolor[HTML]{EFEFEF}19.60 & 79.31         & \cellcolor[HTML]{EFEFEF}8.71  \\ \cline{2-8} 
                                    & 0.2\%                   & 79.62        & \cellcolor[HTML]{EFEFEF}63.76                     & 79.46        & \cellcolor[HTML]{EFEFEF}60.20 & 79.60         & \cellcolor[HTML]{EFEFEF}35.84 \\ \cline{2-8} 
                                    & 0.4\%                   & 79.72        & \cellcolor[HTML]{EFEFEF}87.92                     & 79.19        & \cellcolor[HTML]{EFEFEF}79.41 & 79.36         & \cellcolor[HTML]{EFEFEF}79.60 \\ \cline{2-8} 
                                    & 1.0\%                   & 79.22        & \cellcolor[HTML]{EFEFEF}93.47                     & 79.29        & \cellcolor[HTML]{EFEFEF}93.66 & 79.23         & \cellcolor[HTML]{EFEFEF}73.47 \\ \cline{2-8} 
\multirow{-5}{*}{\textbf{CIFAR100}} & 3.0\%                   & 79.27        & \cellcolor[HTML]{EFEFEF}98.22                     & 79.11        & \cellcolor[HTML]{EFEFEF}95.64 & 78.81         & \cellcolor[HTML]{EFEFEF}90.30 \\ \hline
                                    & 0.1\%                   & 97.70        & \multicolumn{1}{c}{\cellcolor[HTML]{EFEFEF}54.03} & 97.51        & \cellcolor[HTML]{EFEFEF}60.12 & 97.47         & \cellcolor[HTML]{EFEFEF}48.72 \\ \cline{2-8} 
                                    & 0.2\%                   & 97.02        & \multicolumn{1}{c}{\cellcolor[HTML]{EFEFEF}86.44} & 97.70        & \cellcolor[HTML]{EFEFEF}72.10 & 96.84         & \cellcolor[HTML]{EFEFEF}34.58 \\ \cline{2-8} 
                                    & 0.4\%                   & 95.72        & \multicolumn{1}{c}{\cellcolor[HTML]{EFEFEF}80.75} & 96.96        & \cellcolor[HTML]{EFEFEF}97.64 & 93.84         & \cellcolor[HTML]{EFEFEF}80.16 \\ \cline{2-8} 
                                    & 1.0\%                   & 92.15        & \multicolumn{1}{c}{\cellcolor[HTML]{EFEFEF}46.95} & 93.03        & \cellcolor[HTML]{EFEFEF}43.81 & 92.30         & \cellcolor[HTML]{EFEFEF}88.45 \\ \cline{2-8} 
\multirow{-5}{*}{\textbf{GTSRB}}    & 3.0\%                   & 93.12        & \multicolumn{1}{c}{\cellcolor[HTML]{EFEFEF}93.52} & 90.63        & \cellcolor[HTML]{EFEFEF}78.19 & 89.46         & \cellcolor[HTML]{EFEFEF}89.78 \\ \hline
\end{tabular}}
\vspace{10pt}
\caption{\textbf{Augmentation Maintains Performance (CIFAR10, CIFAR100 and GTSRB):} CDA and ASR of backdoored WideResNet34 trained on CIFAR10, CIFAR100 and GTSRB with JPEG compression and Autoencoder augmentation. Both ASR and CDA are maintained even when no preprocessing technique is used.}
\label{table2_b}
\end{table}

\subsection{VGG19}

\begin{table}[htbp!]
\centering
\scalebox{0.75}{
\begin{tabular}{cccccccc}
\cline{2-8}
                                    &                         & \multicolumn{2}{c}{\textbf{Autoencoder}}     & \multicolumn{2}{c}{\textbf{JPEG}}             & \multicolumn{2}{c}{\textbf{JPEG+Autoencoder}} \\ \cline{2-8} 
                                    & \textbf{Poisoning Rate} & \textbf{CDA} & \textbf{ASR}                  & \textbf{CDA} & \textbf{ASR}                   & \textbf{CDA} & \textbf{ASR}                   \\ \hline
                                    & 0.1\%                   & 91.82        & \cellcolor[HTML]{EFEFEF}1.10  & 92.13        & \cellcolor[HTML]{EFEFEF}0.44   & 92.65        & \cellcolor[HTML]{EFEFEF}0.44   \\ \cline{2-8} 
                                    & 0.2\%                   & 92.39        & \cellcolor[HTML]{EFEFEF}0.66  & 92.28        & \cellcolor[HTML]{EFEFEF}0.66   & 92.36        & \cellcolor[HTML]{EFEFEF}0.88   \\ \cline{2-8} 
                                    & 0.4\%                   & 92.30        & \cellcolor[HTML]{EFEFEF}1.10  & 92.16        & \cellcolor[HTML]{EFEFEF}7.47   & 92.43        & \cellcolor[HTML]{EFEFEF}1.76   \\ \cline{2-8} 
                                    & 1.0\%                   & 92.04        & \cellcolor[HTML]{EFEFEF}89.89 & 92.60        & \cellcolor[HTML]{EFEFEF}97.36  & 92.16        & \cellcolor[HTML]{EFEFEF}86.15  \\ \cline{2-8} 
\multirow{-5}{*}{\textbf{CIFAR10}}  & 3.0\%                   & 92.52        & \cellcolor[HTML]{EFEFEF}99.56 & 92.21        & \cellcolor[HTML]{EFEFEF}100.00 & 91.89        & \cellcolor[HTML]{EFEFEF}98.90  \\ \hline
                                    & 0.1\%                   & 68.91        & \cellcolor[HTML]{EFEFEF}0.20  & 68.84        & \cellcolor[HTML]{EFEFEF}0.21   & 68.82        & \cellcolor[HTML]{EFEFEF}0.20   \\ \cline{2-8} 
                                    & 0.2\%                   & 68.24        & \cellcolor[HTML]{EFEFEF}0.59  & 68.71        & \cellcolor[HTML]{EFEFEF}0.40   & 68.76        & \cellcolor[HTML]{EFEFEF}1.19   \\ \cline{2-8} 
                                    & 0.4\%                   & 68.50        & \cellcolor[HTML]{EFEFEF}2.57  & 68.36        & \cellcolor[HTML]{EFEFEF}1.20   & 68.74        & \cellcolor[HTML]{EFEFEF}2.18   \\ \cline{2-8} 
                                    & 1.0\%                   & 68.36        & \cellcolor[HTML]{EFEFEF}8.91  & 68.12        & \cellcolor[HTML]{EFEFEF}4.95   & 68.12        & \cellcolor[HTML]{EFEFEF}7.37   \\ \cline{2-8} 
\multirow{-5}{*}{\textbf{CIFAR100}} & 3.0\%                   & 67.70        & \cellcolor[HTML]{EFEFEF}98.02 & 68.12        & \cellcolor[HTML]{EFEFEF}97.82  & 68.14        & \cellcolor[HTML]{EFEFEF}95.05  \\ \hline
                                    & 0.1\%                   & 96.69        & \cellcolor[HTML]{EFEFEF}0.00  & 96.33        & \cellcolor[HTML]{EFEFEF}0.00   & 96.81        & \cellcolor[HTML]{EFEFEF}0.00   \\ \cline{2-8} 
                                    & 0.2\%                   & 96.39        & \cellcolor[HTML]{EFEFEF}0.20  & 96.43        & \cellcolor[HTML]{EFEFEF}0.20   & 96.62        & \cellcolor[HTML]{EFEFEF}0.79   \\ \cline{2-8} 
                                    & 0.4\%                   & 95.87        & \cellcolor[HTML]{EFEFEF}0.00  & 96.04        & \cellcolor[HTML]{EFEFEF}0.00   & 95.98        & \cellcolor[HTML]{EFEFEF}24.36  \\ \cline{2-8} 
                                    & 1.0\%                   & 95.63        & \cellcolor[HTML]{EFEFEF}88.61 & 96.00        & \cellcolor[HTML]{EFEFEF}91.55  & 96.28        & \cellcolor[HTML]{EFEFEF}89.00  \\ \cline{2-8} 
\multirow{-5}{*}{\textbf{GTSRB}}    & 3.0\%                   & 96.22        & \cellcolor[HTML]{EFEFEF}99.41 & 95.82        & \cellcolor[HTML]{EFEFEF}98.82  & 95.97        & \cellcolor[HTML]{EFEFEF}100.00 \\ \hline
\end{tabular}}
\vspace{10pt}
\caption{\textbf{Augmentation Maintains Performance (CIFAR10, CIFAR100 and GTSRB):} CDA and ASR of backdoored VGG19 trained on CIFAR10, CIFAR100 and GTSRB with JPEG compression and Autoencoder augmentation. Both ASR and CDA are maintained even when no preprocessing technique is used.}
\label{table2_c}
\end{table}

\clearpage
\section{Choice of Additive Filters $\mathcal{A}_{R,G,B}$}
\label{sec3}

The results presented in the manuscript set the values of the additive filters to be the same within the channel but different across the channels.
We now consider different possible design choices for this additive filter, namely, choosing random or same values (within and across channels) for $\mathcal{A}_{R,G,B}$.
Tables \ref{3_a} and \ref{3_b} both show high ASR and CDA for different choices of $\mathcal{A}_{R,G,B}$ illustrating the flexibility of the proposed method in creating backdoor attacks.

\subsection{Random Values for $\mathcal{A}_{R,G,B}$}
\begin{table}[htbp!]
\centering
\scalebox{0.9}{
\begin{tabular}{ccccccc}
\hline
                        & \multicolumn{2}{c}{\textbf{ResNet18}}            & \multicolumn{2}{c}{\textbf{ResNet34}}            & \multicolumn{2}{c}{\textbf{VGG19}}               \\ \hline
\textbf{Poisoning Rate} & \textbf{CDA(\%)} & \textbf{ASR(\%)}              & \textbf{CDA(\%)} & \textbf{ASR(\%)}              & \textbf{CDA(\%)} & \textbf{ASR(\%)}              \\ \hline
0.1\%                   & 92.93            & \cellcolor[HTML]{EFEFEF}2.64  & 93.23            & \cellcolor[HTML]{EFEFEF}0.88  & 92.12            & \cellcolor[HTML]{EFEFEF}2.20  \\ \hline
0.2\%                   & 92.83            & \cellcolor[HTML]{EFEFEF}37.80 & 93.28            & \cellcolor[HTML]{EFEFEF}11.21 & 91.86            & \cellcolor[HTML]{EFEFEF}22.20 \\ \hline
0.4\%                   & 93.16            & \cellcolor[HTML]{EFEFEF}90.11 & 93.49            & \cellcolor[HTML]{EFEFEF}93.19 & 92.18            & \cellcolor[HTML]{EFEFEF}56.68 \\ \hline
1.0\%                   & 93.04            & \cellcolor[HTML]{EFEFEF}97.58 & 93.20            & \cellcolor[HTML]{EFEFEF}98.90 & 92.26            & \cellcolor[HTML]{EFEFEF}95.60 \\ \hline
3.0\%                   & 93.21            & \cellcolor[HTML]{EFEFEF}99.56 & 93.29            & \cellcolor[HTML]{EFEFEF}99.78 & 92.28            & \cellcolor[HTML]{EFEFEF}98.90 \\ \hline
\end{tabular}}
\vspace{10pt}
\caption{\textbf{Random Additive Filter Values.} Evaluating the proposed backdoor attack using random additive filter values shows that our attack can maintain clean data accuracy while reaching high attack success rates with small poisoning rates.}
\label{3_a}
\end{table}

\subsection{Same Value for $\mathcal{A}_{R,G,B}$}
\begin{table}[htbp!]
\centering
\scalebox{0.9}{
\begin{tabular}{ccccccc}
\hline
                        & \multicolumn{2}{c}{\textbf{ResNet18}}            & \multicolumn{2}{c}{\textbf{ResNet34}}            & \multicolumn{2}{c}{\textbf{VGG19}}               \\ \hline
\textbf{Poisoning Rate} & \textbf{CDA(\%)} & \textbf{ASR(\%)}              & \textbf{CDA(\%)} & \textbf{ASR(\%)}              & \textbf{CDA(\%)} & \textbf{ASR(\%)}              \\ \hline
0.1\%                   & 93.01            & \cellcolor[HTML]{EFEFEF}1.76  & 93.55            & \cellcolor[HTML]{EFEFEF}1.32  & 91.93            & \cellcolor[HTML]{EFEFEF}1.54  \\ \hline
0.2\%                   & 92.95            & \cellcolor[HTML]{EFEFEF}66.59 & 93.33            & \cellcolor[HTML]{EFEFEF}22.64 & 92.15            & \cellcolor[HTML]{EFEFEF}29.45 \\ \hline
0.4\%                   & 92.89            & \cellcolor[HTML]{EFEFEF}92.08 & 93.31            & \cellcolor[HTML]{EFEFEF}93.85 & 91.80            & \cellcolor[HTML]{EFEFEF}66.37 \\ \hline
1.0\%                   & 93.18            & \cellcolor[HTML]{EFEFEF}98.24 & 93.42            & \cellcolor[HTML]{EFEFEF}98.46 & 92.10            & \cellcolor[HTML]{EFEFEF}97.14 \\ \hline
3.0\%                   & 92.87            & \cellcolor[HTML]{EFEFEF}99.56 & 93.06            & \cellcolor[HTML]{EFEFEF}99.34 & 91.94            & \cellcolor[HTML]{EFEFEF}99.12 \\ \hline
\end{tabular}}
\vspace{10pt}
\caption{\textbf{Same Additive Filter Values.} Evaluating the proposed backdoor attack using same additive filter values (across and within channels) shows that our attack can maintain clean data accuracy while reaching high attack success rates with small poisoning rates.}
\label{3_b}
\end{table}
\clearpage 
\clearpage
\section{Multi Target Attacks in the Frequency Domain}
\label{sec4}

The manuscript focuses on creating single-target backdoor attacks.
We extend the applicability of the proposed frequency-based backdoor attack to the multitarget regime.
This is done through introducing an additional step to the recipe: 
\begin{enumerate}
    \item Select the top-$k$ frequencies (most sensitive).
    \item Randomly and equally divide the selected frequencies among the poisoned classes creating a binary mask for each.
    \item Create a set of additive filters for each poisoned class.
    \item Poison each class with its corresponding additive filter and binary mask.
    \item Proceed with training.
\end{enumerate}

\begin{figure}[h]
    \centering
    \includegraphics[width=0.8\linewidth]{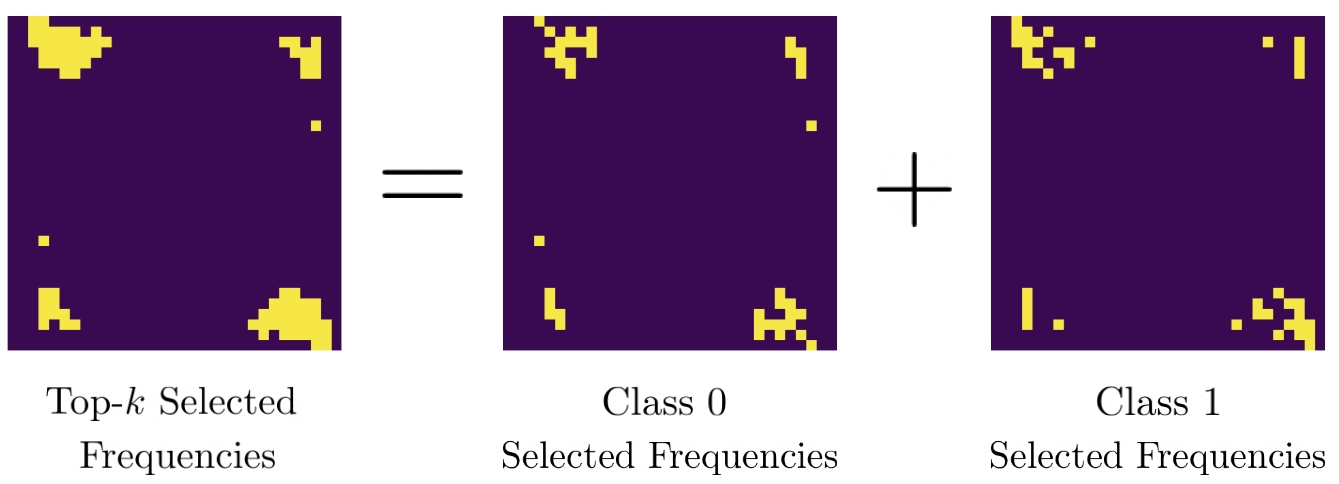}
    \caption{\textbf{Multitarget (2 classes) Binary Filters for ResNet18 on CIFAR10.} The top-$k$ selected frequencies to poison are divided equally and randomly to create two binary masks one for each poisoned class.}
    \label{Fig4_a}
\end{figure}

Figure \ref{Fig4_a} shows the binary masks for the two poisoned classes of ResNet18 trained on CIFAR10; Table \ref{tab4} shows the results for poisoning the first $2$ classes of CIFAR10 for various network architectures.

\begin{table}[h]
\scalebox{0.6}{
\begin{tabular}{cccccccccc}
\hline
                        & \multicolumn{3}{c}{\textbf{ResNet18}}                                            & \multicolumn{3}{c}{\textbf{ResNet34}}                                            & \multicolumn{3}{c}{\textbf{VGG19}}                                               \\ \hline
\textbf{Poisoning Rate} & \textbf{CDA(\%)} & \textbf{ASR-0 (\%)}           & \textbf{ASR-1 (\%)}           & \textbf{CDA(\%)} & \textbf{ASR-0 (\%)}           & \textbf{ASR-1 (\%)}           & \textbf{CDA(\%)} & \textbf{ASR-0 (\%)}           & \textbf{ASR-1 (\%)}           \\ \hline
0.1\%                   & 93.09            & \cellcolor[HTML]{EFEFEF}1.76  & \cellcolor[HTML]{E4E4E4}1.27  & 93.32            & \cellcolor[HTML]{EFEFEF}0.88  & \cellcolor[HTML]{E4E4E4}0.21  & 91.81            & \cellcolor[HTML]{EFEFEF}1.54  & \cellcolor[HTML]{E4E4E4}1.06  \\ \hline
0.2\%                   & 92.76            & \cellcolor[HTML]{EFEFEF}32.31 & \cellcolor[HTML]{E4E4E4}28.23 & 93.24            & \cellcolor[HTML]{EFEFEF}10.32 & \cellcolor[HTML]{E4E4E4}10.82 & 91.97            & \cellcolor[HTML]{EFEFEF}8.13  & \cellcolor[HTML]{E4E4E4}9.98  \\ \hline
0.4\%                   & 92.89            & \cellcolor[HTML]{EFEFEF}96.70 & \cellcolor[HTML]{E4E4E4}87.69 & 93.59            & \cellcolor[HTML]{EFEFEF}86.59 & \cellcolor[HTML]{E4E4E4}72.40 & 91.90            & \cellcolor[HTML]{EFEFEF}61.53 & \cellcolor[HTML]{E4E4E4}69.21 \\ \hline
1.0\%                   & 93.09            & \cellcolor[HTML]{EFEFEF}98.90 & \cellcolor[HTML]{E4E4E4}96.60 & 93.59            & \cellcolor[HTML]{EFEFEF}98.46 & \cellcolor[HTML]{E4E4E4}97.88 & 92.17            & \cellcolor[HTML]{EFEFEF}92.75 & \cellcolor[HTML]{E4E4E4}94.90 \\ \hline
3.0\%                   & 92.87            & \cellcolor[HTML]{EFEFEF}99.86 & \cellcolor[HTML]{E4E4E4}99.57 & 93.18            & \cellcolor[HTML]{EFEFEF}99.78 & \cellcolor[HTML]{E4E4E4}99.79 & 91.93            & \cellcolor[HTML]{EFEFEF}98.68 & \cellcolor[HTML]{E4E4E4}99.15 \\ \hline
\end{tabular}}
\vspace{10pt}
\caption{\textbf{Multitarget Frequency-Based Backdoor Attack.} The proposed multitarget variant of the frequency-based backdoor attack can successfully poison the first two classes of CIFAR10 on various network architectures using small poisoning rate. ASR-0 and ASR-1 denote the attack success rate for triggering classes 0 and 1 respectively. }
\label{tab4}
\end{table}

\clearpage
\clearpage
\section{End-to-End Pipeline: A More Efficient Variant}
\label{Sec:Efficient}

In this section, we present a more efficient variant of the proposed method. The method proposed in the manuscript requires training two models: (1) a clean model ($f_0$) for which the Fourier heatmap is computed for; and (2) a poisoned model ($f$) which utilizes the heatmap generated from the clean model to poison the data and hence embed the backdoor.

Our method could be modified so that only one model is trained, the modified version is summarized below:

\begin{enumerate}
    \item Train a model on clean samples until a reasonable performance is reached. We denote this checkpoint by $C_0$.
    \item Generate the Fourier heatmap for $C_0$ and select the top-$k$ most sensitive frequencies to generate the binary mask $\mathcal{M}$ and the additive filters $\mathcal{A}_{R,G,B}$.
    \item Poison the data using equations (4), (5) and (6) presented in the manuscript and proceed with training $C_0$ on both poisoned and clean samples. The obtained model is the poisoned model $f$.
\end{enumerate}

\begin{figure}[t!]
\centering
    \begin{subfigure}[t]{0.2\textwidth}
      \centering
    \includegraphics[width=\textwidth]{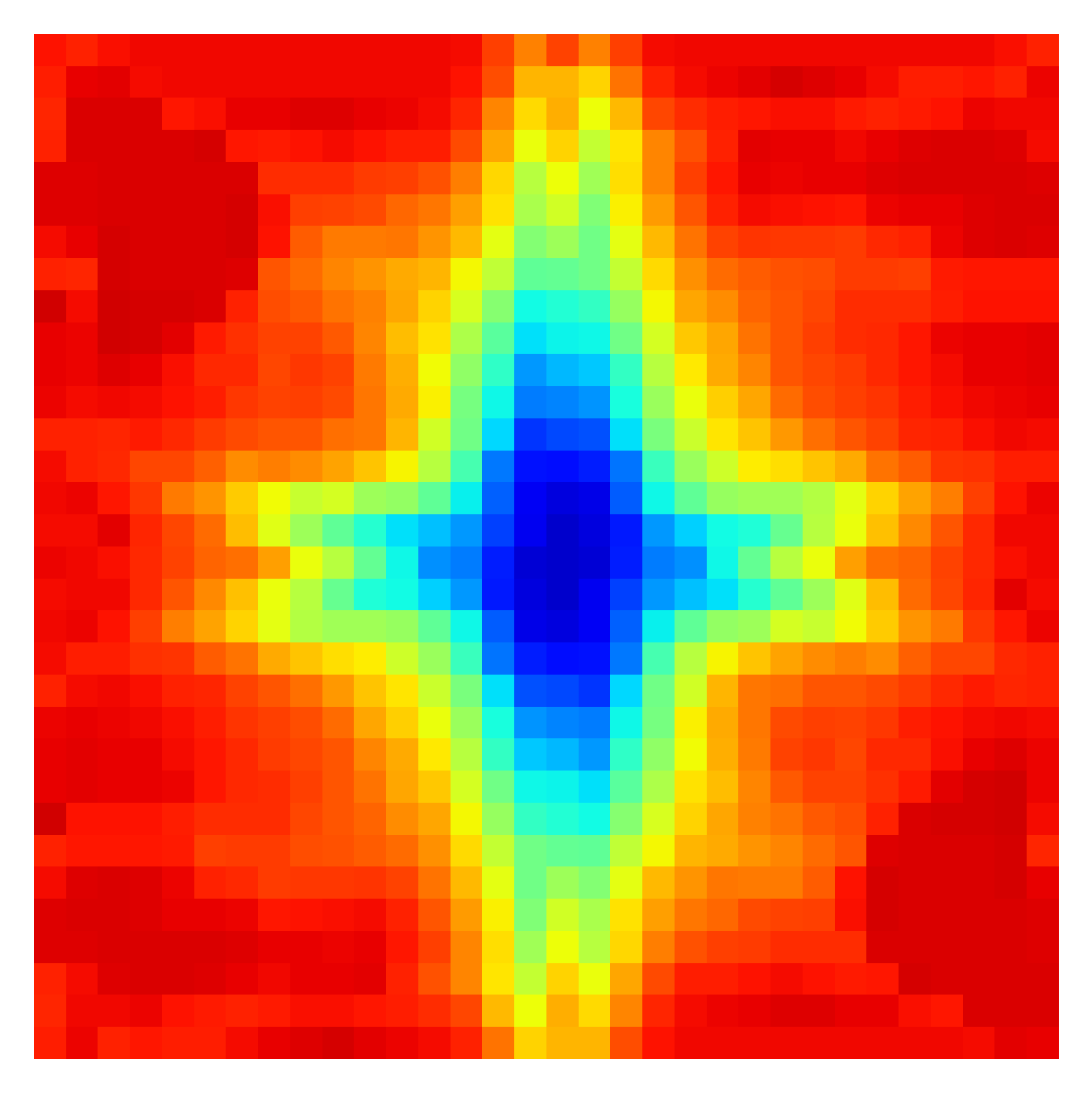}
    \caption{Heatmap of Clean Model}
    \label{fig-b-2}
  \end{subfigure}
  \hspace{2cm}
  \begin{subfigure}[t]{0.2\textwidth}
    \centering
    \includegraphics[width=\textwidth]{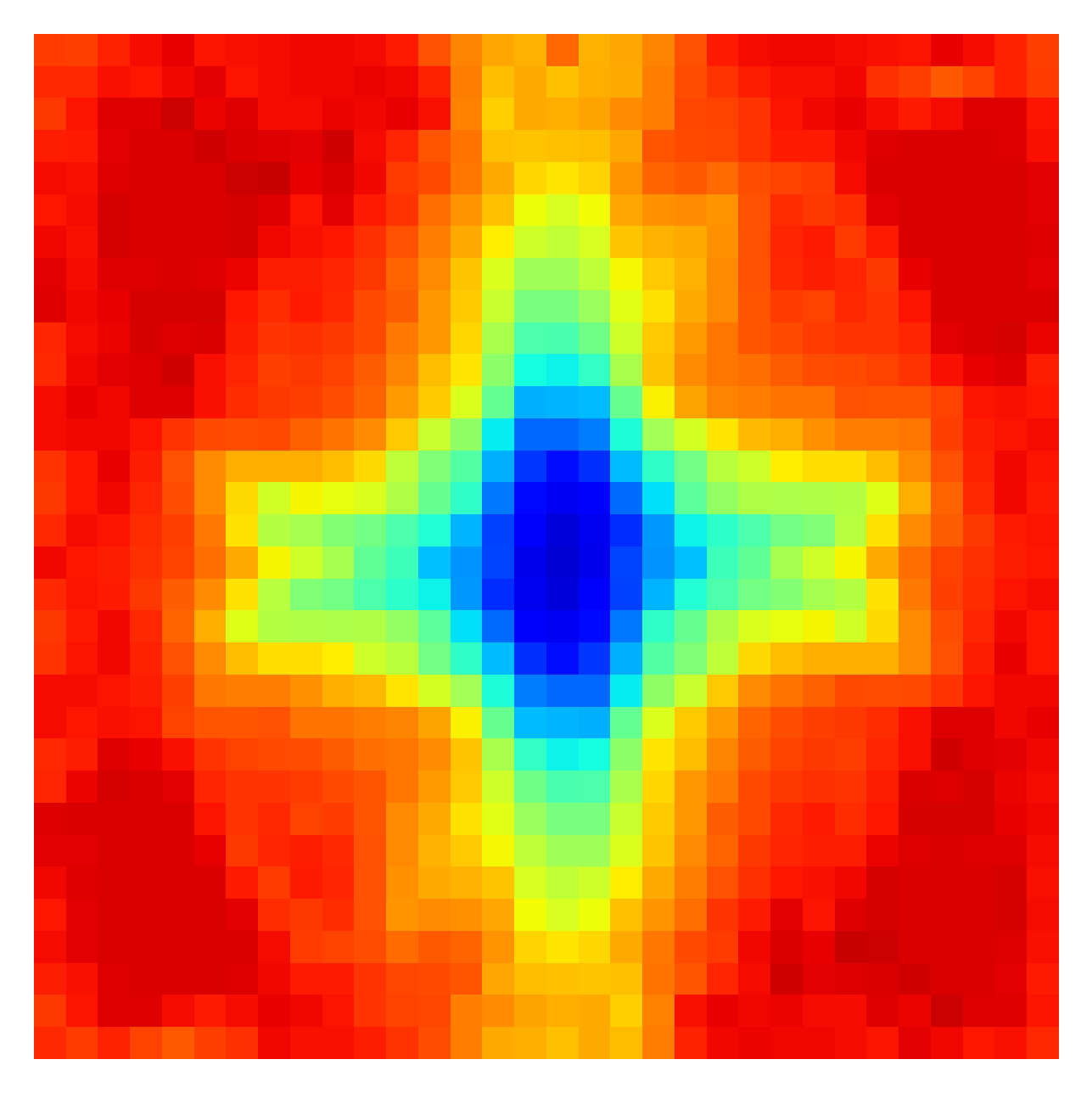}
    \caption{Heatmap of Poisoned Model}
    \label{222}
  \end{subfigure}
\vspace{10pt}
  \caption{\textbf{End-to-End Variant Maintains Fourier Heatmaps.} Utilizing the end-to-end frequency-based backdoor attack allows us to obtain a backdoored model with a Fourier heatmap similar to that of the clean model.}
  \label{10_b}
\end{figure}

Table \ref{10_a} shows the results of using the "end-to-end" variant of the proposed frequency-based backdoor attack. The obtained results are fairly similar to those shown in Table \ref{Table:Main}. Figure \ref{10_b} shows the Fourier heatmaps of the clean and poisoned models for the proposed variant (ResNet18 trained on CIFAR10). As required, the Fourier heatmap of the poisoned model is similar to that of the clean model.

\begin{table}[htbp!]
\centering
\scalebox{0.8}{
\begin{tabular}{ccccccc}
\hline
                        & \multicolumn{2}{c}{\textbf{ResNet18}}            & \multicolumn{2}{c}{\textbf{ResNet34}}            & \multicolumn{2}{c}{\textbf{VGG19}}               \\ \hline
\textbf{Poisoning Rate} & \textbf{CDA(\%)} & \textbf{ASR(\%)}              & \textbf{CDA(\%)} & \textbf{ASR(\%)}              & \textbf{CDA(\%)} & \textbf{ASR(\%)}              \\ \hline
0.1\%                   & 93.19            & \cellcolor[HTML]{EFEFEF}3.08  & 93.82           & \cellcolor[HTML]{EFEFEF}0.88  & 92.24           & \cellcolor[HTML]{EFEFEF}0.44  \\ \hline
0.2\%                   & 93.46            & \cellcolor[HTML]{EFEFEF}48.13 & 93.65            & \cellcolor[HTML]{EFEFEF}3.95 & 91.98            & \cellcolor[HTML]{EFEFEF}5.49 \\ \hline
0.4\%                   & 93.38            & \cellcolor[HTML]{EFEFEF}83.51 & 93.43            & \cellcolor[HTML]{EFEFEF}89.89 & 92.09            & \cellcolor[HTML]{EFEFEF}28.79 \\ \hline
1.0\%                   & 93.25            & \cellcolor[HTML]{EFEFEF}96.04 & 93.45            & \cellcolor[HTML]{EFEFEF}85.05 & 92.45            & \cellcolor[HTML]{EFEFEF}81.75 \\ \hline
3.0\%                   & 93.31           & \cellcolor[HTML]{EFEFEF}98.02 & 93.34            & \cellcolor[HTML]{EFEFEF}99.12 & 92.51            & \cellcolor[HTML]{EFEFEF}96.26 \\ \hline
\end{tabular}}
\vspace{10pt}
\caption{\textbf{End-to-End Pipeline Evaluation.} The proposed end-to-end variant of the frequency-based backdoor attack achieves both a high clean data accuracy and a high attack success rate.}
\label{10_a}
\end{table}

 \clearpage
 
 \ \vfill

\clearpage
\section{Cross Filter Frequency-Based Backdoor Attack}
\label{sec5}
In this section, we show the capability of utilizing binary masks and additive filters generated for one architecture to backdoor attack another. As expected, one can reach a high attack success rate (for a high enough poisoning rate) using such masks and filters (\textit{Check Ablation-Section 5.5 in manuscript}); however, one has no guarantee over maintaining a Fourier heatmap similar to the clean model.   

Table \ref{tab5_a} shows the CDA and ASR of a ResNet18 trained on CIFAR10 poisoned using binary masks and additive filters of WideResNet34, ResNet34, and VGG19.

\begin{table}[htbp!]
\centering
\scalebox{0.8}{
\begin{tabular}{ccccccc}
\hline
                      \textbf{ Filter \& Mask Source}& \multicolumn{2}{c}{\textbf{WideResNet34}}         & \multicolumn{2}{c}{\textbf{ResNet34}}            & \multicolumn{2}{c}{\textbf{VGG19}}               \\ \hline
\textbf{Poisoning Rate} & \textbf{CDA(\%)} & \textbf{ASR(\%)}               & \textbf{CDA(\%)} & \textbf{ASR(\%)}              & \textbf{CDA(\%)} & \textbf{ASR(\%)}              \\ \hline
0.0\%                   & 93.28            & \cellcolor[HTML]{EFEFEF}5.27   & 93.10            & \cellcolor[HTML]{EFEFEF}1.98  & 93.00            & \cellcolor[HTML]{EFEFEF}5.93  \\ \hline
0.1\%                   & 92.93            & \cellcolor[HTML]{EFEFEF}2.86   & 92.75            & \cellcolor[HTML]{EFEFEF}35.82 & 92.81            & \cellcolor[HTML]{EFEFEF}50.77 \\ \hline
0.4\%                   & 93.17            & \cellcolor[HTML]{EFEFEF}93.63  & 93.14            & \cellcolor[HTML]{EFEFEF}95.16 & 92.79            & \cellcolor[HTML]{EFEFEF}95.16 \\ \hline
1.0\%                   & 93.08            & \cellcolor[HTML]{EFEFEF}99.12  & 93.24            & \cellcolor[HTML]{EFEFEF}98.46 & 92.80            & \cellcolor[HTML]{EFEFEF}96.92 \\ \hline
3.0\%                   & 92.74            & \cellcolor[HTML]{EFEFEF}100.00 & 92.94            & \cellcolor[HTML]{EFEFEF}99.78 & 92.85            & \cellcolor[HTML]{EFEFEF}99.34 \\ \hline
\end{tabular}}
\vspace{10pt}
\caption{\textbf{Cross Filter Backdoor Attack Evaluation.} Evaluating different binary masks and additive filters generated for WideResNet34, ResNet34, and VGG19 for attacking ResNet18 on CIFAR10.}
\label{tab5_a}
\end{table}
\clearpage
\newpage
\section{Spatial Visualization of the Proposed Frequency-based Backdoor \\ Attack}
\label{sec6}

\begin{figure}[htbp!]
    \centering

    \begin{minipage}{0.2\linewidth}
    
    \vspace{-7cm}
    
    \begin{flushright}

        \normalsize$\mathcal{I}$ \vspace{1.2cm}
        
        $\mathcal{I}_P$ (ResNet18)\vspace{1.2cm}
        
        $\mathcal{D}$ (ResNet18) \vspace{1.2cm}
        
        $\mathcal{I}_P$ (ResNet34)\vspace{1.2cm}
        
        $\mathcal{D}$ (ResNet34)
            \end{flushright}
    \end{minipage}
     \begin{minipage}[b]{0.78\linewidth}
     \hfill
    \includegraphics[width=0.9\linewidth]{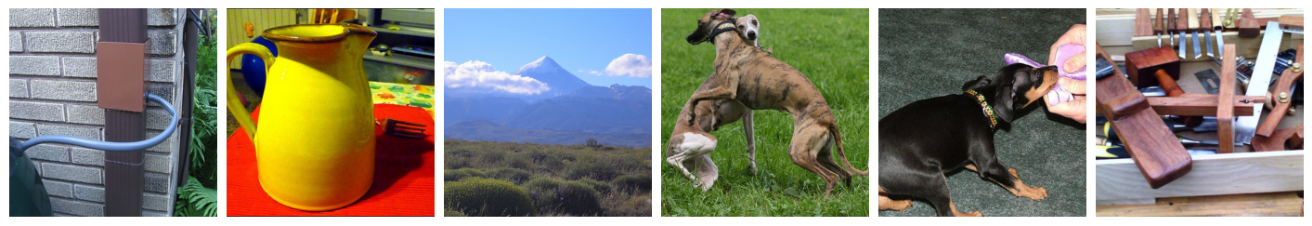}   
    
    \hfill  
    \includegraphics[width=0.9\linewidth]{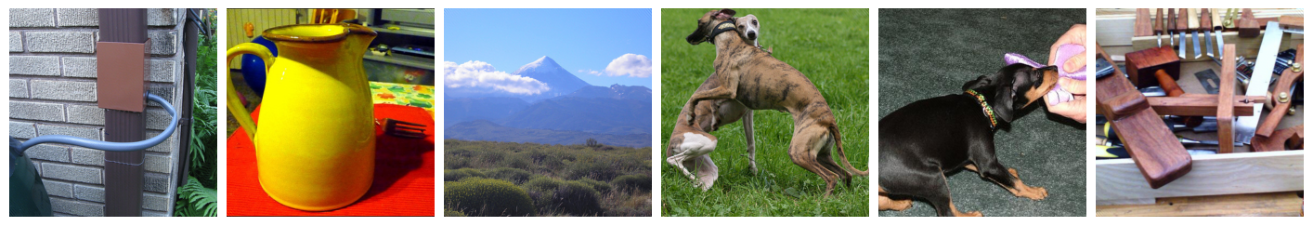} 
    
    \hfill
    \includegraphics[width=0.9\linewidth]{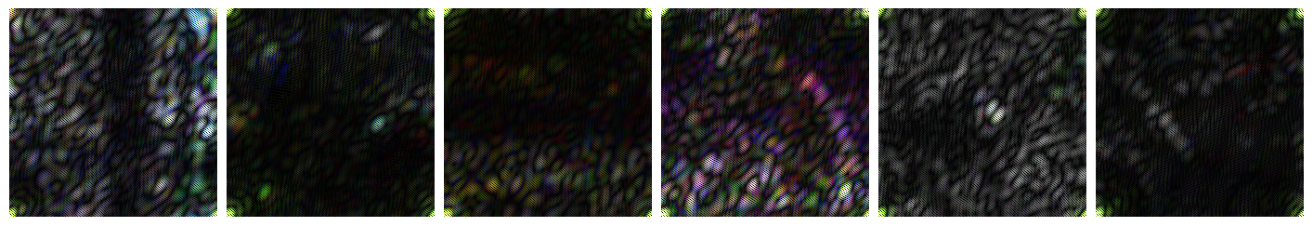} 
    
    \hfill
 \includegraphics[width=0.9\linewidth]{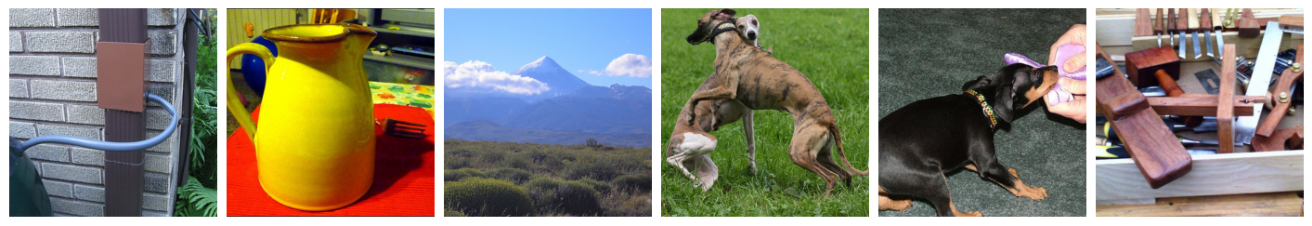} 

 \hfill
    \includegraphics[width=0.9\linewidth]{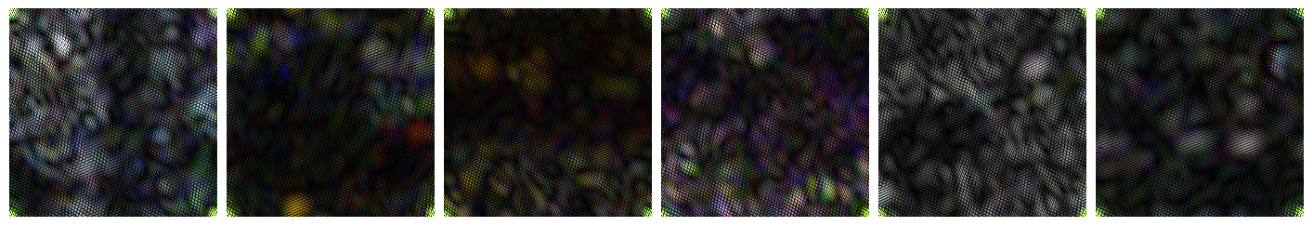}
    \end{minipage}
            \caption{\textbf{Spatial Visualization of Proposed Attack.} Visualization of the absolute scaled difference shows how dynamic the proposed attack. The poisoned images show the imperceptibility of the attack.  }
    \label{Fig6_a}
\end{figure}

\begin{figure}[tp!]
    \centering

    \begin{minipage}{0.2\linewidth}
    
    \vspace{-7cm}
    
    \begin{flushright}

        \normalsize$\mathcal{I}$ \vspace{1.2cm}
        
        $\mathcal{I}_P$ (ResNet18)\vspace{1.2cm}
        
        $\mathcal{D}$ (ResNet18) \vspace{1.2cm}
        
        $\mathcal{I}_P$ (ResNet34)\vspace{1.2cm}
        
        $\mathcal{D}$ (ResNet34)
            \end{flushright}
    \end{minipage}
     \begin{minipage}[b]{0.78\linewidth}
     \hfill
    \includegraphics[width=0.9\linewidth]{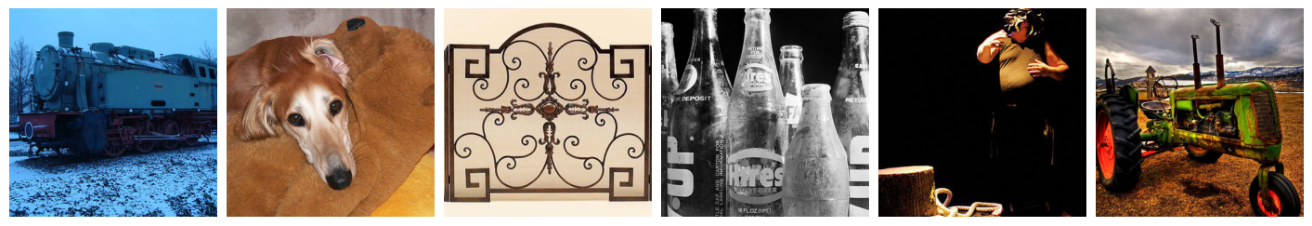}   
    
    \hfill  
    \includegraphics[width=0.9\linewidth]{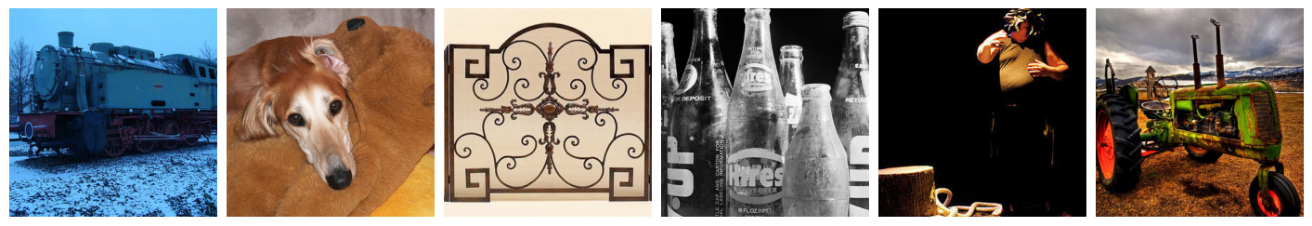} 
    
    \hfill
    \includegraphics[width=0.9\linewidth]{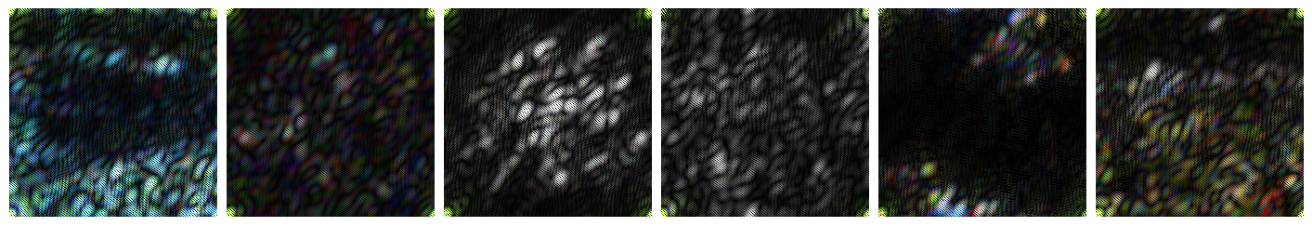} 
    
    \hfill
 \includegraphics[width=0.9\linewidth]{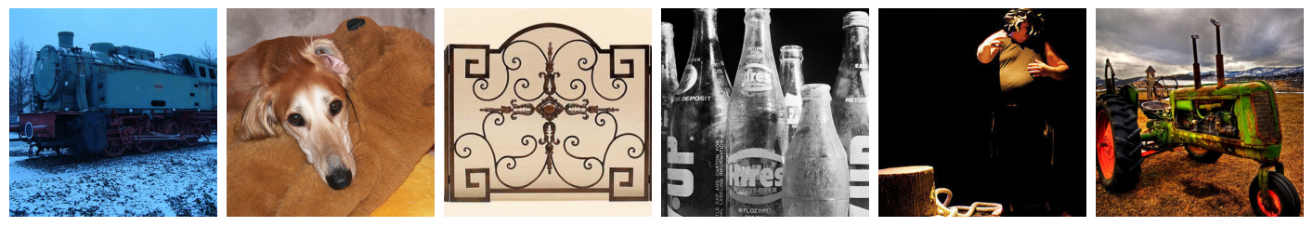} 

 \hfill
    \includegraphics[width=0.9\linewidth]{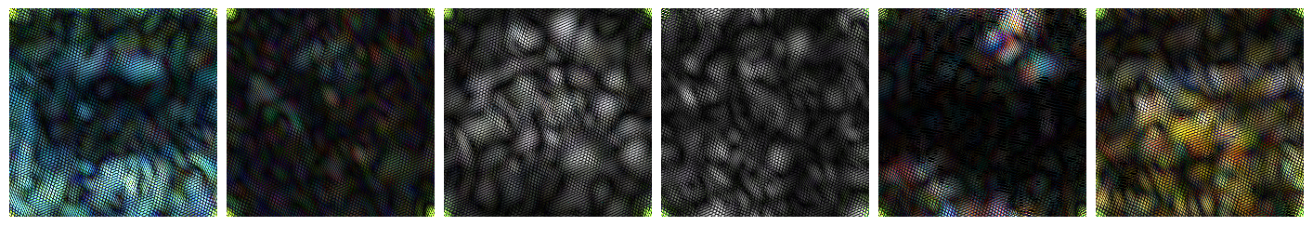}
    \end{minipage}
            \caption{\textbf{Spatial Visualization of Proposed Attack.} Visualization of the absolute scaled difference shows how dynamic the proposed attack. The poisoned images show the imperceptibility of the attack.  }
    \label{Fig6_b}
\end{figure}

In this section, we visualize the scaled absolute difference ($\mathcal{D}$) of non-poisoned images ($\mathcal{I}$) and poisoned images ($\mathcal{I}_P$) defined as:
\begin{align}
    \mathcal{D} = \gamma |\mathcal{I} - \mathcal{I}_P|
\end{align}

where $|.|$ of a matrix denotes element-wise absolute value operation and $\gamma$ is a scalar multiplier in $\mathbb{R}$. Figures \ref{Fig6_a} and \ref{Fig6_b} visualize two sets of non-poisoned images, their poisoned counterparts, and the absolute scaled difference with $\gamma = 50$ for ResNet18 and ResNet34 (recall that our attack is model dependent).

\clearpage

\ \hfill
\clearpage

\section{Fourier Heatmaps}
\label{sec7}

\let\thefootnote\relax\footnotetext{WideResNet34 was not included for ImageNet experiments as there is no official implementation of this model in \textit{torchvision.models} .}

\begin{figure*}[htbp!]

    \centering

\includegraphics[width=\textwidth]{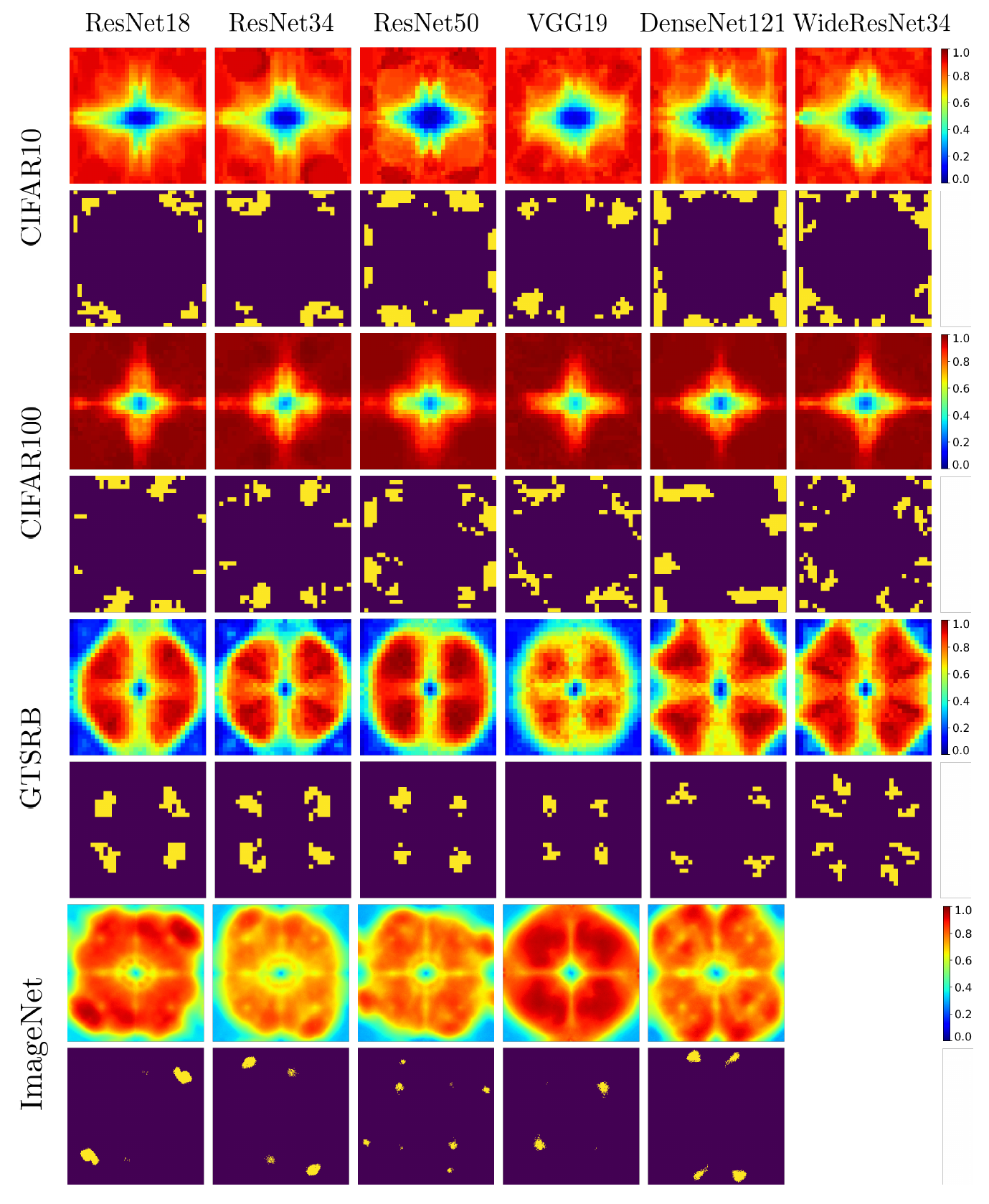}
\caption{\textbf{Fourier Heatmaps and Top-$k$ Filters.} Fourier heatmaps for various architectures and datasets along with their top-$k$ selected frequencies for the binary mask.}
 \label{sec4_allfigs}
\end{figure*}

\clearpage

\begin{table}[]
\centering
\scalebox{0.9}{
\begin{tabular}{ccccccc}
\cline{2-7}
                  & \textbf{ResNet18} & \textbf{ResNet34} & \textbf{ResNet50} & \textbf{VGG19} & \textbf{DenseNet121} & \textbf{WideResNet34} \\ \hline
\textbf{CIFAR10}  & 9.77\%            & 9.77\%            & 11.72\%           & 9.77\%         & 16.61\%              & 14.65\%               \\ \hline
\textbf{CIFAR100} & 15.62\%           & 19.53\%           & 9.77\%            & 19.53\%        & 15.63\%              & 15.63\%               \\ \hline
\textbf{GTSRB}    & 8.79\%            & 8.79\%            & 2.92\%            & 3.90\%         & 4.88\%               & 8.80\%                \\ \hline
\textbf{ImageNet} & 1.99\%            & 1.59\%            & 0.99\%            & 0.90\%         & 1.99\%               & -                     \\ \hline
\end{tabular}}
\vspace{10pt}
\caption{\textbf{Percentage of Poisoned Frequencies.} Different models and datasets require poisoning different percentages of the Fourier bases to achieve a balance between stealthiness, attack success rate and clean data accuracy.}
\label{sec4_table}
\end{table}

Figure \ref{sec4_allfigs} shows all the Fourier heatmaps and the binary masks generated for poisoning the different models on all datasets. Table \ref{sec4_table} shows the percentage of poisoned frequencies for each binary mask. 

\clearpage
\section{Fourier Heatmap as a Backdoor Detector}
\label{sec8}

\begin{figure*}[htbp!]

    \centering

  \begin{subfigure}[t]{0.3\textwidth}
    \centering
    \includegraphics[width=0.8\textwidth]{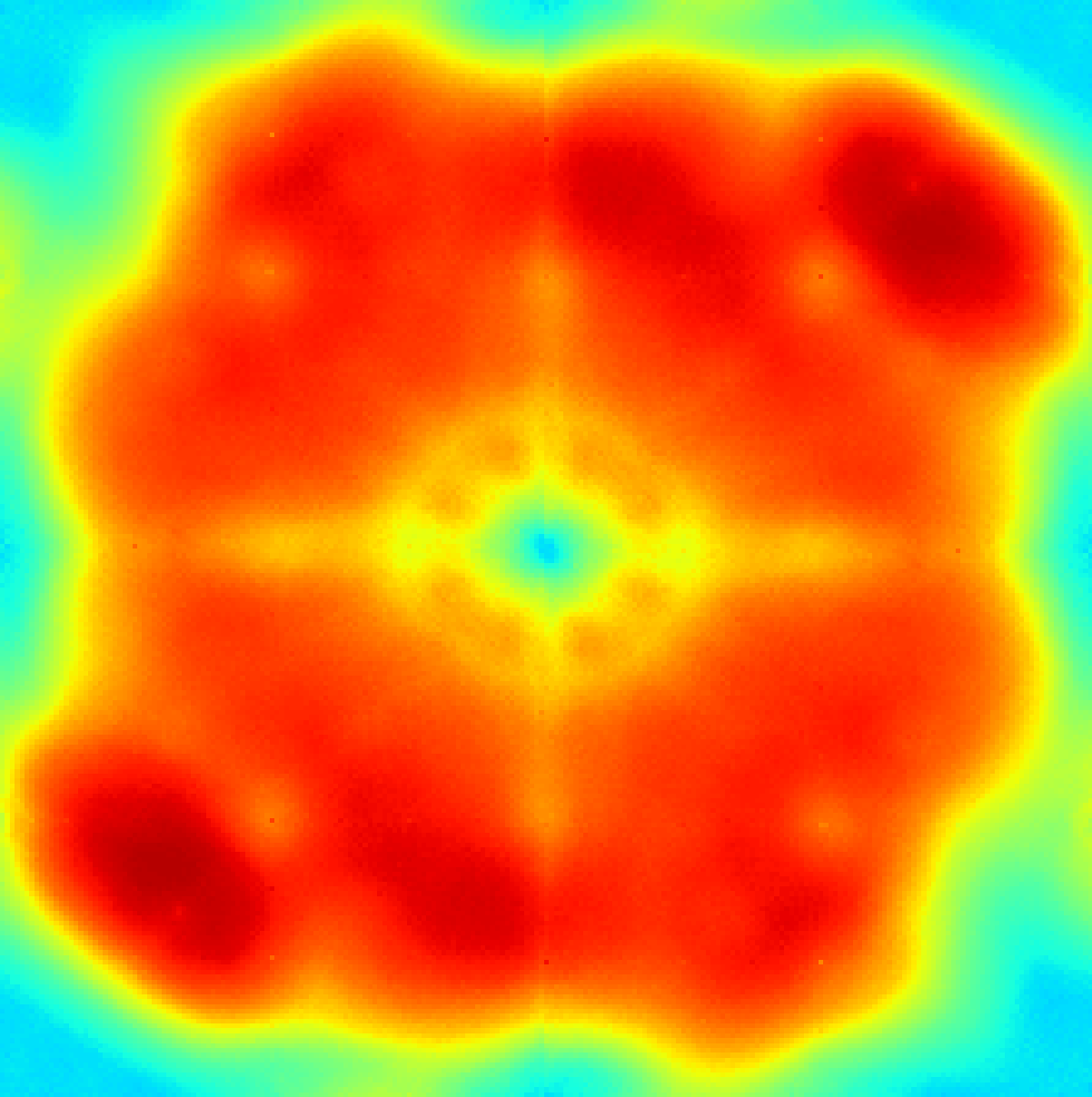}
    \caption{Clean Model Heatmap}
    \label{fig8_a}
  \end{subfigure}
  \hspace{1cm}
  \begin{subfigure}[t]{0.3\textwidth}
      \centering
    \includegraphics[width=0.8\textwidth]{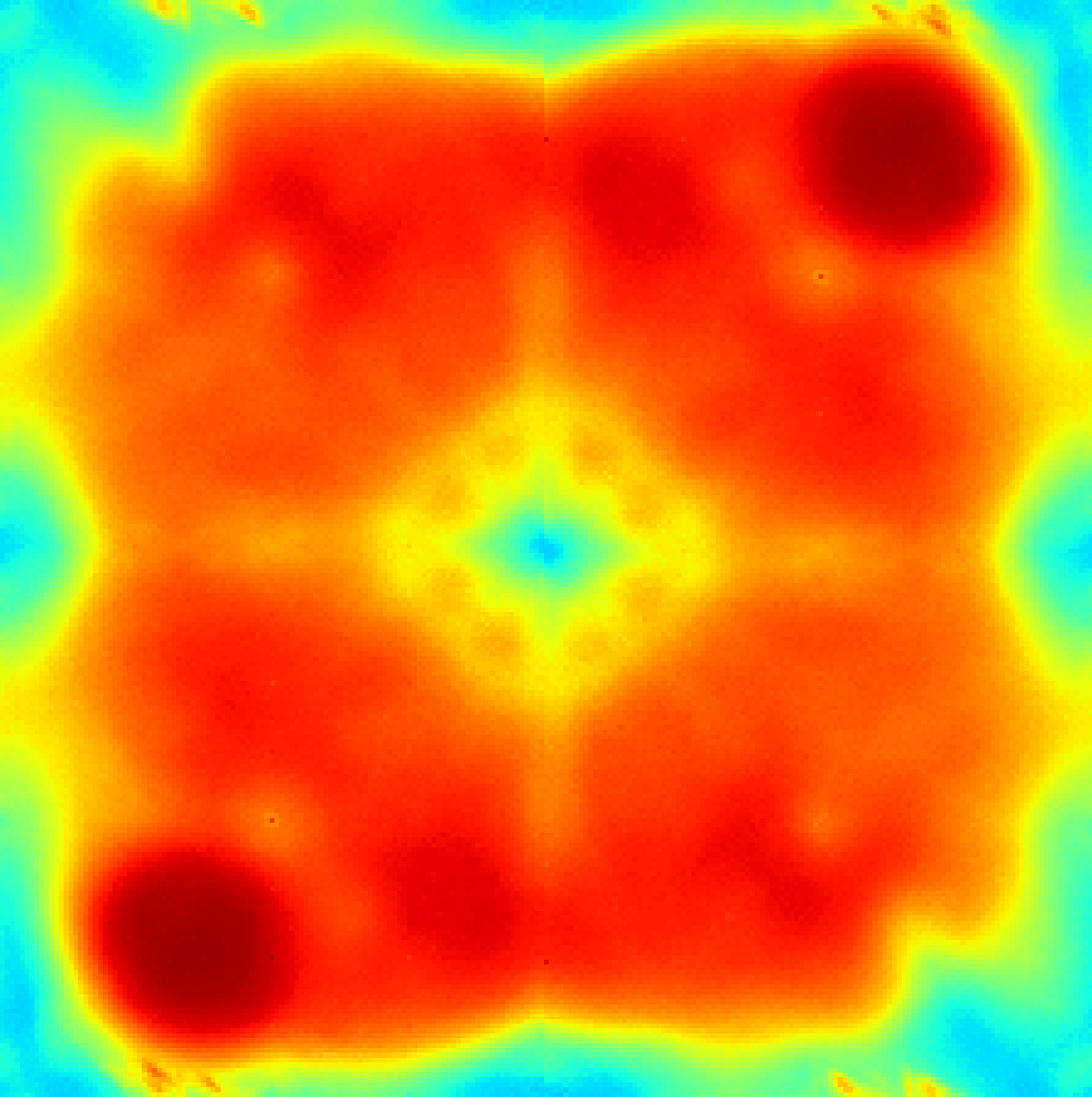}    
    \caption{Top-$k$ Poisoned Model Heatmap}
    \label{fig8_b}
  \end{subfigure}

  \begin{subfigure}[t]{0.3\textwidth}
      \centering
    \includegraphics[width=0.8\textwidth]{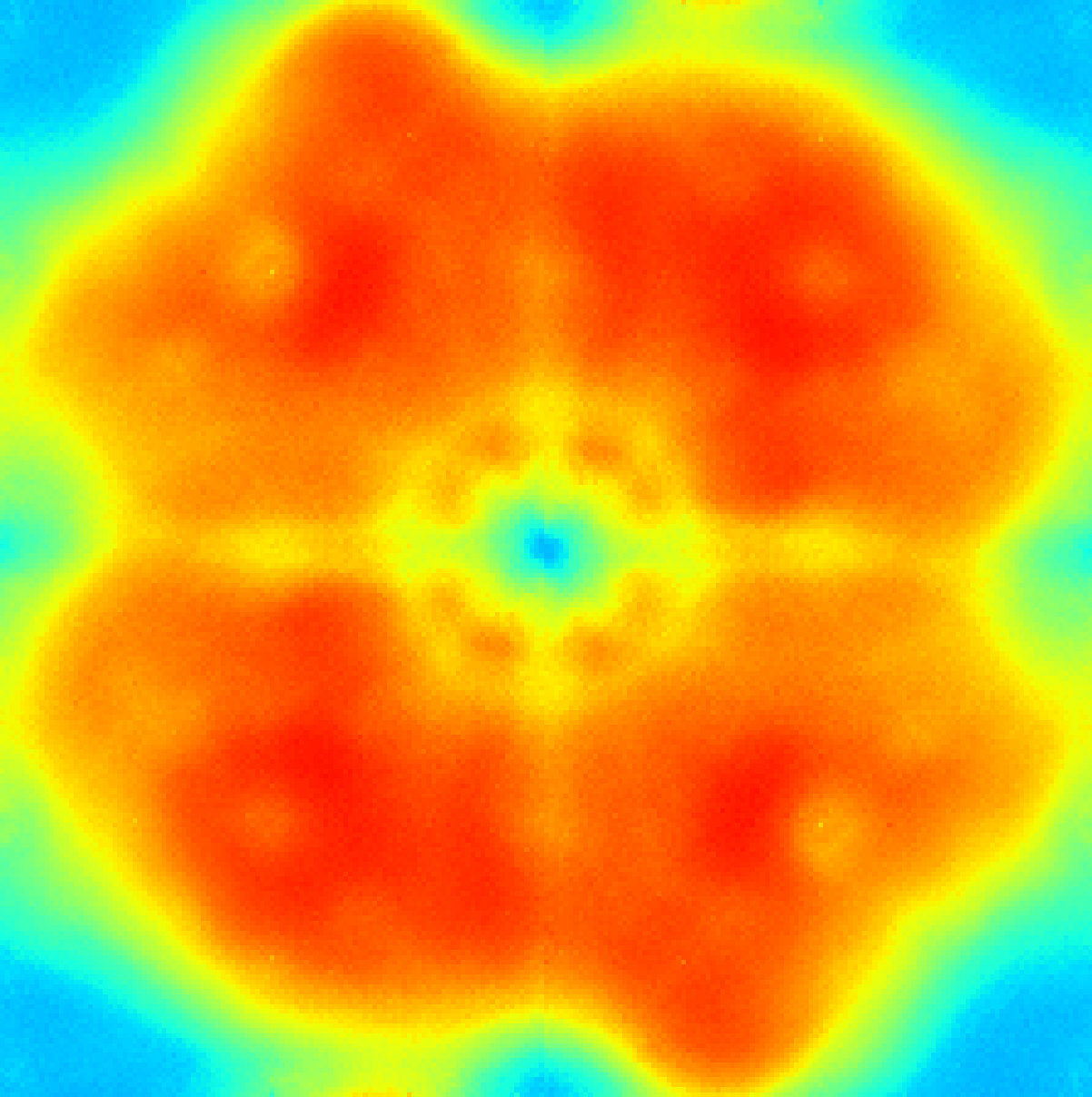}
    \caption{BadNet \cite{Gu2019BadNetsEB} Model Heatmap}
    \label{fig8_c}
  \end{subfigure}
   \hspace{1cm} 
  \begin{subfigure}[t]{0.3\textwidth}
      \centering
    \includegraphics[width=0.8\textwidth]{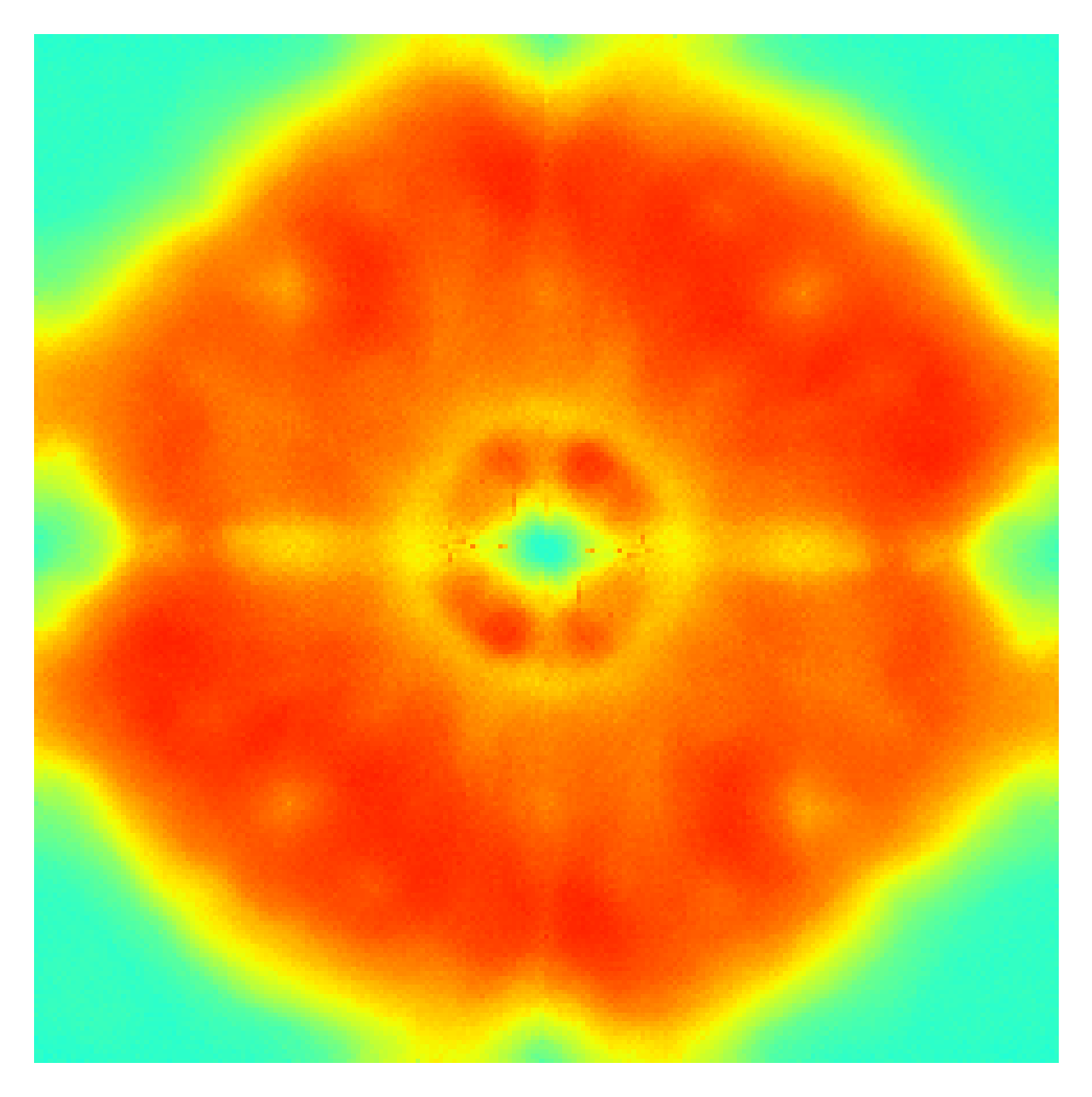}
    \caption{Blend \cite{Chen2017TargetedBA} Model Heatmap}
    \label{fig8_d}
  \end{subfigure}  
\vspace{10pt}
\caption{\textbf{Fourier Heatmap As a Backdoor Detector.} BadNet and Blend poisoned models introduce more significant changes to the clean heatmap as compared to the proposed top-$k$ frequency-based backdoor attack. These heatmaps could be exploited as means to detecting whether a model is poisoned or not. }
\label{fig8}
  \end{figure*}

As shown in the manuscript, if the choice of poisoned frequencies is not carried out properly, a simple check on the Fourier heatmap of the obtained model could expose the attacker (an abnormal trend is observed in the heatmap). Figures \ref{fig8_a}, \ref{fig8_b}, \ref{fig8_c} and \ref{fig8_d} show the Fourier heatmaps of clean ResNet18, top-$k$ poisoned ResNet18, BadNet \cite{Gu2019BadNetsEB} poisoned ResNet18, and Blend \cite{Chen2017TargetedBA} poisoned ResNet18, respectively. BadNet represents the first backdoor attack in the literature and is based on poisoning data by applying a white patch to the corner of a subset of the training set. Blend on the other hand, was the first to recognize the importance of imperceptibility and suggested blending images with the poison trigger for a more stealthy attack.
Figures \ref{fig8_c} and \ref{fig8_d} show that both BadNets and Blend tend to highly change the frequency sensitivity of the attacked model compared to the clean one and hence could be detected as poisoned models by inspecting their heatmaps. The proposed frequency-based backdoor attack is more conservative and introduces only mild changes to the clean model heatmap and therefore are less detectable as poisoned.

  \clearpage

\begin{figure*}[t!]

    \centering

  \begin{subfigure}[t]{0.3\textwidth}
    \centering
    \includegraphics[width=0.8\textwidth]{figures/heatmaps/Clean.png}
    \caption{Clean Model Heatmap}
    \label{fig8_e}
  \end{subfigure}
  \begin{subfigure}[t]{0.3\textwidth}
      \centering
    \includegraphics[width=0.8\textwidth]{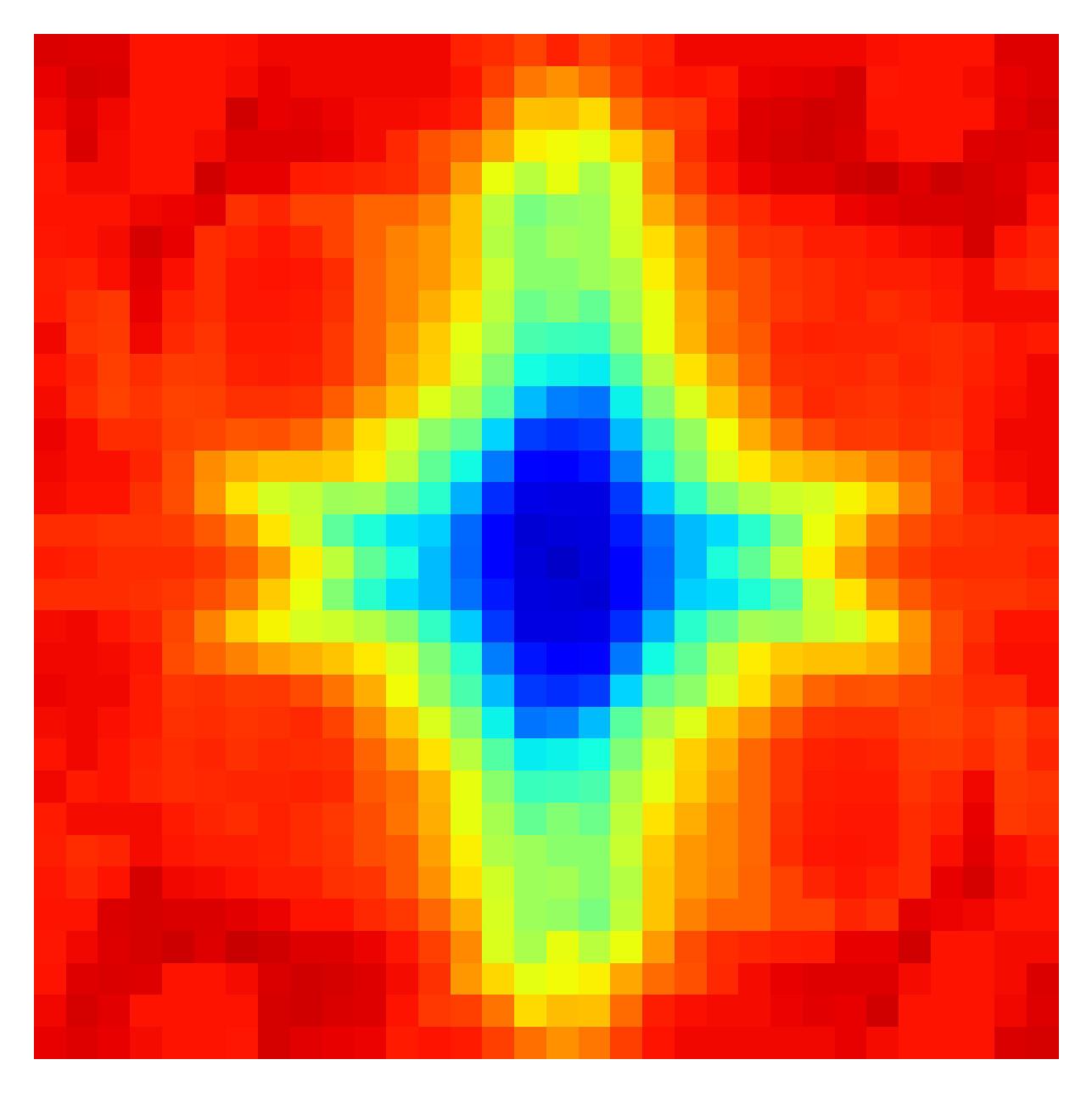}    
    \caption{Top-$k$ Poisoned Model Heatmap (Ours)}
    \label{fig8_f}
  \end{subfigure}
  \begin{subfigure}[t]{0.3\textwidth}
      \centering
    \includegraphics[width=0.8\textwidth]{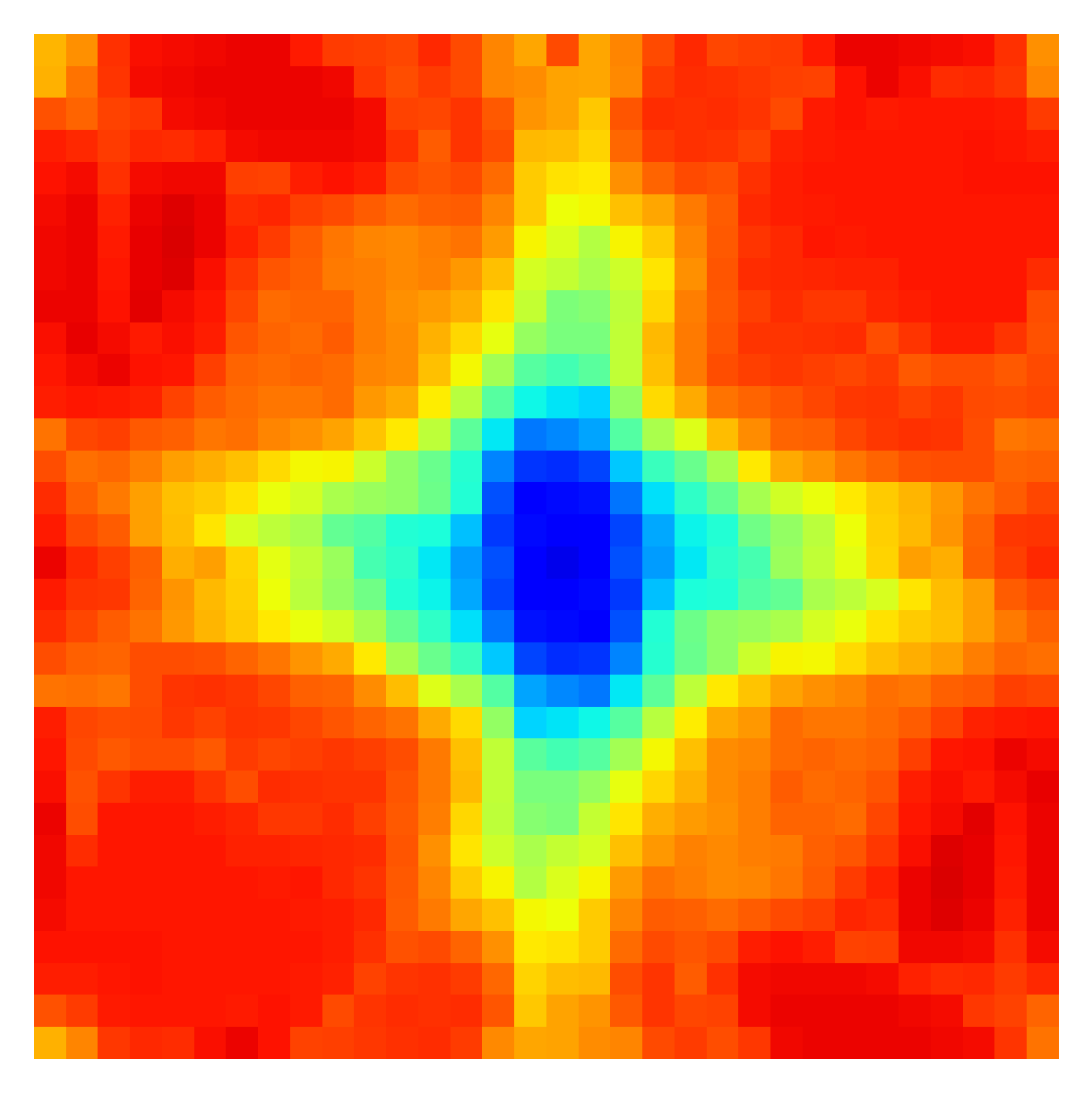}
    \caption{BadNet \cite{Gu2019BadNetsEB} Model Heatmap}
    \label{fig8_g}
  \end{subfigure}
  
  \begin{subfigure}[t]{0.3\textwidth}
      \centering
    \includegraphics[width=0.8\textwidth]{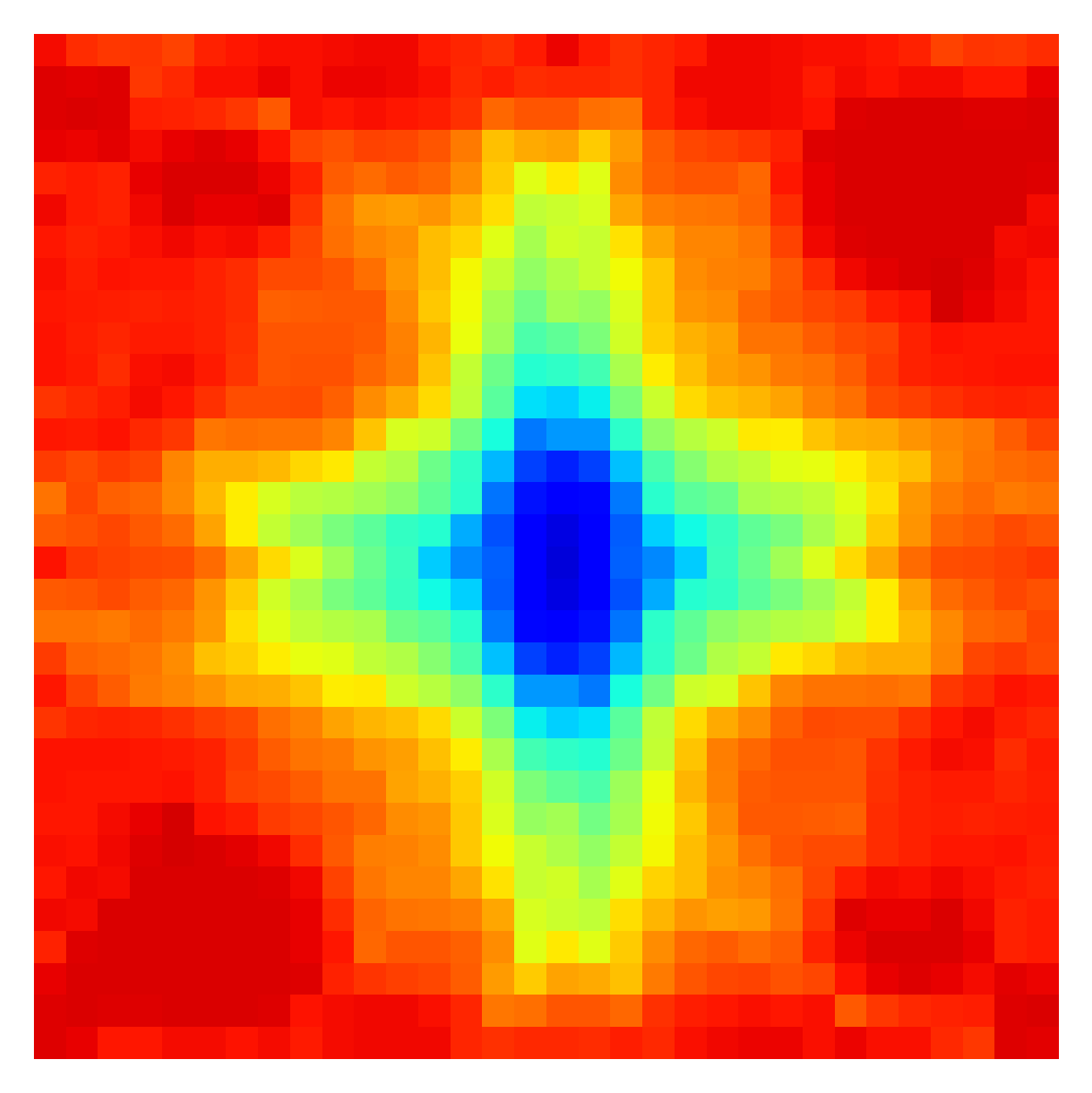}
    \caption{Blend \cite{Chen2017TargetedBA} Model Heatmap}
    \label{fig8_h}
  \end{subfigure}  
   \begin{subfigure}[t]{0.3\textwidth}
      \centering
    \includegraphics[width=0.8\textwidth]{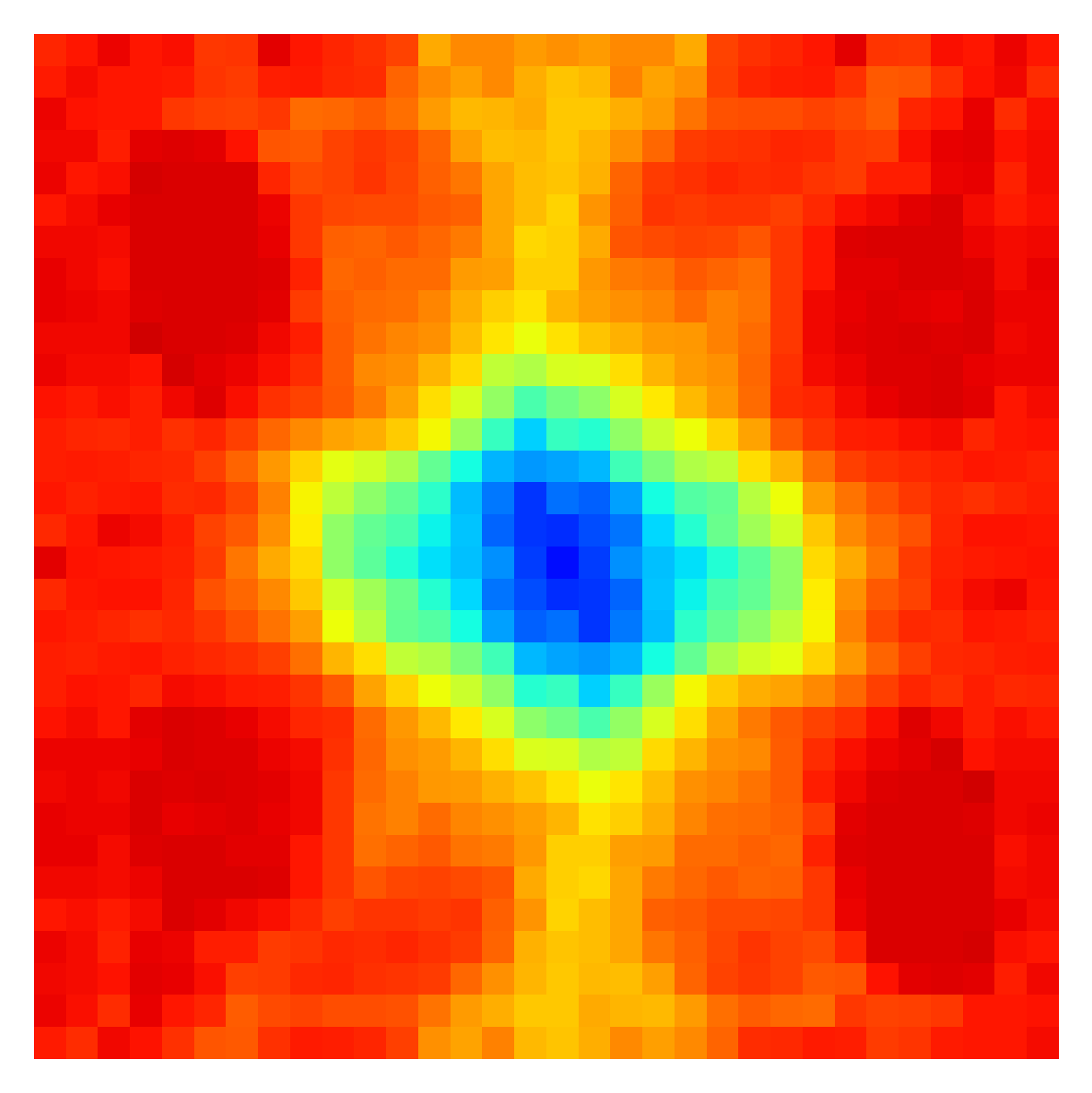}
    \caption{Clean Label \cite{Zhao2020CleanLabelBA} Model Heatmap}
    \label{fig8_i}
  \end{subfigure}   
     \begin{subfigure}[t]{0.3\textwidth}
      \centering
    \includegraphics[width=0.8\textwidth]{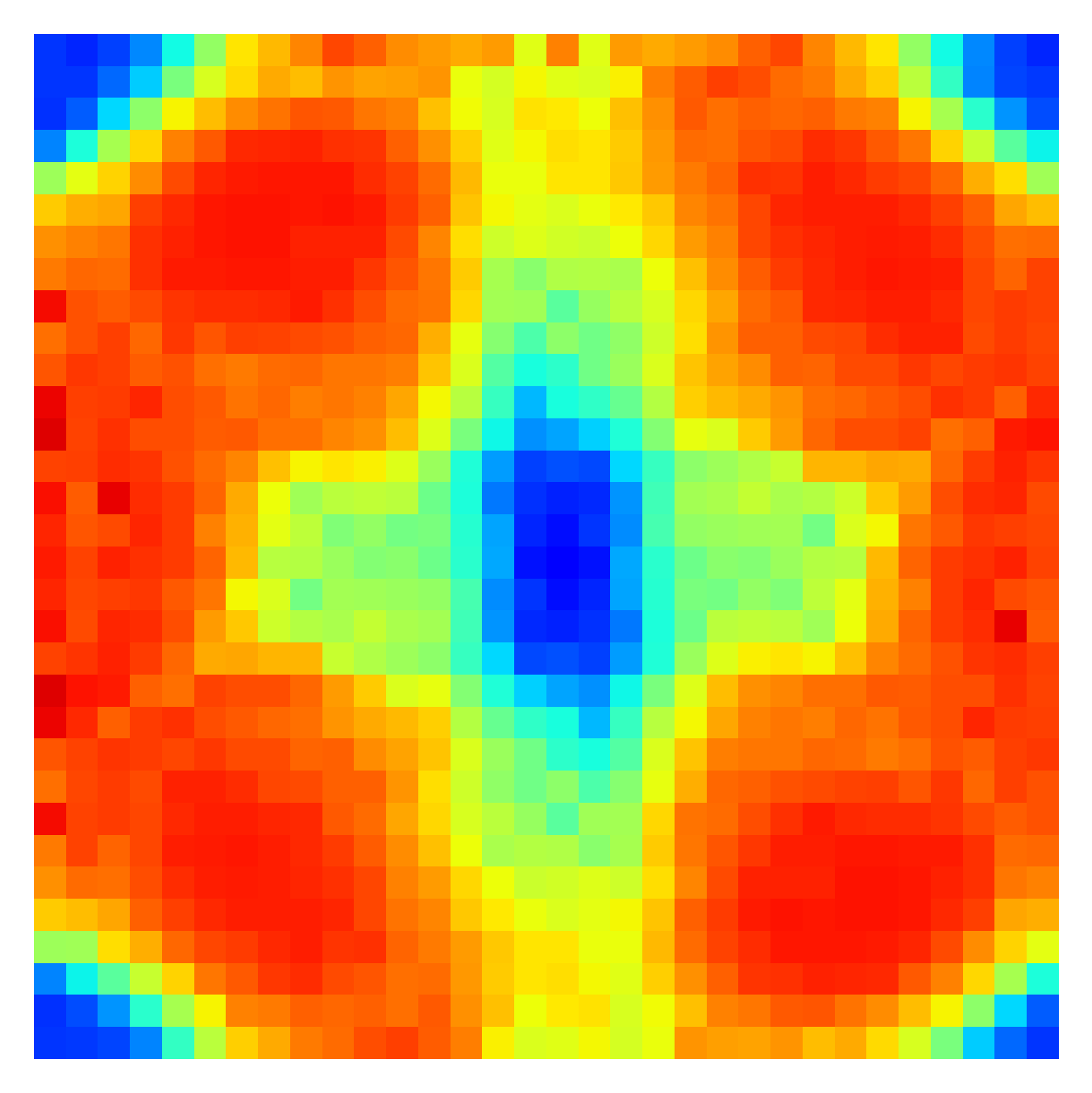}
    \caption{RE \cite{Xiang2021ReverseEI} Model Heatmap}
    \label{fig8_j}
  \end{subfigure}   
\caption{\textbf{Fourier Heatmap As a Backdoor Detector.} Various spatial backdoor attacks introduce more significant changes to the clean heatmap as compared to the proposed top-$k$ frequency-based backdoor attack. These heatmaps could be exploited as means to detecting whether a model is poisoned or not. }
\label{fig8}
  \end{figure*}

  Similarly, this inspection could be applied for models trained on small image datasets such as CIFAR10. Figures \ref{fig8_e}, \ref{fig8_f}, \ref{fig8_g}, \ref{fig8_h}, \ref{fig8_i} and  \ref{fig8_j} show the Fourier heatmaps for ResNet18 trained on CIFAR10 with different poisoning strategies. Our proposed method maintains the highest similarity to the clean model's Fourier heatmap as compared to other methods.
  
\ \vfill
\clearpage

\section{Evaluation Against Backdoor Defenses}
\label{sec9}

In this section, we provide a further evaluation of the spatial defenses presented in the manuscript. We also evaluate our method against additional defenses, namely, STRIP \cite{Gao2019STRIPAD}, Activation Clustering \cite{Chen2019DetectingBA} and Spectral Signatures \cite{Tran2018SpectralSI}. 

Recall that Grad-CAM uses gradients of a particular class to visualize where the network is looking/focusing at to make its prediction. \cite{Liu2018FinePruningDA} prunes the
least active neurons (on clean samples) and then fine-tunes the network on clean samples. STRong Intentional Pertubation (STRIP) \cite{Gao2019STRIPAD} intentionally perturbs the input through blending it with clean samples. The authors rely on the realization that blending a poisoned sample with a clean sample would still activate the backdoor attack and therefore studying the entropy of the prediction vectors could be used for backdoor detection. Activation Clustering (AC) \cite{Chen2019DetectingBA}  analyzes the neural network's representation layer activation to determine whether the data has been poisoned. Since a poisoned model assigns poisoned and clean data to the target class based on a different feature representation, one can cluster the representations of the poisoned class into two distinct clusters. Similar to AC, Spectral Signatures (SS) \cite{Tran2018SpectralSI} operates on feature representations to detect backdoor attacks. SS detects the poisoned samples using robust statistics and SVD methods.

\begin{figure*}[h!]

    \centering
  \begin{subfigure}[t]{0.12\textwidth}
      \centering
    \includegraphics[width=\textwidth]{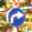}
    \label{fig-b-2}
  \end{subfigure}
  \begin{subfigure}[t]{0.12\textwidth}
    \centering
    \includegraphics[width=\textwidth]{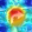}
    \label{fig-a-1}
  \end{subfigure}
  \begin{subfigure}[t]{0.12\textwidth}
      \centering
    \includegraphics[width=\textwidth]{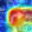}
    \label{fig-a-3}
  \end{subfigure}
    \begin{subfigure}[t]{0.12\textwidth}
      \centering
    \includegraphics[width=\textwidth]{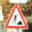}
    \label{fig-b-2}
  \end{subfigure}
  \begin{subfigure}[t]{0.12\textwidth}
    \centering
    \includegraphics[width=\textwidth]{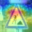}
    \label{fig-a-1}
  \end{subfigure}
  \begin{subfigure}[t]{0.12\textwidth}
      \centering
    \includegraphics[width=\textwidth]{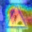}
    \label{fig-a-3}
  \end{subfigure}

    \begin{subfigure}[t]{0.12\textwidth}
      \centering
    \includegraphics[width=\textwidth]{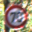}
    \label{fig-b-2}
  \end{subfigure}
  \begin{subfigure}[t]{0.12\textwidth}
    \centering
    \includegraphics[width=\textwidth]{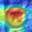}
    \label{fig-a-1}
  \end{subfigure}
  \begin{subfigure}[t]{0.12\textwidth}
      \centering
    \includegraphics[width=\textwidth]{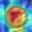}
    \label{fig-a-3}
  \end{subfigure}
    \begin{subfigure}[t]{0.12\textwidth}
      \centering
    \includegraphics[width=\textwidth]{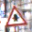}
    \label{fig-b-2}
  \end{subfigure}
  \begin{subfigure}[t]{0.12\textwidth}
    \centering
    \includegraphics[width=\textwidth]{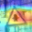}
    \label{fig-a-1}
  \end{subfigure}
  \begin{subfigure}[t]{0.12\textwidth}
      \centering
    \includegraphics[width=\textwidth]{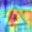}
    \label{fig-a-3}
  \end{subfigure}

    \begin{subfigure}[t]{0.12\textwidth}
      \centering
    \includegraphics[width=\textwidth]{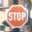}
    \label{fig-b-2}
  \end{subfigure}
  \begin{subfigure}[t]{0.12\textwidth}
    \centering
    \includegraphics[width=\textwidth]{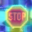}
    \label{fig-a-1}
  \end{subfigure}
  \begin{subfigure}[t]{0.12\textwidth}
      \centering
    \includegraphics[width=\textwidth]{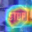}
    \label{fig-a-3}
  \end{subfigure}
    \begin{subfigure}[t]{0.12\textwidth}
      \centering
    \includegraphics[width=\textwidth]{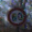}
    \label{fig-b-2}
  \end{subfigure}
  \begin{subfigure}[t]{0.12\textwidth}
    \centering
    \includegraphics[width=\textwidth]{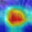}
    \label{fig-a-1}
  \end{subfigure}
  \begin{subfigure}[t]{0.12\textwidth}
      \centering
    \includegraphics[width=\textwidth]{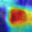}
    \label{fig-a-3}
  \end{subfigure}

  \caption{\textbf{Grad-CAM on GTSRB.} The proposed Frequency-based backdoor attack allows the network to focus on similar regions when classifying poisoned images as compared to clean network operating on the clean version of the images. Methods that focus on Grad-CAM based image-reconstruction fail to remove the poison.  }
  \label{fig9_a}
\end{figure*}

\begin{figure*}[h!]

    \centering
  \begin{subfigure}[t]{0.12\textwidth}
      \centering
    \includegraphics[width=\textwidth]{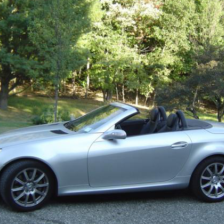}
    \label{fig-b-2}
  \end{subfigure}
  \begin{subfigure}[t]{0.12\textwidth}
    \centering
    \includegraphics[width=\textwidth]{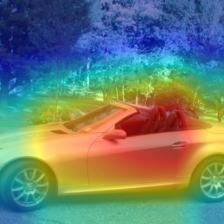}
    \label{fig-a-1}
  \end{subfigure}
  \begin{subfigure}[t]{0.12\textwidth}
      \centering
    \includegraphics[width=\textwidth]{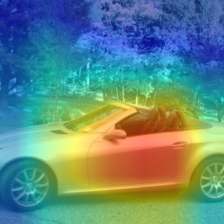}
    \label{fig-a-3}
  \end{subfigure}
  \begin{subfigure}[t]{0.12\textwidth}
      \centering
    \includegraphics[width=\textwidth]{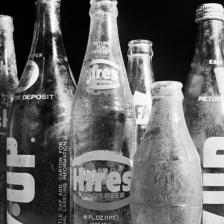}
    \label{fig-b-2}
  \end{subfigure}
  \begin{subfigure}[t]{0.12\textwidth}
    \centering
    \includegraphics[width=\textwidth]{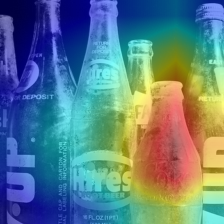}
    \label{fig-a-1}
  \end{subfigure}
  \begin{subfigure}[t]{0.12\textwidth}
      \centering
    \includegraphics[width=\textwidth]{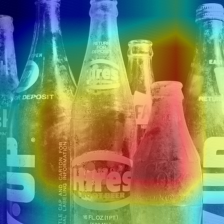}
    \label{fig-a-3}
  \end{subfigure}

    \begin{subfigure}[t]{0.12\textwidth}
      \centering
    \includegraphics[width=\textwidth]{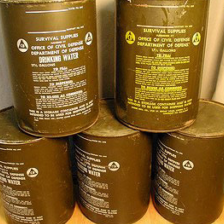}
    \label{fig-b-2}
  \end{subfigure}
  \begin{subfigure}[t]{0.12\textwidth}
    \centering
    \includegraphics[width=\textwidth]{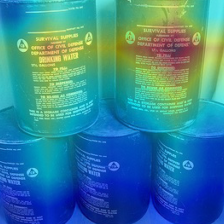}
    \label{fig-a-1}
  \end{subfigure}
  \begin{subfigure}[t]{0.12\textwidth}
      \centering
    \includegraphics[width=\textwidth]{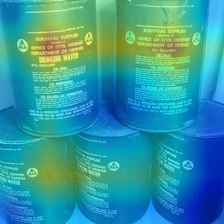}
    \label{fig-a-3}
  \end{subfigure}
  \begin{subfigure}[t]{0.12\textwidth}
      \centering
    \includegraphics[width=\textwidth]{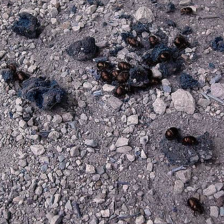}
    \label{fig-b-2}
  \end{subfigure}
  \begin{subfigure}[t]{0.12\textwidth}
    \centering
    \includegraphics[width=\textwidth]{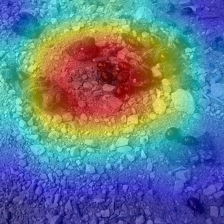}
    \label{fig-a-1}
  \end{subfigure}
  \begin{subfigure}[t]{0.12\textwidth}
      \centering
    \includegraphics[width=\textwidth]{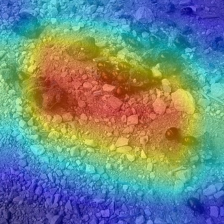}
    \label{fig-a-3}
  \end{subfigure}

    \begin{subfigure}[t]{0.12\textwidth}
      \centering
    \includegraphics[width=\textwidth]{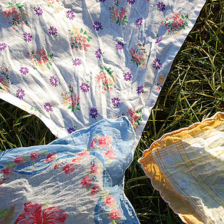}
    \label{fig-b-2}
  \end{subfigure}
  \begin{subfigure}[t]{0.12\textwidth}
    \centering
    \includegraphics[width=\textwidth]{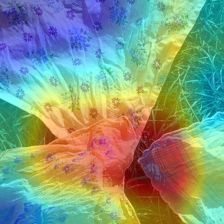}
    \label{fig-a-1}
  \end{subfigure}
  \begin{subfigure}[t]{0.12\textwidth}
      \centering
    \includegraphics[width=\textwidth]{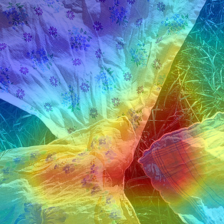}
    \label{fig-a-3}
  \end{subfigure}
  \begin{subfigure}[t]{0.12\textwidth}
      \centering
    \includegraphics[width=\textwidth]{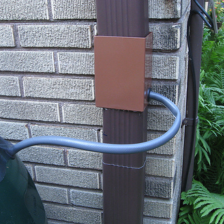}
    \label{fig-b-2}
  \end{subfigure}
  \begin{subfigure}[t]{0.12\textwidth}
    \centering
    \includegraphics[width=\textwidth]{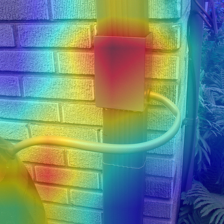}
    \label{fig-a-1}
  \end{subfigure}
  \begin{subfigure}[t]{0.12\textwidth}
      \centering
    \includegraphics[width=\textwidth]{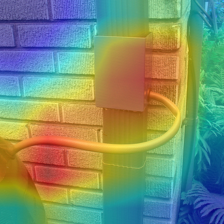}
    \label{fig-a-3}
  \end{subfigure}

  \caption{\textbf{Grad-CAM on ImageNet.} The proposed Frequency-based backdoor attack allows the network to focus on similar regions when classifying poisoned images as compared to clean network operating on the clean version of the images. Methods that focus on Grad-CAM based image-reconstruction fail to remove the poison.  }
  \label{fig9_b}
\end{figure*}

Figures \ref{fig9_a} and \ref{fig9_b} show a set of images visualizing the original image, the Grad-CAM for a clean network evaluated on the clean sample, and the Grad-CAM for a poisoned network evaluated on the poisoned sample (left to right) for GTSRB and ImageNet respectively (the network architecture is ResNet18). As shown in the manuscript, the network focus regions are relatively unchanged when the frequency-based poison is applied.

\begin{figure}[t!]
    \centering
    \includegraphics[width=0.5\textwidth]{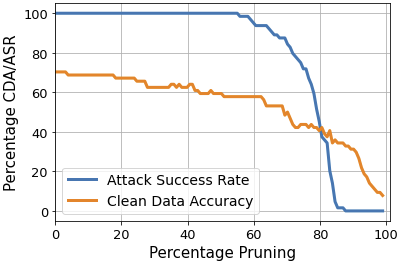}
    \caption{\textbf{Pruning ResNet18 Trained on ImageNet.} Frequency-based backdoors are successfully maintained across high pruning rates that significantly drop the clean data accuracy.   }
    \label{fig9_c}
\end{figure}

Figure \ref{fig9_c} shows the result of pruning a ResNet34 trained on ImageNet. Again, as observed in the manuscript (for CIFAR10), the attack success rate of the frequency-based backdoor is maintained for large pruning rates that highly drop the clean data accuracy.

\begin{figure}[h!]
\centering
    \begin{subfigure}[t]{0.33\textwidth}
      \centering
    \includegraphics[width=\textwidth]{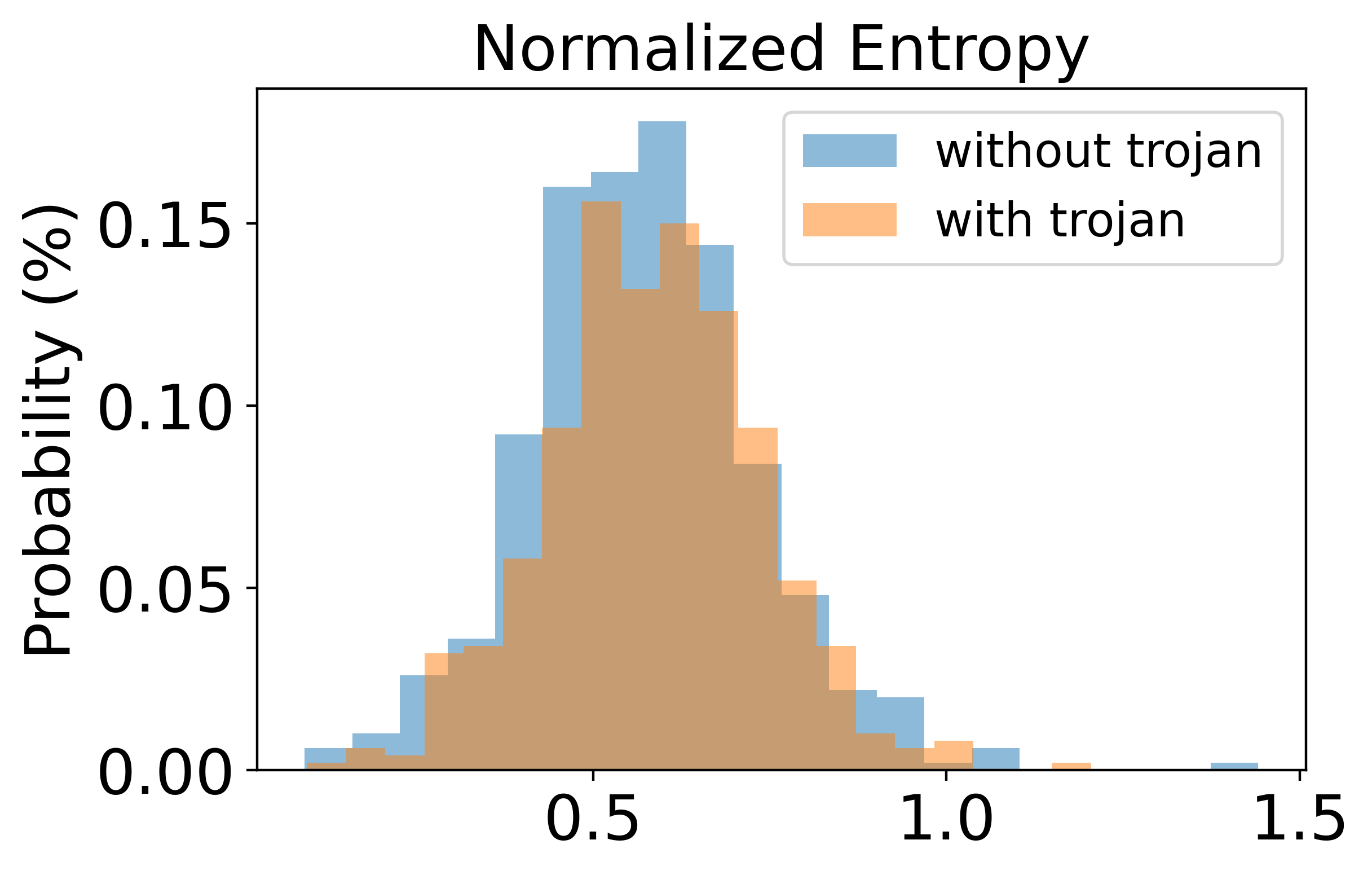}
    \label{fig-b-2}
  \end{subfigure}
  \begin{subfigure}[t]{0.32\textwidth}
    \centering
    \includegraphics[width=\textwidth]{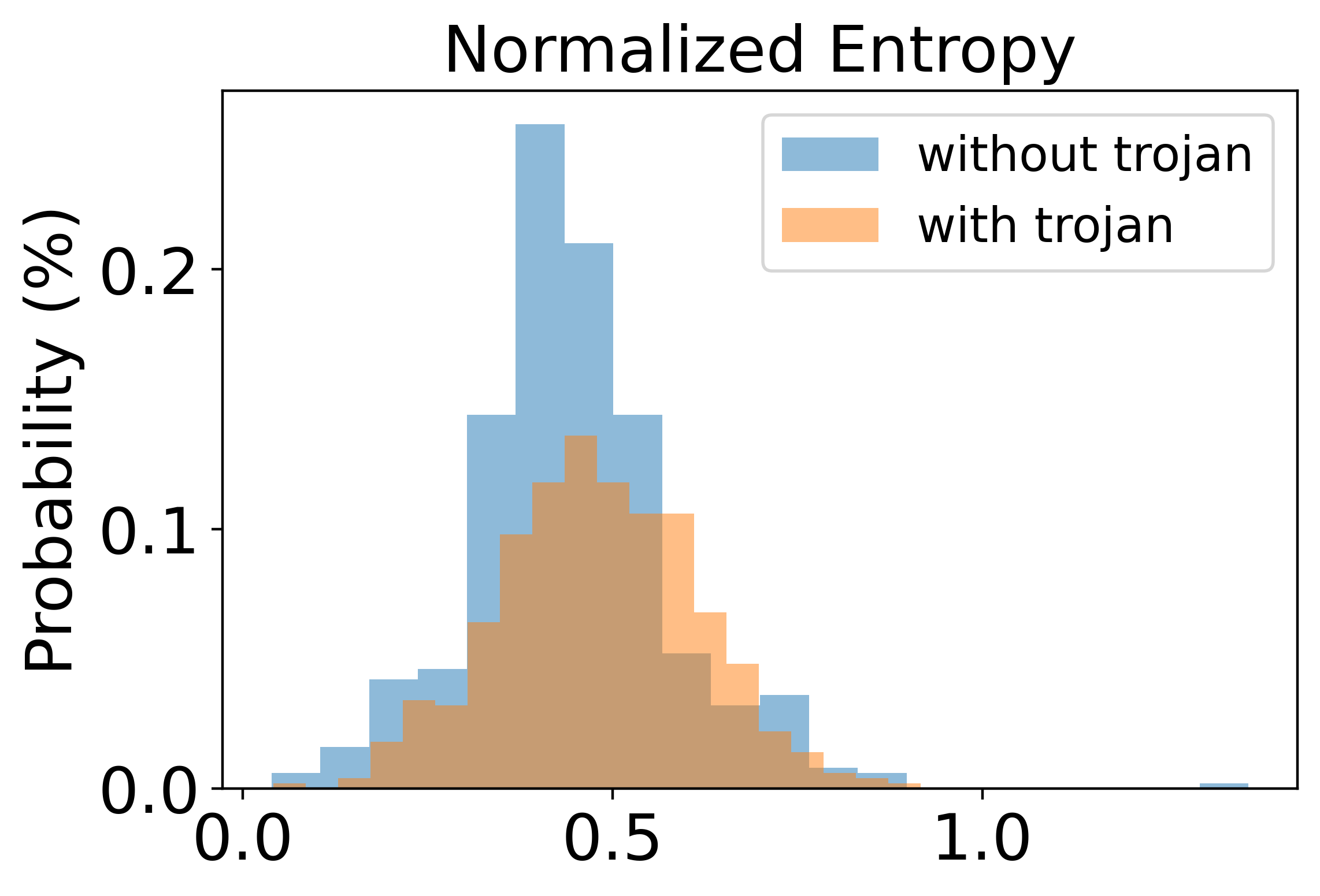}
    \label{fig-a-1}
  \end{subfigure}
  \begin{subfigure}[t]{0.32\textwidth}
      \centering
    \includegraphics[width=\textwidth]{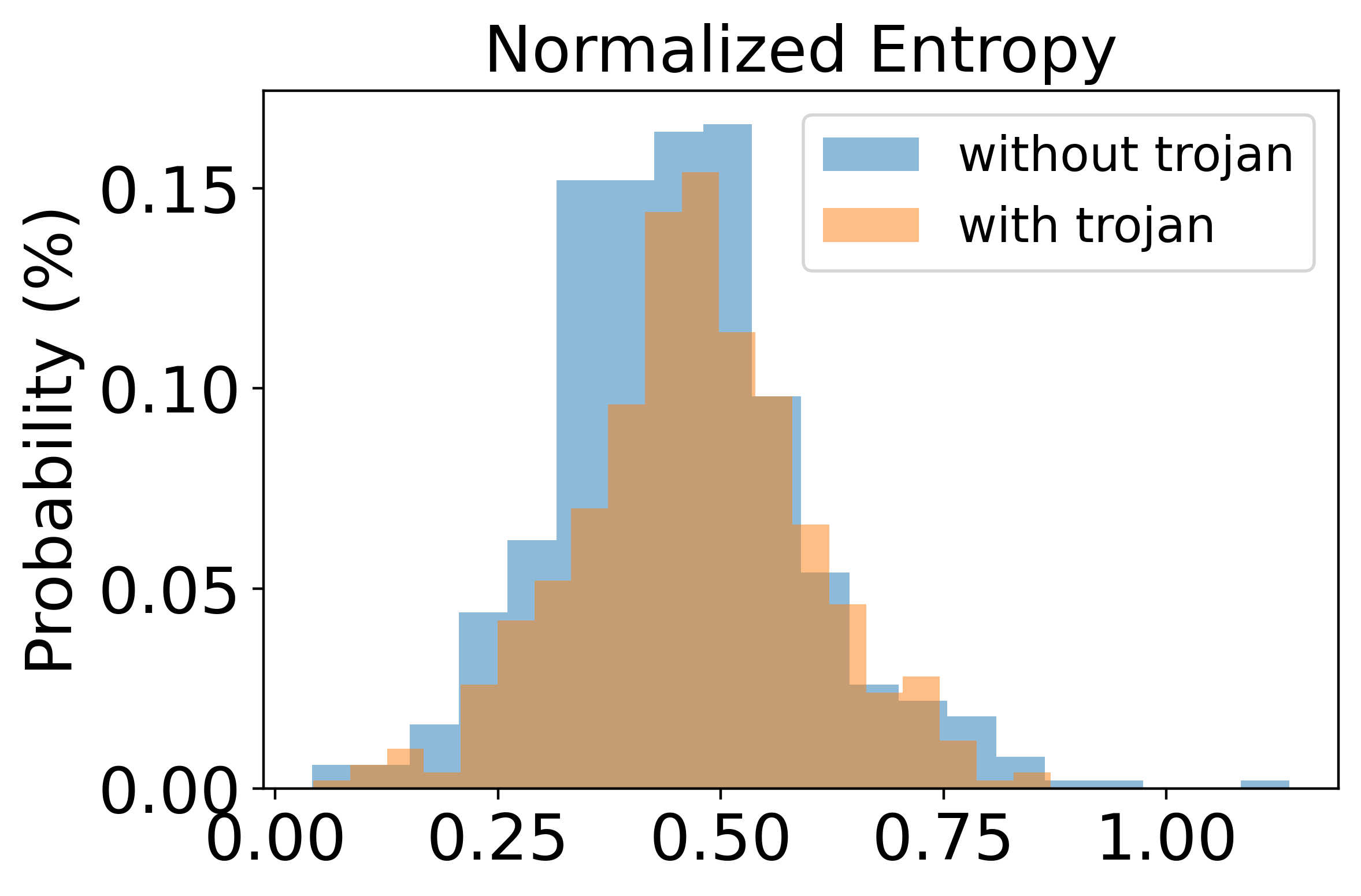}
    \label{fig-a-3}
  \end{subfigure}
  \begin{subfigure}[t]{0.32\textwidth}
      \centering
    \includegraphics[width=\textwidth]{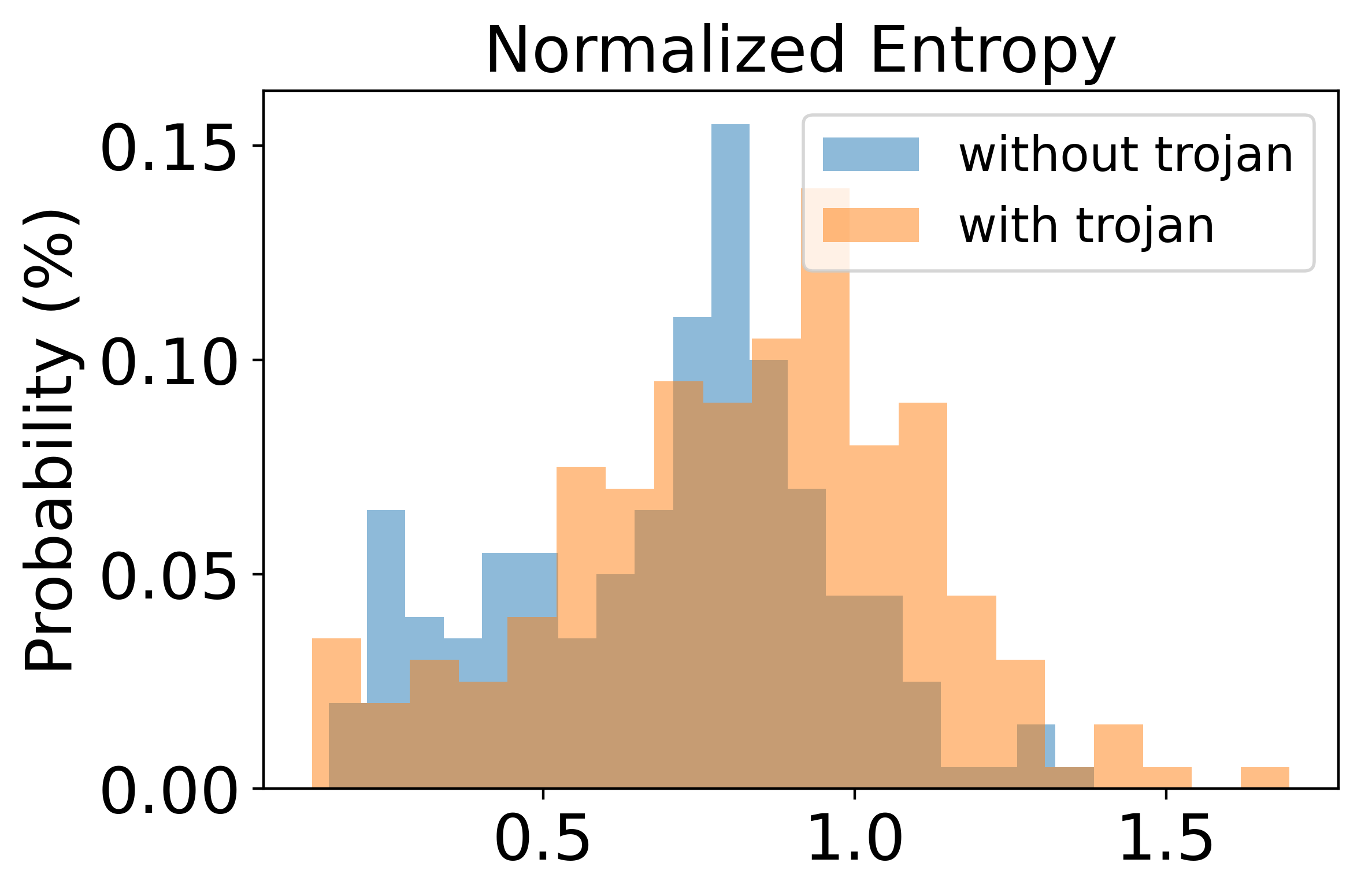}
    \label{fig-b-2}
  \end{subfigure}
  \begin{subfigure}[t]{0.32\textwidth}
    \centering
    \includegraphics[width=\textwidth]{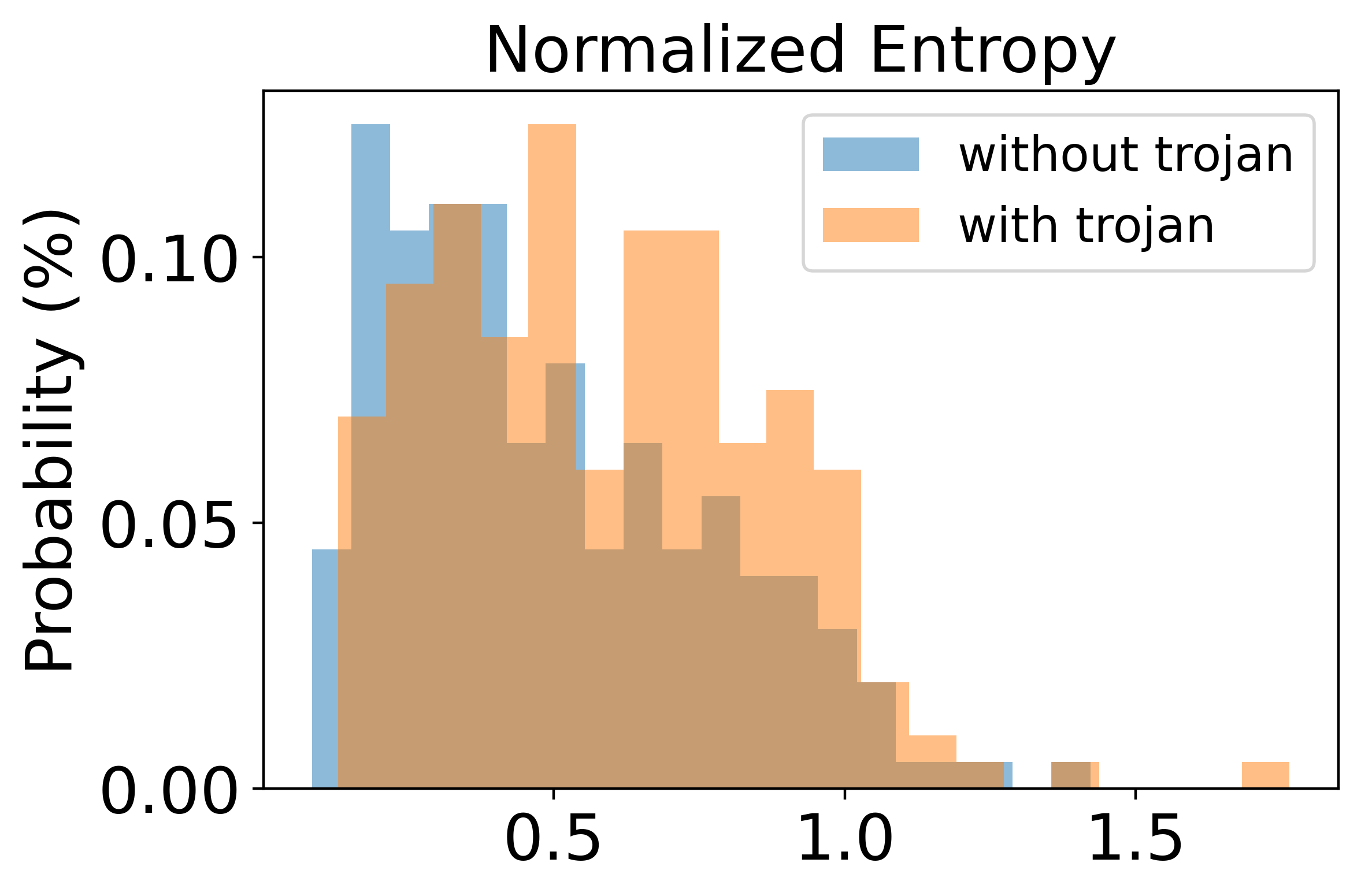}
    \label{fig-a-1}
  \end{subfigure}
  \begin{subfigure}[t]{0.32\textwidth}
      \centering
    \includegraphics[width=\textwidth]{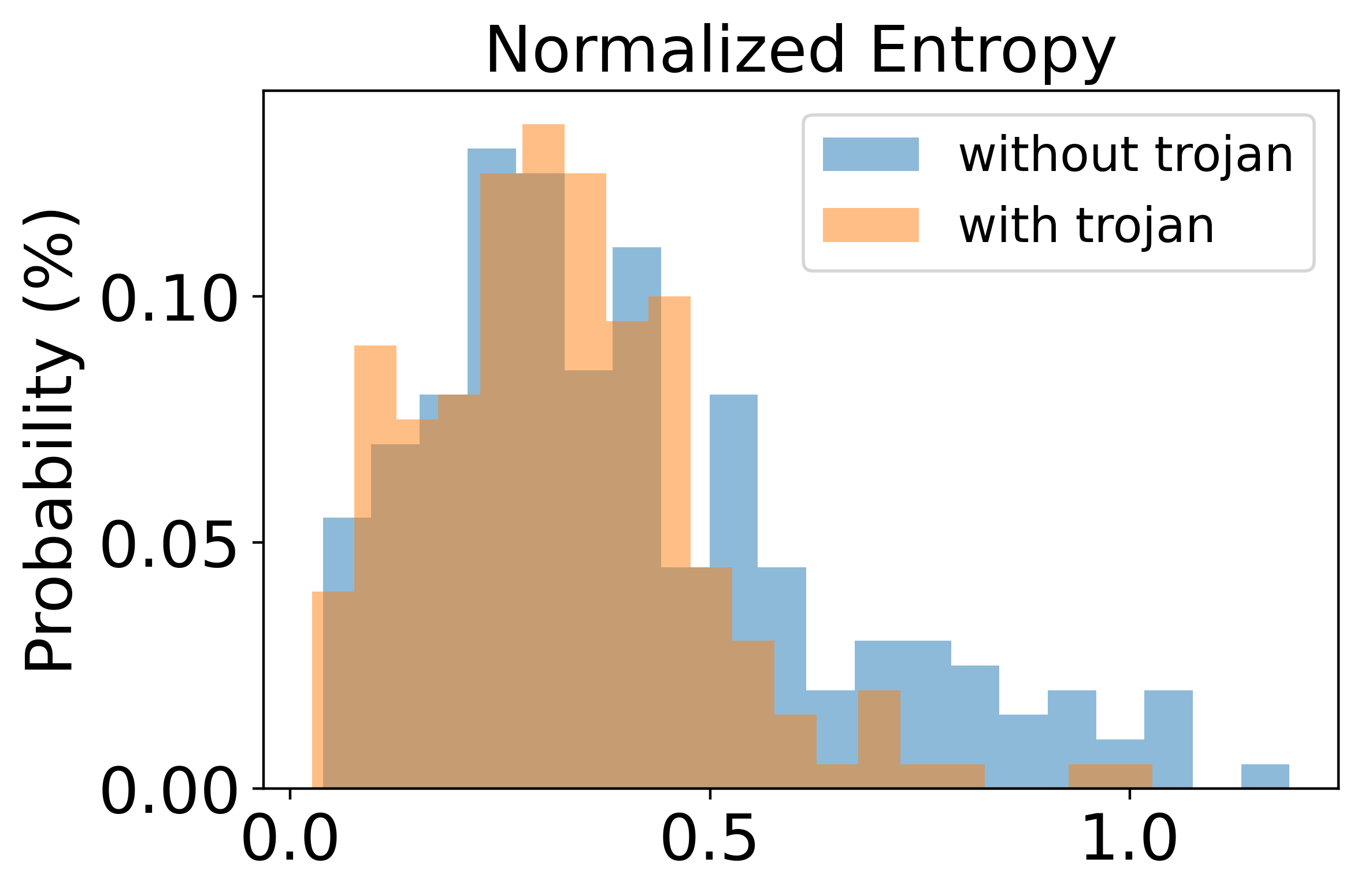}
    \label{fig-a-3}
  \end{subfigure}

  \caption{\textbf{STRIP On Various Datasets.} The proposed frequency backdoor attack is not detectable by STRIP defense mechanism. Rows 1 and 2 show the results for VGG19 trained on CIFAR10 and GTSRB, respectively, with poisoning rates of 0.4\%, 1.0\% and 3.0\% (left to right).   }
  \label{fig9_d}
\end{figure}

Figure \ref{fig9_d} shows the results of applying STRIP defense to a poisoned VGG19 model trained on CIFAR10 and GTSRB with various poisoning rates. Our method causes no significant distributional shift in the prediction vector's entropy therefore is not detectable by STRIP.

\begin{figure}[h!]
\centering
    \begin{subfigure}[t]{0.4\textwidth}
      \centering
    \includegraphics[width=\textwidth]{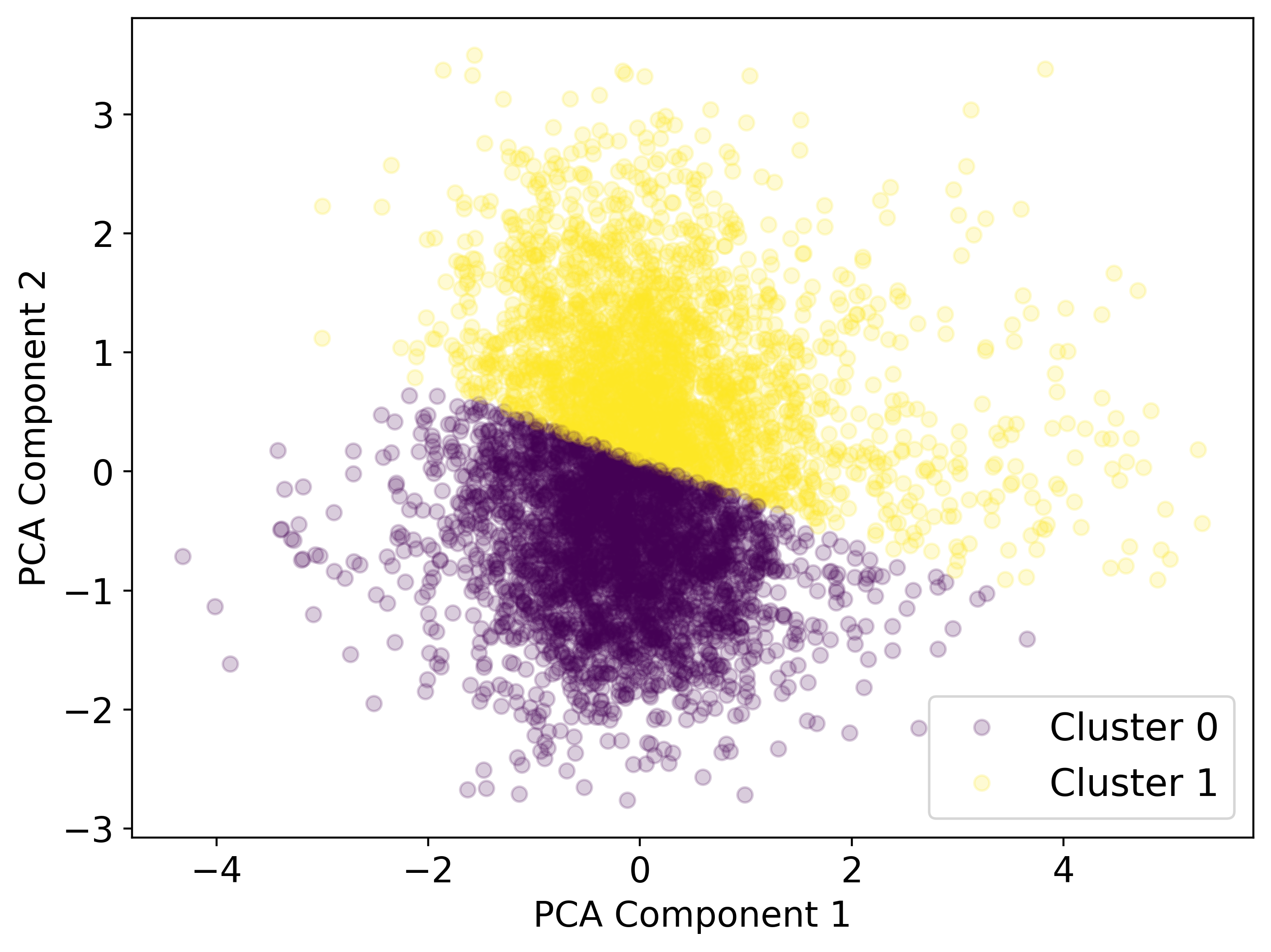}
    \label{fig-b-2}
  \end{subfigure}
  \begin{subfigure}[t]{0.4\textwidth}
    \centering
    \includegraphics[width=\textwidth]{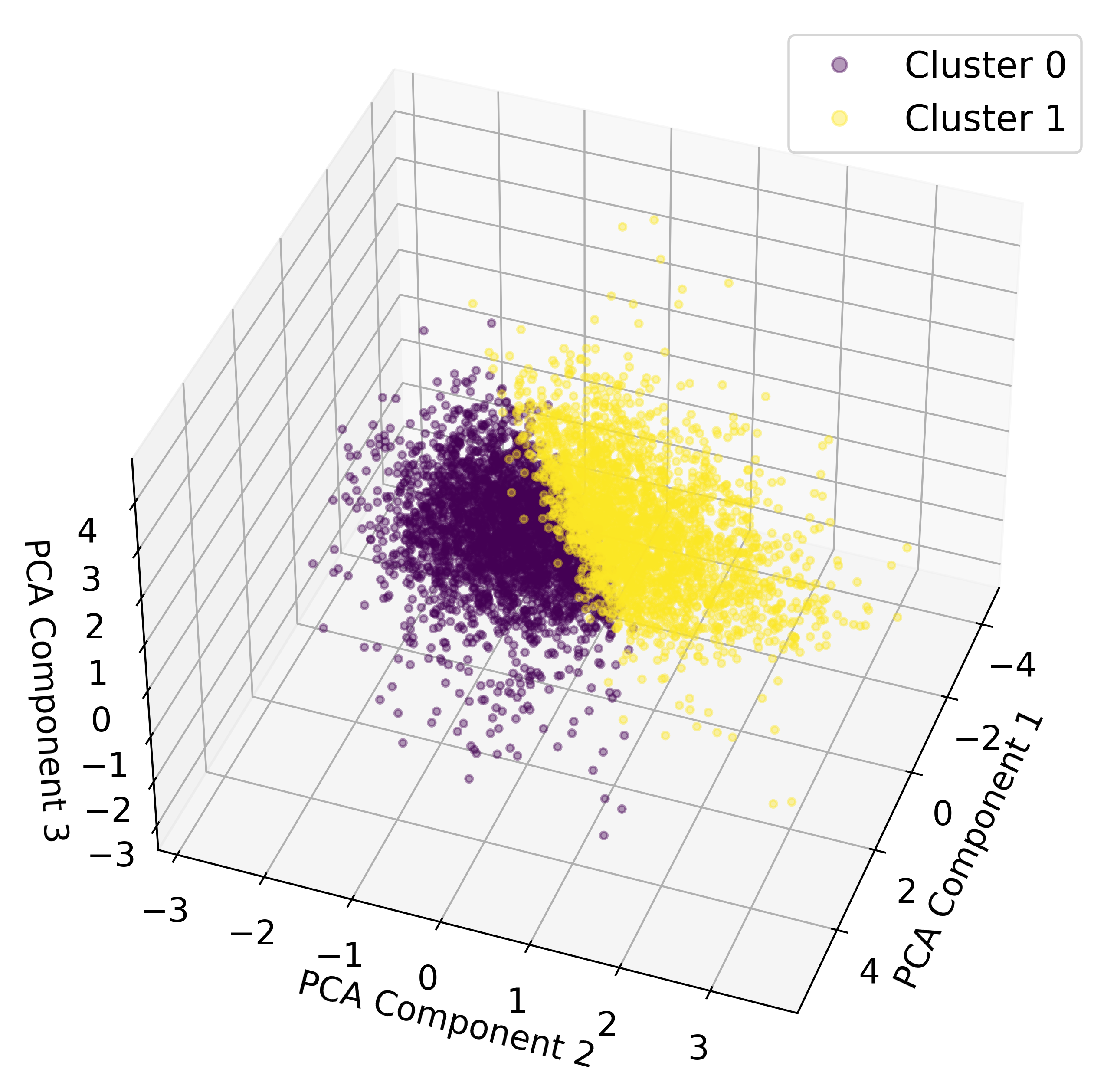}
    \label{222}
  \end{subfigure}

  \caption{\textbf{Activation Clustering on ResNet18 (1.0\% Poisoning Rate).} Activation clustering fails to find two distant clusters in both cases of 2 and 3 principal components.  Under the assumption that less than 50\% of the data is poisoned, we generally consider the smaller cluster as poisonous (in our case no cluster smaller than the other exists for the poisoned class).   }
  \label{fig9_e}
\end{figure}

\begin{figure}[h!]
\centering
    \begin{subfigure}[t]{0.4\textwidth}
      \centering
    \includegraphics[width=\textwidth]{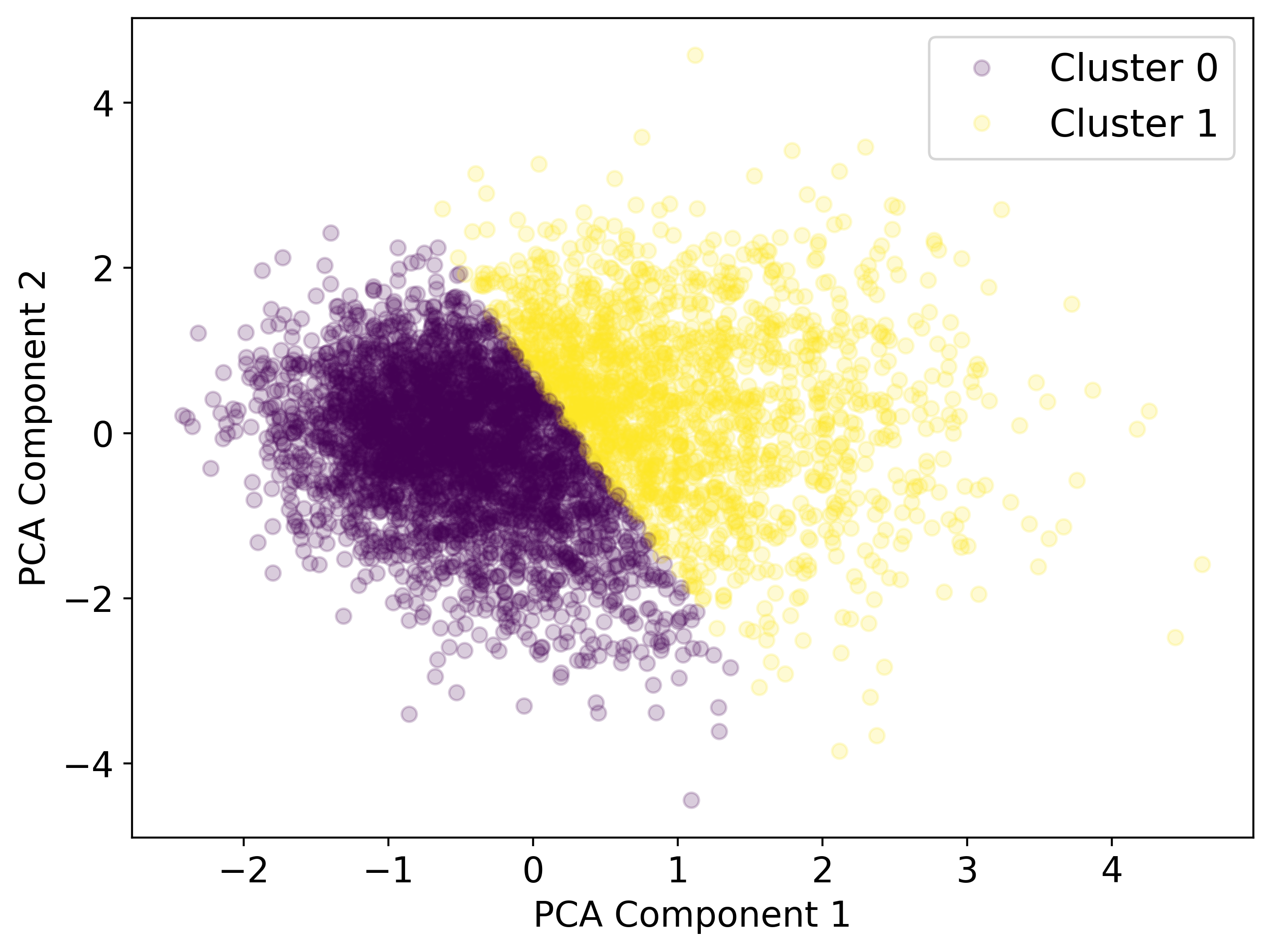}
    \label{fig-b-2}
  \end{subfigure}
  \begin{subfigure}[t]{0.4\textwidth}
    \centering
    \includegraphics[width=\textwidth]{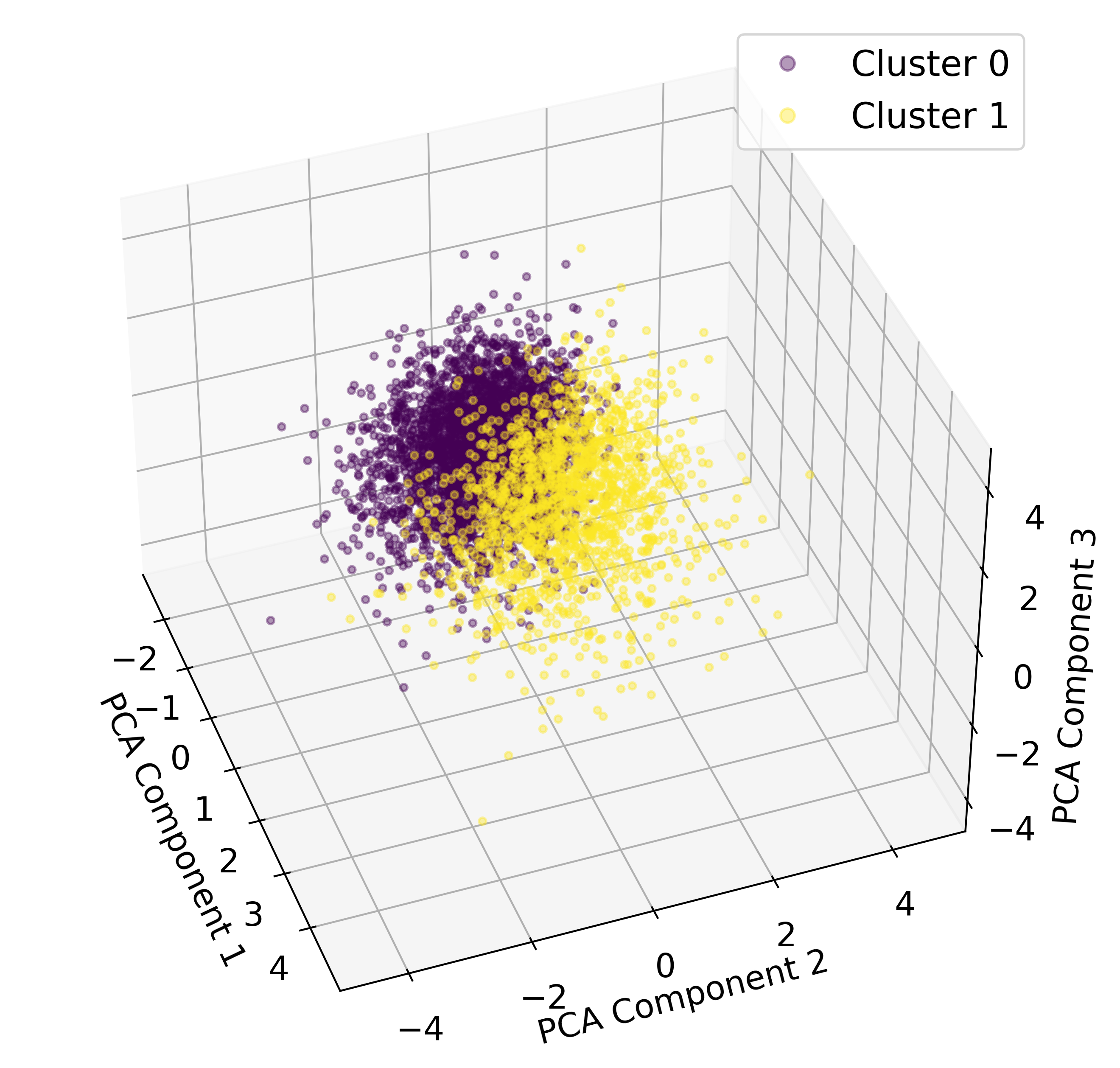}
    \label{222}
  \end{subfigure}

  \caption{\textbf{Activation Clustering on ResNet34 (0.4\% Poisoning Rate).} Activation clustering fails to find two distant clusters in both cases of 2 and 3 principal components.  Under the assumption that less than 50\% of the data is poisoned, we generally consider the smaller cluster as poisonous (in our case no cluster smaller than the other exists for the poisoned class).   }
  \label{fig9_e2}
\end{figure}

\begin{figure}[h!]
\centering
    \begin{subfigure}[t]{0.4\textwidth}
      \centering
    \includegraphics[width=\textwidth]{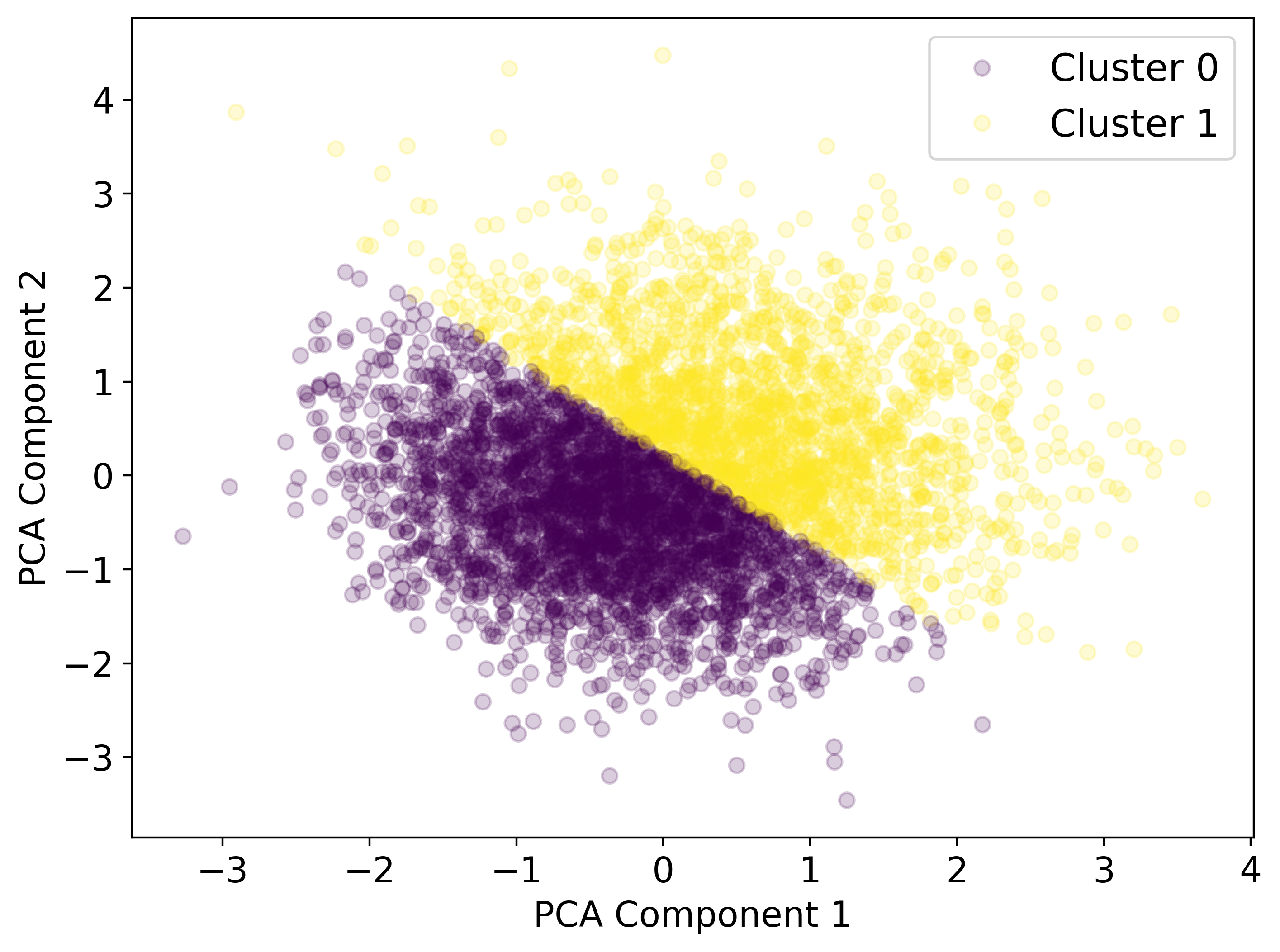}
    \label{fig-b-2}
  \end{subfigure}
  \begin{subfigure}[t]{0.4\textwidth}
    \centering
    \includegraphics[width=\textwidth]{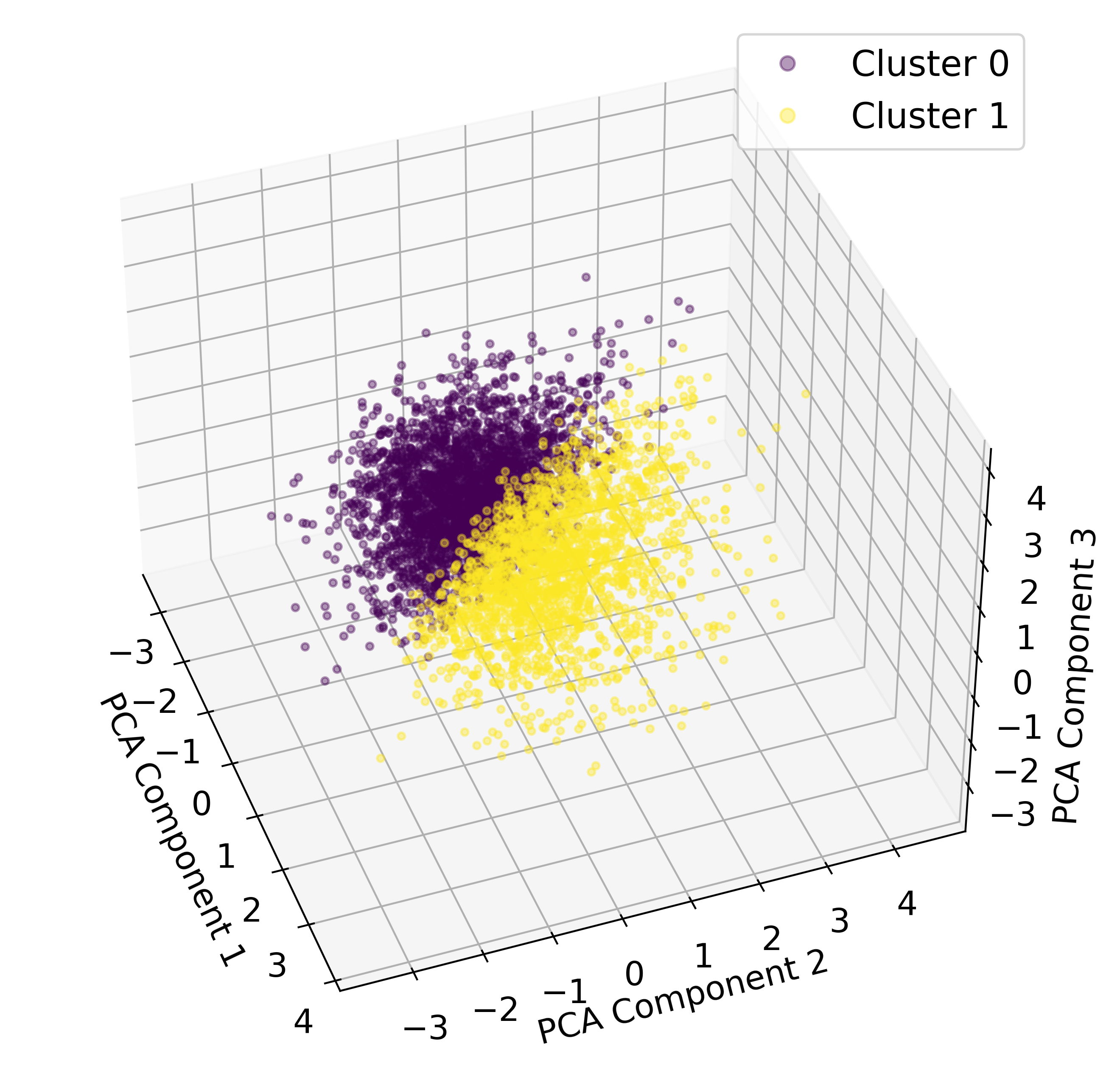}
    \label{222}
  \end{subfigure}

  \caption{\textbf{Activation Clustering on ResNet50 (0.4\% Poisoning Rate).} Activation clustering fails to find two distant clusters in both cases of 2 and 3 principal components.  Under the assumption that less than 50\% of the data is poisoned, we generally consider the smaller cluster as poisonous (in our case no cluster smaller than the other exists for the poisoned class).   }
  \label{fig9_e3}
\end{figure}

Figures \ref{fig9_e}, \ref{fig9_e2}, and \ref{fig9_e3} show the results of Activation Clustering defense method applied to various models with different poisoning rates, namely, ResNet18 (1.0\% poisoning rate), ResNet34 (0.4\% poisoning rate), and ResNet50 (0.4\% poisoning rate) respectively. Visually, AC fails to find two distinct and separable clusters and therefore fails to detect the backdoor attack. Numerically, in terms of silhouette scores, Tables \ref{tablesil}, \ref{tablesil2}, and \ref{tablesil3} show that no score is significantly higher than the other scores for the three considered models.

\begin{figure}[h!]
\centering
    \begin{subfigure}[t]{0.45\textwidth}
      \centering
    \includegraphics[width=\textwidth]{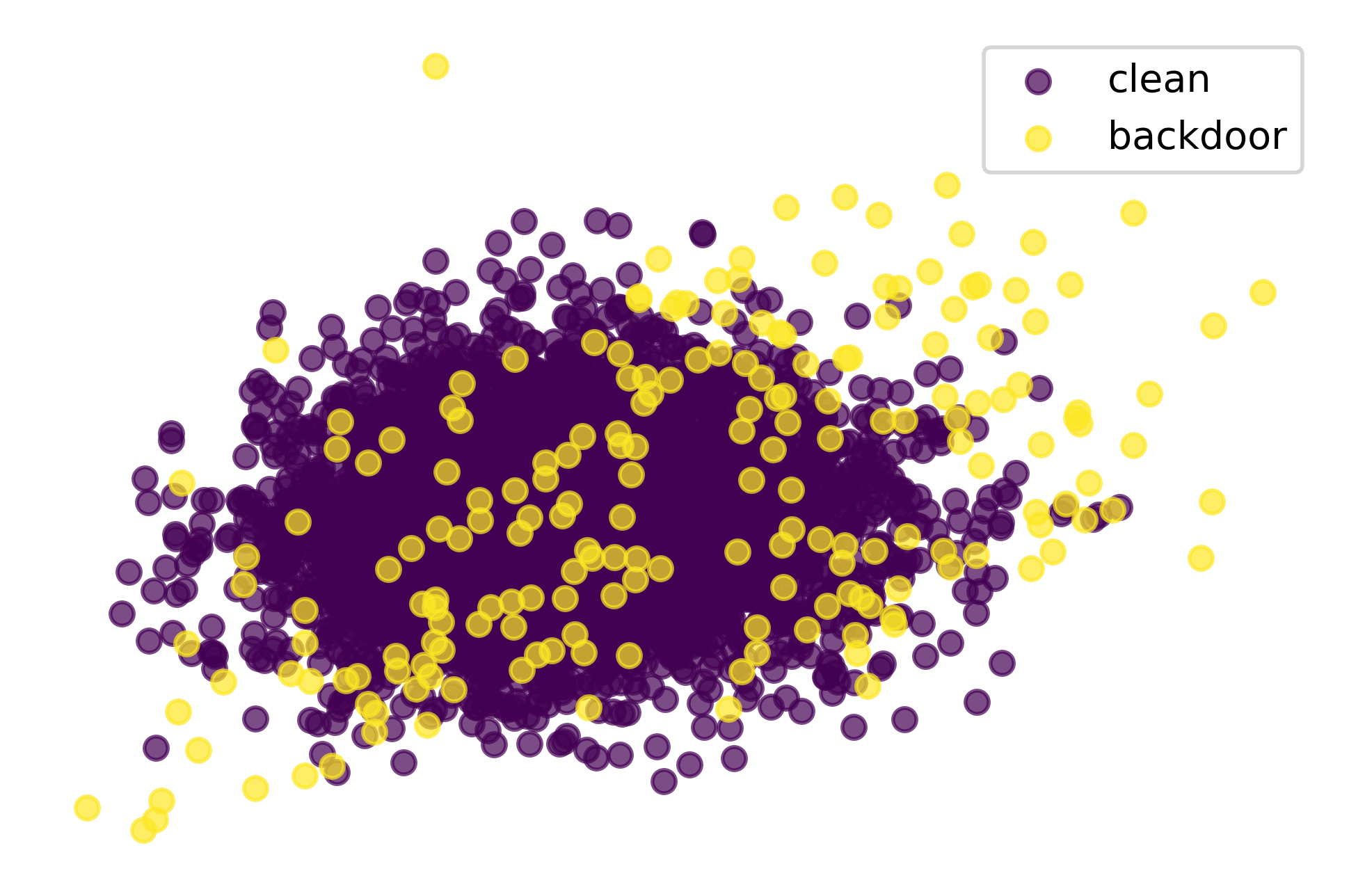}
    \label{123123}
  \end{subfigure}
  \caption{\textbf{Spectral Signatures Defense on DenseNet121 (0.4\% Poisoning Rate).} Spectral Signatures (SS) backdoor defense method fails to find two separate clusters for clean and backdoored samples. The true positive (TP) and false positive (FP) detection rates are 6\% and 7.3\% respectively and hence SS fails to detect our method.}
  \label{fig9_f}
\end{figure}

\begin{figure}[h!]
\centering
    \begin{subfigure}[t]{0.45\textwidth}
      \centering
    \includegraphics[width=\textwidth]{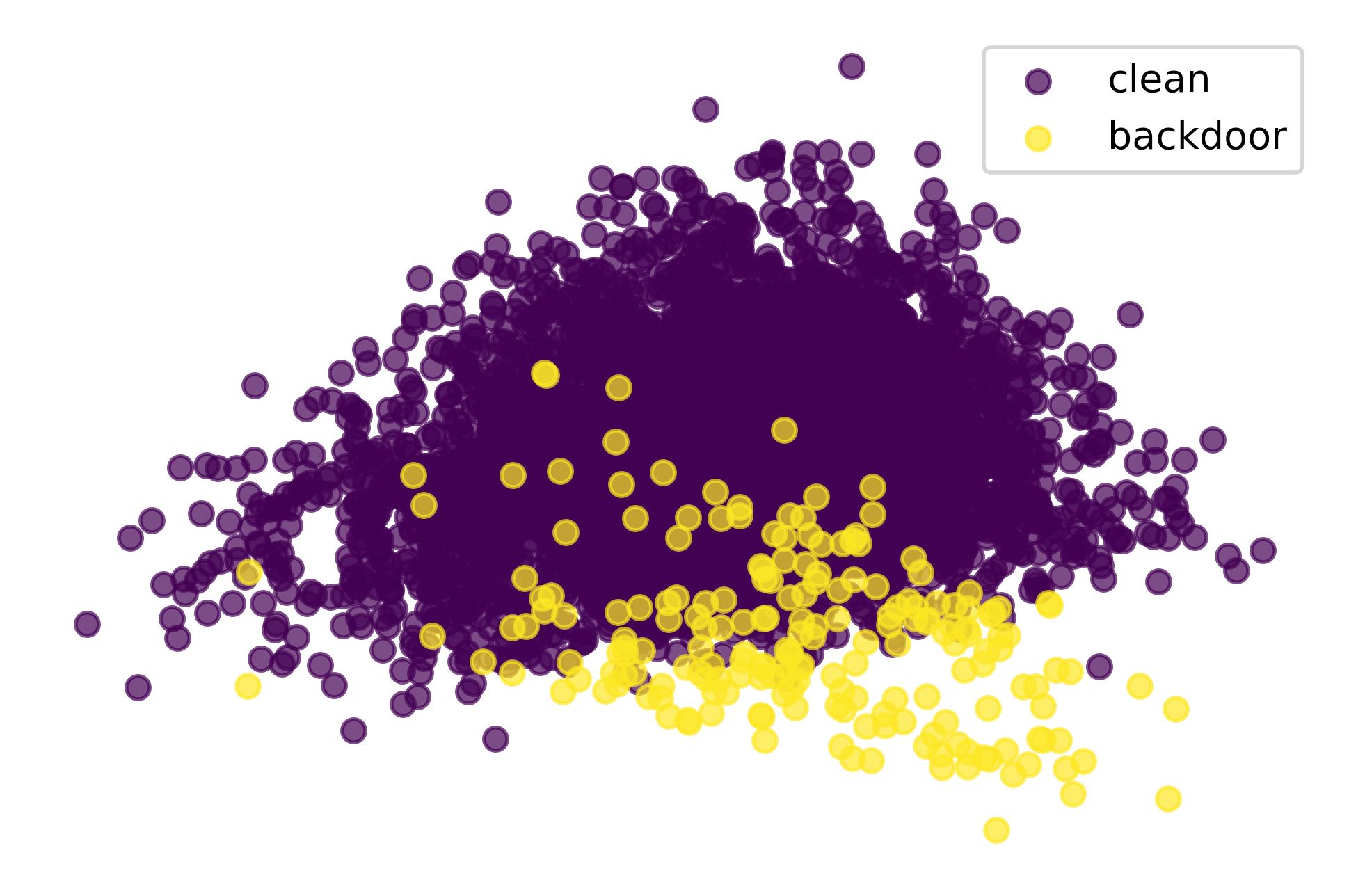}
    \label{123123}
  \end{subfigure}
  \caption{\textbf{Spectral Signatures Defense on ResNet34 (0.4\% Poisoning Rate).} Spectral Signatures (SS) backdoor defense method fails to find two separate clusters for clean and backdoored samples. The true positive (TP) and false positive (FP) detection rates are 45\% and 6.3\% respectively and hence SS fails to detect our method.}
  \label{fig9_f2}
\end{figure}

\begin{figure}[t!]
\centering
    \begin{subfigure}[t]{0.45\textwidth}
      \centering
    \includegraphics[width=\textwidth]{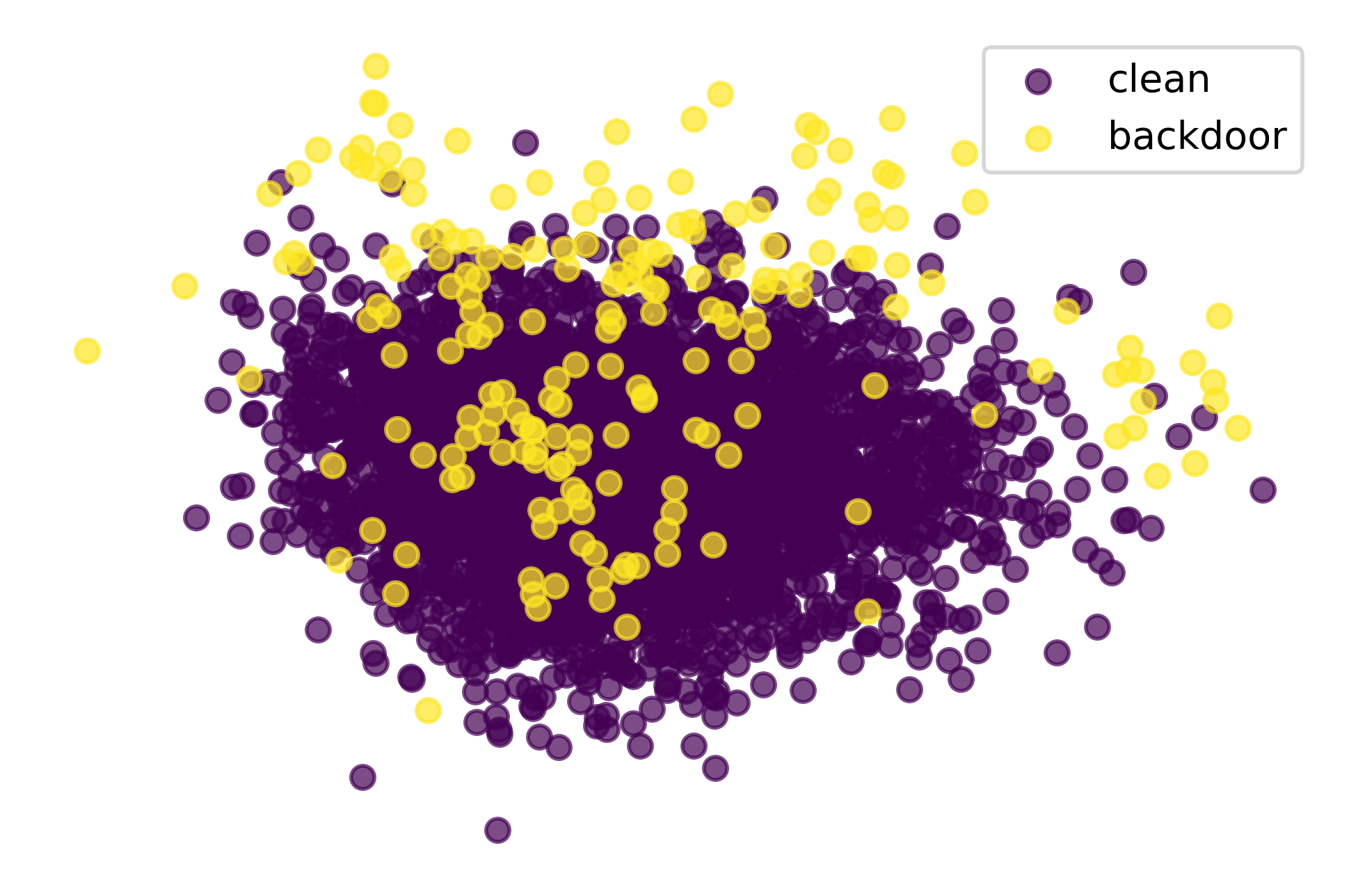}
    \label{123123}
  \end{subfigure}
  \caption{\textbf{Spectral Signatures Defense on ResNet50 (0.4\% Poisoning Rate).} Spectral Signatures (SS) backdoor defense method fails to find two separate clusters for clean and backdoored samples. The true positive (TP) and false positive (FP) detection rates are 33\% and 7.2\% respectively and hence SS fails to detect our method.}
  \label{fig9_f3}
\end{figure}

Figures \ref{fig9_f}, \ref{fig9_f2}, and \ref{fig9_f3} show the results for Spectral Signatures defense method applied to various models with different poisoning rates, namely, DenseNet121 (0.4\% poisoning rate), ResNet34 (0.4\% poisoning rate) and ResNet50 (0.4\% poisoning rate) respectively. Visually, the method fails to find spectrally separable clusters and therefore the backdoor is not detected. Numerically, the true positive rates (6\%, 45\%, and 33\%) are lower than the threshold (90\%) \cite{Tran2018SpectralSI} required for the defense to be deemed successful.

\ 

\clearpage

\begin{table}[h!]
\centering
\scalebox{0.85}{
\begin{tabular}{|l|l|l|l|l|l|l|l|l|l|l|}
\hline
\textbf{Label}            & \textbf{0} & \textbf{1} & \textbf{2} & \textbf{3} & \textbf{4} & \textbf{5} & \textbf{6} & \textbf{7} & \textbf{8} & \textbf{9} \\ \hline
\textbf{Silhouette Score} & 0.322      & 0.343      & 0.319      & 0.318      & 0.321      & 0.328      & 0.321      & 0.363      & 0.337      & 0.326      \\ \hline
\end{tabular}}
\vspace{10pt}
\caption{\textbf{Activation Clustering on ResNet18 (1.0\% Poisoning Rate - 2 PCA Components).} Silhouette scores indicate how well the clustering fits the data. The higher the score the better the clusters fit the data. AC's silhouette scores on our method are similar hence it fails to detect the backdoor.}
\label{tablesil}
\end{table}

\begin{table}[h!]
\centering
\scalebox{0.85}{
\begin{tabular}{|l|l|l|l|l|l|l|l|l|l|l|}
\hline
\textbf{Label}            & \textbf{0} & \textbf{1} & \textbf{2} & \textbf{3} & \textbf{4} & \textbf{5} & \textbf{6} & \textbf{7} & \textbf{8} & \textbf{9} \\ \hline
\textbf{Silhouette Score} & 0.321      & 0.357      & 0.321      & 0.319      & 0.325      & 0.313      & 0.311      & 0.324      & 0.346      & 0.347      \\ \hline
\end{tabular}}
\vspace{5pt}
\caption{\textbf{Activation Clustering on ResNet34 (0.4\% Poisoning Rate - 2 PCA Components).} Silhouette scores indicate how well the clustering fits the data. The higher the score the better the clusters fit the data. AC's silhouette scores on our method are similar hence it fails to detect the backdoor.}
\label{tablesil2}
\end{table}

\clearpage

\begin{table}[h!]
\centering
\scalebox{0.85}{
\begin{tabular}{|l|l|l|l|l|l|l|l|l|l|l|}
\hline
\textbf{Label}            & \textbf{0} & \textbf{1} & \textbf{2} & \textbf{3} & \textbf{4} & \textbf{5} & \textbf{6} & \textbf{7} & \textbf{8} & \textbf{9} \\ \hline
\textbf{Silhouette Score} & 0.307      & 0.343      & 0.320      & 0.309      & 0.311      & 0.314      & 0.313      & 0.329      & 0.329      & 0.325      \\ \hline
\end{tabular}}
\vspace{5pt}
\caption{\textbf{Activation Clustering on ResNet50 (0.4\% Poisoning Rate - 2 PCA Components).} Silhouette scores indicate how well the clustering fits the data. The higher the score the better the clusters fit the data. AC's silhouette scores on our method are similar hence it fails to detect the backdoor.}
\label{tablesil3}
\end{table}

\clearpage

\clearpage

\hfill

\hfill

\clearpage

\section{Evaluation of Proposed Defenses on Other Frequency Attacks}
\label{sec92}

Table \ref{freqdefense} shows the clean data accuracy and the attack success rate of models poisoned by our method, FIBA and FTrojan. The defense shows promising results against two of three frequency backdoor attacks. 
\begin{table}[h]
\centering
\scalebox{0.8}{
\begin{tabular}{@{}ccccc@{}}
\toprule
                                  & \textbf{}    & \textbf{No Defense} & \textbf{JPEG Compression} & \textbf{Autoencoder} \\ \midrule
\multirow{2}{*}{\textbf{FIBA}}    & \textbf{CDA(\%)} & 92.51 & 83.80                     & 82.33                \\
                                  & \textbf{ASR(\%)} & 96.54 & 92.85                     & 71.43                \\ \midrule
\multirow{2}{*}{\textbf{FTrojan}} & \textbf{CDA(\%)} & 92.84 & 85.05                     & 0.00                 \\
                                  & \textbf{ASR} & 100.00 & 82.05                     & 0.00                 \\ \midrule
\multirow{2}{*}{\textbf{Ours}}    & \textbf{CDA(\%)} & 94.38 &  86.91                     & 0.00                 \\
                                  & \textbf{ASR(\%)} & 99.44 &  83.15                     & 0.00                 \\ \bottomrule
\end{tabular}}
\vspace{10pt}
\caption{\textbf{Proposed Frequency Backdoor Defenses:} The proposed backdoor defenses, JPEG compression and autoencoder, could break FTrojan and the proposed method. FIBA poisons low-frequency content that is usually not removed through either compression techniques. The results are reported for CIFAR10 - ResNet18 model.}
\label{freqdefense}
\end{table}
\section{Comparing Backdoor Attacks}
\label{sec11}

Table \ref{table:compare} compares the proposed method with other spatial backdoor attacks. As observed, our method surpasses the performance of spatial backdoor attacks in clean data accuracy, attack success rate, and invisibility metrics (PSNR and LPIPS). Least Significant Bit (LSB) attack is not included in the comparison as it fails to achieve a high attack success rate and hence fails to create a backdoor attack in the first place.

\begin{table*}[h!]
\scalebox{0.55}{
\begin{tabular}{cc|c|c|c|c|c|c|c|c}
\hline
\multicolumn{1}{l|}{\textbf{Metric}}                   & \textbf{Ratio} & \textbf{BadNets} & \textbf{Blend} & \textbf{SIG} & \textbf{Refool} & \textbf{SPM}   & \cellcolor[HTML]{F2D7D5} \textbf{LSB}    & \textbf{Poison Ink} & \textbf{Ours} \\ \hline
\multicolumn{1}{l|}{\multirow{3}{*}{\textbf{CDA/ASR}}} & 3\%            & 87.38 / 66.55      & 89.89 / 89.39    & 89.74 / 99.23  &  89.20 / 87.16     & 88.89 / 58.53    &88.18 / 10.91     & 89.65 / 94.22         &     \textbf{92.31 / 99.43}          \\ \cline{2-10} 
\multicolumn{1}{l|}{}                                  & 5\%            & 87.13 / 65.36      & 89.60 / 90.99    & 89.64 / 99.47  & 89.16 / 89.79     & 88.90 / 57.69    & 86.98 / 11.67     & 89.69 / 93.58         &      \textbf{ 91.88 / 99.88}        \\ \cline{2-10} 
\multicolumn{1}{l|}{}                                  & 10\%           & 85.61 / 68.01      & 89.77 / 93.11    & 89.45 / 99.40  & 88.80 / 92.80     & 89.07 / 57.33    &83.69 / 15.76     & 89.47 / 93.67         &       \textbf{92.10 / 99.97}        \\ \hline
\multicolumn{2}{l|}{\textbf{PSNR$\uparrow$/LPIPS$\downarrow$}}                                & 25.68 / 0.0009     & 21.29 / 0.0240   & 25.12 / 0.0400 & 19.38 / 0.0397    & 38.94 / 0.0001 &51.13 / 0.00001& 42.95 / 0.0001      &     \textbf{43.15} / \textbf{0.00001}          \\ \hline
\end{tabular}}
\vspace{5pt}
\caption{\textbf{Comparison between the Proposed Attack and Backdoor Attacks in the Literature.} Our proposed frequency-based technique provides the best trade off as compared to spatial attacks. It achieves SOTA ASR, CDA, PSNR, and LPIPS metrics. The results shown are for VGG19 trained on CIFAR10. The LSB method is dropped as it fails to create a backdoor with good ASR.}
\label{table:compare}
\end{table*}

\section{Learning Capacity vs Poisoning Capabilities} 
\label{sec12}
Based on our experiments (check Table \ref{Table:Main}), a particularly interesting yet expected trend is noticed. Networks like VGG19, which lack any skip connections, tend to be harder to backdoor attack. This is because the poison information dilutes as we move deeper and deeper in the network architecture. Low norm invisible attacks tend to be particularly influenced by this, and hence, 
non-residual networks require a higher poisoning rate for embedding a backdoor. On the other hand, networks like ResNets, WideResNets, and DenseNets seem to be capable of maintaining the poison information through their skip connections and hence can be backdoored with a fairly small amount of poisoned data.

\end{document}